\documentclass[a4paper,12pt,nocenter,sfbold,oneside,noindent,chapterbib,longtable]{thesis}%
\usepackage{graphicx}
\usepackage{amsmath}%
\usepackage{amsfonts}%
\usepackage{amssymb}
\title{Torsion Balance Investigation of the Casimir Effect}

\begin{document}
\pagestyle{empty}


\begin{center}
{\huge \bf \textsf{Torsion Balance Investigation of the Casimir
Effect}}
\end{center}
\vspace{1.0cm}
\begin{center}
\textsf{A thesis}

\vspace{-0.3cm}%
{\large \textsf{submitted for the Degree of}}

{\LARGE \textsf{Doctor of Philosophy}}
\end{center}

\begin{center}
\textsf{in}

\vspace{-0.3cm}%
{\large \textsf{The Faculty of Science}}

\textsf{{\LARGE Bangalore University}}
\end{center}
\vspace{2cm}
\begin{center}
{\large \textsf{by}}
\end{center}
\begin{center}
{\LARGE \textsf{G RAJALAKSHMI}}
\end{center}

\vspace{3cm}
\begin{figure}[h]
\centerline{\includegraphics[width=3cm]{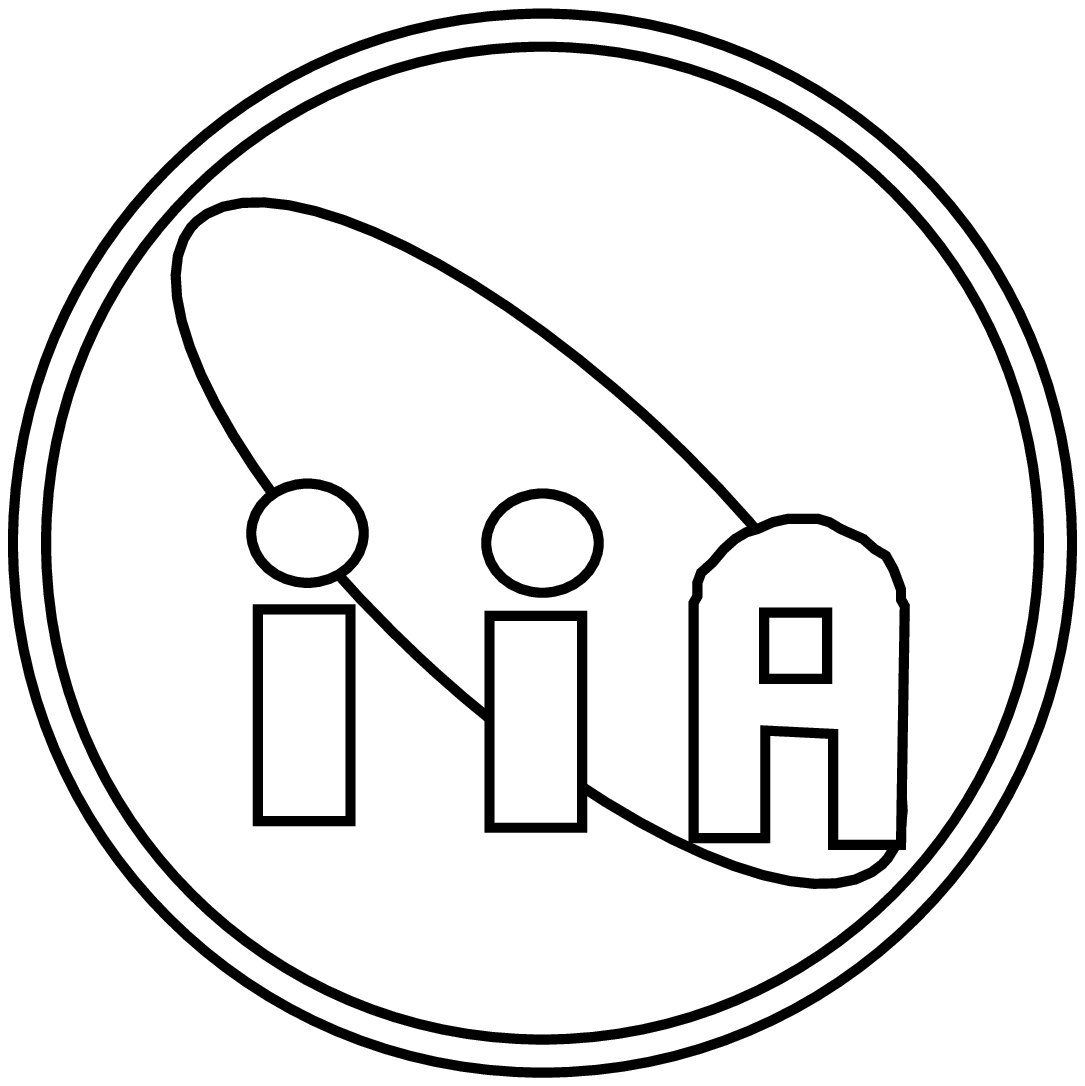}}
\end{figure}

\begin{center}
{\LARGE \textsf{Indian Institute of Astrophysics}}

{\Large \textsf{Bangalore 560 034, India}}
\end{center}
\begin{center}
{\Large \textsf{May 2004}}
\end{center}



\begin{center}
{\Large \bf Declaration}
\end{center}

\vspace{2cm}%
I hereby declare that the matter contained in this thesis is the
result of the investigations carried out by me at the Indian
Institute of Astrophysics, Bangalore, under the supervision of
Prof. R. Cowsik. This work has not been submitted for the award of
any degree, diploma, associateship, fellowship etc. of any
university or institute.

\vspace{2.0cm}
\begin{center}
\begin{tabbing}
space............................... \hspace*{2.5in} :\=     \kill
\>  \\
\>  \\
\>  \\
Prof. R. Cowsik     \>   G Rajalakshmi       \\
(Thesis Supervisor) \>  (Ph.D. Candidate)      \\
\end{tabbing}

Indian Institute of Astrophysics

Bangalore 560 034, India

May, 2004
\end{center}


\begin{center}
{\Large \bf Certificate}
\end{center}

\vspace{2cm}

This is to certify that the thesis entitled ` Torsion Balance
Investigation of the Casimir Effect' submitted to the Bangalore
University by Ms. G. Rajalakshmi for the award of the degree of
Doctor of Philosophy in the faculty of Science, is based on the
results of the investigations carried out by her under my
supervision and guidance, at the Indian Institute of Astrophysics,
Bangalore. This thesis has not been submitted for the award of any
degree, diploma, associateship, fellowship etc. of any university
or institute.\\ \\ \vspace{2cm}

\begin{tabular}{lc}
Indian Institute of Astrophysics,&~~~~~~~~~~~~~~~~~~Prof. R Cowsik\\
Bangalore                       &~~~~~~~~~~~~~~~~~~(Thesis Supervisor)\\
May, 2004                  &                                     \\
\end{tabular}

\pagenumbering{roman}

\begin{center}
{\Large \bf Acknowledgements}
\end{center}

I thank Prof. R. Cowsik for his guidance and continuous support as
my thesis advisor during the course of this work. I have learnt
many aspects of research from discussion with him.

I thank Dr. C. S. Unnikrishnan for his support and encouragement
throughout - as a colleague and as a friend. His never ending
stream of crazy-sounding ideas are a constant source of
inspiration.

I have gained a lot from discussions with Dr. N. Krishnan and
Prof. S. N Tandon. Their open criticisms have put me on the right
track many a times.

I thank Prof. R. Srinivasan, for the technical help and support he
provided in building the experiment, especially the CCD optical
lever.

I thank Dr. B. R. Prasad and Dr. Pijush Bhattacharjee, who as
doctoral committee members helped in advancing the thesis.

I thank Prof. B. P. Das, Dr. Sharath Ananthamurthy, Prof. C.
Sivaram, Dr. A. K. Pati and Dr. Andal Narayanan for the
interesting discussions I have had with them.

I thank Suresh for his patient support and companionship during
the years it took to build the experiments.

I thank Ms. A. Vagiswari, Ms. Christina Birdie and all the other
library staff for the excellent library facility provided at the
Institute.

I thank the present and past members of the Board of Graduate
Studies for their help and guidance.

I thank Mr. J. P. A. Samson, Mr. Thimmiah, Mr. Sagayanathan and
Mr. Periyanayagam of the mechanical workshop for their help with
the design and fabrication of the various mechanical parts of the
experimental apparatus.

Building a lab requires support and advice from many people on
aspects ranging from instrumentation to administration, it is
impossible to acknowledge everybody's contribution individually. I
thank all the staff of IIA who helped in the setting up of the
`Zerolab'. I thank the staff of the computer center, electrical
section, photonics and the administration for their continuous
support.

I thank the present and past Chairmen of the Physics Department of
Bangalore University and the members of staff in the University
for their help with the administrative matters related to the
thesis.

 The several years I have spend in
IIA was made pleasant by the many wonderful friends that I was
lucky to have. I thank - Ramesh, Krishna, Sankar, Sivarani,
Geetha, Sridharan and Rajesh Kumapuaram for introducing me into
IIA; Mogna and Holger for being there during hard times; Arun,
Rana and Pratho for those numerous discussion - scientific and
otherwise; Dharam for being a nag and for the care; Manoj for the
enumerable arguments on wide ranging topics; Appu for being such a
patient listener; Ravi for help in a many a things- personal and
technical; Preeti, Latha, Sivarani and Shalima for those many
early morning chais; Geetanjali for being a wonderful roommate.  I
also thank E Reddy, Sujan, Annapurani, Srikanth, Rajguru, Pandey,
Swara, Bhargavi, Sanjoy, Pavan, Mangala, Jana, Ramachandra,
Ambika, Shanmugham, Kathiravan, Reji, Sahu, Malai, Jayendra,
Nagaraj and Maiti for making my stay at IIA enjoyable.

I thank Pramila, Rani and Samson for their hospitality and care.

I thank my loving parents whose guidance and encouragement has
made me what I am. My brother, Sankar for his silent support and
affection. I thank Raja for enduring me through these years and
being a constant source of  encouragement and companionship. Athai
and Athember for their affection and support. Last but not the
least, I thank my grand fathers - Chandrasekharan,
Ananthanarayanan and Krishnamurthy - who at a very young age
guided me in developing a personality of my own. To them I
dedicate this thesis.

\pagestyle{headings}

\tableofcontents%
\clearpage %
\setcounter{page}{0}%
\pagestyle{headings}%
\pagenumbering{arabic}



\chapter{Introduction}

\emph{Abstract:  This chapter presents an overview of the
theoretical background and the motivations for the experiment. The
chapter begins with a general description of Casimir force, with
discussions on the effect of finite temperature and finite
conductivity. This is followed by a short historical review of the
earlier experiments to measure Casimir force. The recent
motivations to study Casimir force are then presented. }

\stepcounter{footnote}

\section{Casimir Force - an introduction }

    Even Schwinger, who was awarded the Nobel Prize for his pioneering
contribution to Quantum Electrodynamics, has remarked that one of
the least intuitive consequences of Quantum electrodynamics is the
existence of a force of attraction between two perfectly
conducting uncharged plates \cite{Schwin78}. In 1948, H. B.
Casimir~\cite{Cas48}, showed that for two infinite parallel
plates, separated by a distance $d$, this force per unit area is
given by,
\begin{eqnarray}
F_c(d)& = &-\frac{\pi^2 \hbar c}{240d^4} \label{cas_pres}\\
      & = & -\frac{0.013}{d_\mu^4} \quad \mathrm{dyn.\ cm}^{-2}\ \
    \mathrm{where} \ d_\mu \equiv d\ \mathrm{in\ microns}
\end{eqnarray}
     The force is independent of the charge or mass of the
plate. For plates of $1$ cm$^2$ area separated by $1~\mu$m, the
force is comparable to the gravitational attraction of two $400$~g
masses separated by $1$~cm or the Coloumb force on the electron in
a hydrogen atom.

\subsection{Casimir force as a manifestation of zero point energy}
  When electromagnetic field is described quantum mechanically, it
has properties similar to an assembly of quantized harmonic
oscillators. Each mode of the electromagnetic field defined by a
set of parameters like frequency and polarization, is represented
by one oscillator. The allowed energy levels of an electromagnetic
wave of angular frequency $\omega$ are given by the Planck
relation $E_n\ =\ (n+\frac{1}{2})\hbar \omega, n\ =\ 0,1,2,3
\ldots$ . The integer $n$, for the electromagnetic field,
corresponds to the number of photons. Thus, even the zero quanta
or the `vacuum' state of the field still contains field
fluctuations which result in a non-zero energy. In free space, all
modes (frequencies) of the electromagnetic wave are possible and
each mode has a finite energy. As a consequence the vacuum or the
zero photon state of the electromagnetic field in free space has
infinite energy and infinite energy density. However, physically
real effects arising from quantum fluctuations in the vacuum turn
out to be finite and``renormalized". One example is the Lamb
shift of atomic energy levels. \\
\begin{figure}
\begin{center}
\includegraphics{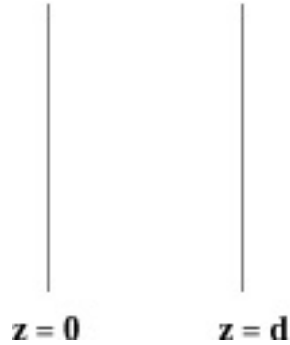}
\end{center}
\caption{Two perfectly conducting infinite plates, placed distance
d apart.} \label{plates}
\end{figure}

These vacuum field fluctuations also give rise to measurable
mechanical effects on macroscopic systems which manifest as
Casimir forces~\cite{Rey2002}. If an infinite, planar cavity
comprising of perfect conducting plates separated by a distance,
d, is placed in `vacuum', it imposes boundary conditions on the
zero-point electromagnetic fluctuations.  As a result, the
possible modes of the electromagnetic field is restricted within
the cavity. Thus there is a finite difference in the energy of the
vacuum field outside and inside the cavity. This results in a
quantum vacuum pressure that attracts the cavity plates together.
For the zero photon state, this pressure is given by
Eqn.~\ref{cas_pres} (see Appendix A for details).

\subsection{Effect of Finite Temperature}
    The non-vacuum state of the electromagnetic field also has a
Casimir force associated with it. In general at any finite
temperature, T, thermal fields are also present in addition to the
vacuum electromagnetic fields and the possible energy levels are
given by the Planck's spectrum,

\begin{equation}
E_{n} = \left(  n(\omega) + \frac{1}{2}\right)  \hbar\omega,\ \mathrm{where}%
\ n(\omega) = \frac{1}{e^{\frac{\hbar\omega}{k_{{\tiny B}} T}}-1}.
\label{planck}
\end{equation}
\newline
     The Casimir vacuum pressure (force per unit area) is then
given by~\cite{Miloni94},
\begin{align}
F_{c}^T(d)  &  = -\frac{k_{B} T}{4\pi d^{3}}\sum_{n=0}^{n}\hspace*{-1.0mm}%
{}^{\prime}\int_{nx}^{\infty}\frac{dyy^{2}}{e^{y} - 1}
\quad\mathrm{where}\;
\; x \equiv4\pi k_{B} Td/\hbar c \footnotemark{}\label{Pgen_eqn}\\
F_c(d)  &  \simeq-\frac{\pi^2\hbar c}{240d^4}\quad\text{at\;low}%
\;T\;(i.e.\;x\ll1)\nonumber\\
F_c^T(d)  &  \simeq-\frac{\zeta(3) k_BT}{4\pi d^3}\quad\text{at\;high\;}%
T\;(i.e.\;x\gg1) \label{P_eqn}\\ %
        &\quad \quad \mathrm{with\ }\zeta(3)= 1.20206
\end{align}
\footnotetext{The prime over the summation symbol means that a
factor half should be inserted for the  $n=0$ term. See for
example~\cite{Miloni94}}
     From these results note that \textit{the distance
dependence changes from $1/d^{4}$ at low temperatures to $1/d^{3}$
at high temperatures.}\newline

     Considering Eqn.~\ref{P_eqn}, it is obvious that the
important non-dimensional parameter, that distinguishes the
domains of \textit{high} and \textit{low} temperature, is $x=4\pi
k_{B}Td/\hbar c.$ \emph{Moreover, it is interesting to note that
high temperature also corresponds to larger separation} $d$\emph{\
between the plates and vice versa.} Thus at any given temperature,
the law which governs the vacuum pressure will depend on the
distance between the plates. No experiment till date has been able
to observe these finite temperature corrections to Casimir force,
due to limitations in sensitivity in the distance range where such
effects start to become significant.\\

  The primary aim of the work described in this thesis is to
observe the Casimir force in the distance range where this change
in the distance dependance of the force occurs. At a temperature
of $\sim300^{\circ}$ K, the change from $1/d^{4}$ to $1/d^{3}$ is
expected to occur at about $2~\mu$m to $4~\mu$m. Our experiments
scan separations from about $1~\mu$m to $10~\mu$m and thus will be
able to probe this change over from the low temperature to the
high temperature domain.

\subsection{Effect of Finite Conductivity}
The discussions so far assumed that the cavity plates are
perfectly  conducting, i.e, they have infinite conductivity at all
frequencies of the electromagnetic field.  In experimental
situations, this simplified assumption is unrealistic and the
dielectric properties of the cavity plates should also be
considered.\\

\begin{figure}
\begin{center}
\includegraphics{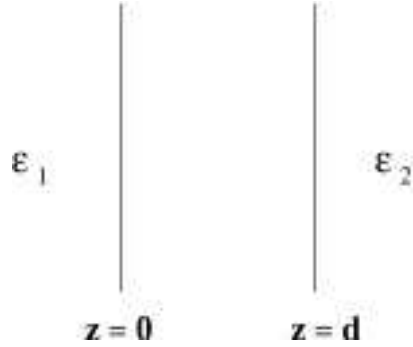}
\end{center}
\caption{Two semi-infinite dielectric slabs of dielectric
constants $\epsilon_1$ and $\epsilon_2$, placed distance $d$ apart
in vacuum.} \label{dielectric2}
\end{figure}

     Liftshitz~\cite{Lift56} , developed the first macroscopic
theory of forces between dielectrics. His results reduce to the
Casimir force given by Eqn.~\ref{cas_pres} for the case of perfect
conductors. For finite size plates with dielectric constants
$\epsilon_1(\omega)$ and $\epsilon_2(\omega)$
[Fig.~\ref{dielectric2}], the Casimir force per unit area is given
by the Liftshitz formula,

\begin{eqnarray}
F_c^p(d)& = &-\frac{\hbar}{2\pi^2c^3}
\int_1^\infty dp\ p^2 \int_{0}^\infty d\xi\  \xi^3  \nonumber \\
&&\left[ \left\{ \frac{ s_1 + \epsilon_1 p}{ s_1 - \epsilon_1 p}.
        \frac{ s_2 + \epsilon_2 p}{ s_2 - \epsilon_1 p}.
        e^{2\xi p d/c} - 1 \right\}^{-1}  \right. \nonumber \\
&& + \left. \left\{ \frac{s_1 + p}{s_1 - p} .\frac{s_2 + p}{s_2 -
p}.
     e^{2\xi p d/c} - 1 \right\}^{-1} \right]
\label{Lift-P(d)}
\end{eqnarray}
where $s_i,\ p$ are variables that depend on the dielectric
constants $\epsilon_i$ of the medium and the wave vector $k$ of
the electromagnetic field. (see Appendix A for details).

\subsection{Importance of these effects for experiments}
The experiments on Casimir force are typically carried out at room
temperature using metals with finite electrical conductivity.  In
order to make a comparison between the experiments and the theory,
it is essential to quantify the effects of temperature and
conductivity. As summarized above the effects of the thermal field
fluctuations on the Casimir force are known to become important
when the spacing between the boundaries is of the order of the
characteristic length, $\lambda_T = \frac{2 \pi}{\omega_T} $.
$\omega_T$ is the dominant thermal angular frequency at the
temperature $T$. Similarly, the plasma frequency $\omega_p$ of the
metal determines the length scale, $\lambda_p$ at which the finite
conductivity effects are appreciable, i.e, $\lambda_p = \frac{2
\pi}{\omega_p}$. \\

    These effects have been calculated by several methods
(see for example, ~\cite{Genet-T, Reynaud2000, Bordag2000}).
Fig.~\ref{Rey} represents one such calculation
~\cite{Reynaud2000}. The figure shows a plot of the correction
factors estimated using,
\begin{eqnarray}
  \eta^T &=& \frac{F_c^T(d)}{F_c(d)} \label{etaT} \\
  \eta^p &=&  \frac{F_c^p(d)}{F_c(d)} \label{etap}\\
  \eta &=& \eta^T \eta^p \label{eta}
\end{eqnarray}
Thus $\eta^T$ and $\eta^p$ are the estimates of the thermal and
the conductivity effects respectively. $\eta$ is the combined
corrections. As is clear from Fig.~\ref{Rey}, the thermal effects
and finite reflectivity of the metals are significant at quite
different distance scales. At the sub-micron distances, the finite
temperature effects are negligible, while the conductivity effects
are considerable. Above a few microns the reflectivity effects are
small as compared to the appreciable thermal effects. Thus, to
study the effects due to finite temperature on the Casimir force,
experiments need to be performed in the distance range of a few
microns.
\begin{figure}
  \begin{center}
  \resizebox{!}{7cm}{
  \includegraphics*[11.6cm,14.2cm][19.3cm,20.3cm]
  {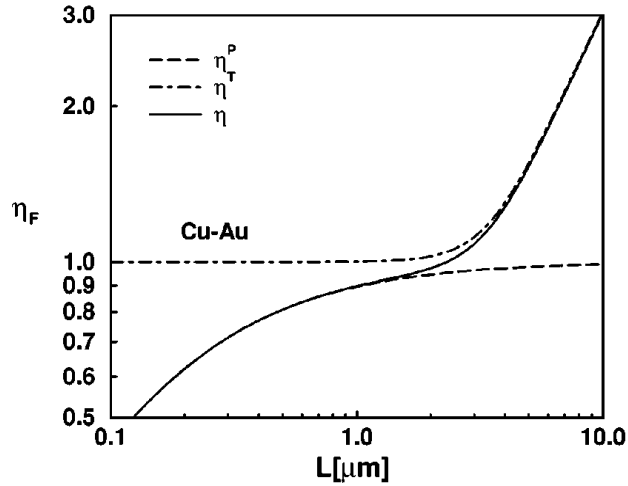}}\end{center}
  \vspace*{-1cm}
\caption{Force correction factors as defined by Eqns. \ref{etaT},
\ref{etap} and \ref{eta} are plotted as a function of spacing
between plates at a temperature of $300^{\circ}$ K
(from,~\cite{Reynaud2000}).}
  \label{Rey}
\end{figure}

\section{Historical Review of Experimental Status}

Experiments to detect forces between flat plates have been
attempted since the $17^{th}$ century \cite{Sparnaay89}. These
early experiments were aimed at studying the existence or
non-existence of the physical ``vacuum". They studied the adhesion
forces between two plates in `evacuated' containers. In the
$20^{th}$ century, with the emergence of theories concerning
long-range, London - van der Waals interactions and Casimir force,
renewed experiments were carried out. (For a review of the
experiments until 2001, see~\cite{Bordag01}) \\

     The earliest attempt to measure Casimir force was by
Overbeek and Sparnaay~\cite{Sparnaay52} in 1952. They tried to
measure the force between two parallel polished flat glass plates
with a surface area of $1$~cm$^2$, in the distance range of $0.6\
\mu$m to $1.5\ \mu$m. The measurements at $1.2\ \mu$m, `pointed to
the existence of a force which was of the expected order of
magnitude'~\cite{Sparnaay89}.\\

     Derjaguin and Abrikossova~\cite{Derj54,Derj60} were the
first to obtain results in the distance range $0.1~\mu$m -
$1.0~\mu$m that were in agreement with Lifshitz's theory.
Sparnaay~\cite{Sparnaay57} repeated his measurements with metal
plates in 1957. He measured the force between chromium plates and
chromium-steel plates. The measurements did not `contradict' the
expected force per unit area from Casimir's relation. \\

     The next major set of improved measurements with
metallic surfaces were performed by van Blokland and
Overbeek~\cite{Overbeek78} nearly 20 years later, in 1978. They
measured the forces between  a lens and a flat plate coated with
chromium using a spring balance at distances between $0.13~\mu$m
and $0.67~\mu m$. This measurement can be considered as the first
unambiguous demonstration of the Casimir force between metallic
surfaces.\\

     In the last decade, attempts to understand the nature of
quantum fluctuations at macroscopic scales and predictions of new
long range forces at the sub-millimeter scales by theories that
unify the fundamental forces have rekindled interest in Casimir
force measurements. The earliest of these was by Lamoreaux in
1997~\cite{Lamor97}, who measured the Casimir force between a lens
and a plate using a torsion balance in the range $0.6~\mu$m -
$6~\mu$m. In a later experiment, Mohideen \emph{et
al.}~\cite{Mohi98} measured the Casimir force  for separations
from  $0.1~\mu$m to $0.9~\mu$m using an atomic force microscope.
Experiments at Bell-Labs by Chan \emph{et al.}~\cite{Chan01}
indicate that Casimir type forces play an important role in
micro-electromechanical systems.  Recently Bressi \emph{et
al.}~\cite{Ruoso02} have carried out high precision experiments
between parallel plates in the range $0.5~\mu$m - $3~\mu$m and the
related force coefficient was determined at the $15\%$ precision
level. The most recent experiment by Decaa \emph{et. al.} measures
the Casimir force between two dissimilar metals for separations of
$0.2~\mu$m - $2~\mu$m.

  A more detailed review and a discussion of data from the recent
experiments on Casimir force will be presented in a later chapter.
All the experiments performed so far to measure Casimir force were
carried out at room temperatures and probed the distance range of
$0.1~\mu$m to $3~\mu$m. As the separation increases, the force
decreases rapidly as $d^{-4}$ to start with and as $d^{-3}$ in the
finite temperature regime [Eqn.~\ref{P_eqn}]. It is important to
measure the force at $d > 3~\mu$m to detect and characterize the
finite temperature corrections to the Casimir force.

\section{Motivations to study Casimir force}
Historically Casimir derived his results while attempting to
explain the inter-molecular interactions seen in experiments on
colloidal suspensions. Casimir force was looked upon as the effect
of finite speed of light (retardation effect) on the London - van
der Waals interaction. The early experiments that measured the
force of attraction between surfaces where aimed at understanding
the inter atom interactions and to see the change over from the
van der Waals force at very small separations to the retarded van
der Waals or Casimir force as the separations increased. The
interest slowly waned once the macroscopic theories of
interactions were verified by experiments.\\

In modern times, interest in Casimir force has been aroused by the
crucial role it plays in theories of fundamental physics. Casimir
force provides explicit evidence for the existence of vacuum
fluctuations and for the interplay between the
microscopic~(quantum) and the macroscopic worlds. Also, in the
last couple of decades, several new theories have been proposed
that predict new physics in the sub-millimeter distance range.
Casimir force is the dominant background in this range and hence
it becomes essential to understand all aspects of Casimir force
before looking for new forces at these scales.

\subsection{Understanding the Quantum Vacuum}
The existences of electromagnetic field fluctuations in vacuum
presents problems due to the amount of energy it carries. General
Relativity states that all forms of matter and energy should
gravitate. The large energy density, $\rho_v$ associated with the
vacuum fluctuations should induce very large gravitational
effects, much larger than that allowed by observations. This
``vacuum catastrophe'' is related to the famous cosmological
constant problem (see \cite{Wein1989,Wein2001, Sahni2002,
Paddy2003} for a
review). \\

In 1917, when Einstein first attempted to apply his new theory to
relativity to the Universe, the Universe was believed to be
static. Einstein could not construct a static universe if there
was only matter and curvature, so he introduced a free parameter
$\Lambda$, the cosmological constant, into his theory which was a
form of energy with negative pressure. With the discovery of the
expansion of the Universe, Einstein suggested that this could be
dropped. But it was retained alive in discussions of cosmology,
and has been used time and again to explain observations that did
not fit into the standard scenarios in cosmology \cite{Stein2000,
Straumann2002}. Currently, there is strong observational evidence
\cite{Riess1998,Perlm1999} for an accelerated expansion of the
Universe. This can be explained by a non-vanishing cosmological
constant. Thus, in the present scenario, the geometry of the
Universe is determined by the energy density of matter, $\rho_m$,
the energy density due to vacuum, $\rho_v$ and that from the
cosmological constant, $\Lambda/(8\pi G)$. The quantum vacuum has
properties similar to those attributed to a positive cosmological
constant, the most important property being an effective repulsive
gravity. This is because the acceleration is proportional, in
General Relativity, to the term $-(\rho+ 3 p)$ of the matter,
where $\rho$ is the energy density and $p$ the pressure. For
normal matter, the term $(\rho + 3 p)$ is positive and therefore,
the universe is expected to decelerate as it expands. The diagonal
elements of the energy momentum tensor are energy density and the
three components of pressure. Therefore, an energy momentum tensor
that is proportional to ``vacuum" with diagonal elements $(1, -1,
-1, -1)$ will have its equation of state $\rho=-p$, and the
effective acceleration of the Universe with such a source term
will be positive. It is this property that allowed Einstein to
construct a static Universe, balancing the gravitational
attraction of normal matter with the effective repulsion of the
cosmological constant. The total energy density in the Universe,
$\rho_c$ can be determined from the expansion rate of the Universe
and is given by $3H_0^2/(8\pi G)$, where $H_0$ is the Hubble
constant. Cosmological data indicates that $\rho_v = 0.7\ \rho_c\
\sim 4$~keV/cm$^3$. A naive estimate of $\rho_v$ calculated for
all modes of the zero point field with a high frequency cut-off at
the Planck scale is 120 orders of magnitude larger than $\rho_c$.
Therefore, there is a need to find a fine-tuned suppression
mechanism that will bring this large number close to zero. If a
small vacuum energy as well as a cosmological constant are
present, then their values need to be fine tuned to the small
number $\rho_c$, which amounts to a fine cancelation to $120$
decimal places. This bizarre coincidence is the present
cosmological constant problem. To find a solution to this problem,
the contributions from the vacuum fields have to be understood and
estimated better. Since Casimir force is a direct manifestation of
the vacuum fluctuations, it provides a tool to comprehend the
`vacuum'.

\subsection{``The Hierarchy Problem"}
Interactions in nature have been identified to be of four
fundamental type: gravitational, electromagnetic, strong and weak.
General Theory of relativity and Standard Model of particles
physics are the two most successful theories that explain these
interactions. Attempts to link these two theories are plagued with
difficulties due to the vast differences in the strength of
gravity as compared to others. The Standard Model of electroweak
and strong interactions unifies electromagnetic and weak
interactions at the characteristic energy scale, $M_{ew}$ of
$10^3$ GeV ($=1$ TeV). The energy scale of gravity is the Planck
energy, $M_{pl}\ (=10^{19}$ GeV), where the Compton wavelength of
a particle becomes equal to its Schwarzchild radius. In natural
units ($\hbar = c =1$), the gravitational constant $G =
1/M_{pl}^2$. The vast difference between these two energy scales
is the ``hierarchy problem". Several frameworks have been put
forward to solve this problem. The most popular among them are
string theories, M-theory and theories with large extra dimensions
(see \cite{Adel2003} for a recent review).

\subsubsection{Large extra dimensions and Warped geometries}

A model that attempts to reconcile gravity and quantum theory is
one in which the fundamental objects that constitute our Universe
are not particles but very tiny extended objects: strings.
However, to have a consistent string theory that can explain all
known phenomena, spatial dimensions larger than 3 are required
(for a review, see \cite{Randall02, Hamed02, Adel2003}). The
question then is, why do we not see these additional dimensions?\\

Two scenarios have been proposed to explain this. One that follows
the original idea by Kaluza \cite{Kulza21}, which tries to unify
the fundamental forces at Planck scale. In this picture, besides
the three spatial dimensions of infinite extent, there are
additional dimensions of finite size, $r_c$, that are curled up as
compactified circular extra dimensions. The typical size of the
circle would be determined by the Planck length, $10^{-33}$cm.
This means that physics at short distances appears to be higher
dimensional and forces go as $1/r^{-(n+2)}$, where $n$ is the
number of extra dimensions. At distances much larger than $r_c$,
$n=0$ and one observes the $1/r^2$ law. Around the compatification
scale, the forces can be modelled as $e^{-r/r_c}$ and would show
up as deviations from the standard law.\\

The other idea proposed by Arkani-Hamed and others
\cite{ADD1,ADD2} , tries to unify forces at the TeV scale, thereby
removing the hierarchy problem. The $10^{32}$ times weakness of
gravity as compared to the other forces at this  electroweak scale
is explained by the presence of the extra-dimensions of finite
size $R_*$. They propose that the three spatial dimensions in
which we live are perhaps just a membrane (3-brane) embedded in a
higher dimensional bulk of ($3+n$) spatial dimensions. The
Standard Model fields are confined to the 3-brane while gravity
can propagate into the bulk. Thus, at distances greater than
$R_*$, gravity spreads in all $3+n$ dimensions and goes as
$1/r^{-(n+2)}$ while the strength of the other forces, still falls
as $1/r^2$. At $r > R_*$, gravity too reverts back to $1/r^2$.
This scenario is distinguished by the term, ``large extra
dimensions scenario", as the additional dimensions in this theory
could even be macroscopic. If there were only one extra dimension,
its size would have to be of the order of $10^{10}$ km to account
for the weakness of gravity. Such an extra dimension would change
the dynamics of the Solar system and is eliminated by known
experimental results. With two equal extra dimensions, the scale
length would be of the order of a $0.3$ mm. This is inconsistent
with laboratory experiments~\cite{Hoyle2001} and astrophysical
bounds~\cite{ADD1999,Cullen1999, Hanhart2001-1,Hanhart2001-2}. For
$n\geq3$, the scale length is less than about a nanometer. This
does not imply that the new dimensions will not show observable
effects in experiments at sub-millimeter scales. A single large
dimension of size $1$ mm with several much smaller extra
dimensions is still allowed. Experiments have shown that the scale
length of the `largest' extra dimension has to be $< 200~\mu$m
\cite{Hoyle2001,Adel2002}. This would give rise to observable
changes to inverse square law of gravity at these scales. At such
distances, the strengths of gravity and Casimir forces are
comparable and in order to look for new corrections to inverse
square law of gravity, Casimir background
has to be first understood and eliminated.\\

\subsubsection{New particles}
    The super-symmetric extensions to standard
model unify the electro-weak and strong interactions at energy
scales of $10^{16}$ GeV, which is very close to the Planck scale,
with the additional assumption that there are no charged particles
between the TeV and the Planck scale (see for ex.
\cite{Hamed02,Randall02} and references therein). A trade mark of
super-string theories is the occurrence of scalar partners of the
graviton (dilaton), gravitationally coupled massless scalers
called moduli, and other light scalers like axions. The exchange
of these particles could give rise to Yukawa type interactions,
that would appear in experiments that measure forces. The range
$\lambda$, of these interactions depends on the mass of the
elementary particle. For super-symmetric theories with low energy
(few TeV) symmetry breaking, these scalar particles would produce
effects in the sub-millimeter scales \cite{Dimo96,Kap00}. The
predictions on the strength of these effects is less precise than
those of the extra-dimension scenarios.
\begin{figure}[h]
 \begin{center}
 \resizebox{10cm}{!}{
 \includegraphics*[3.2cm,13.7cm][16.2cm,23.8cm]
 {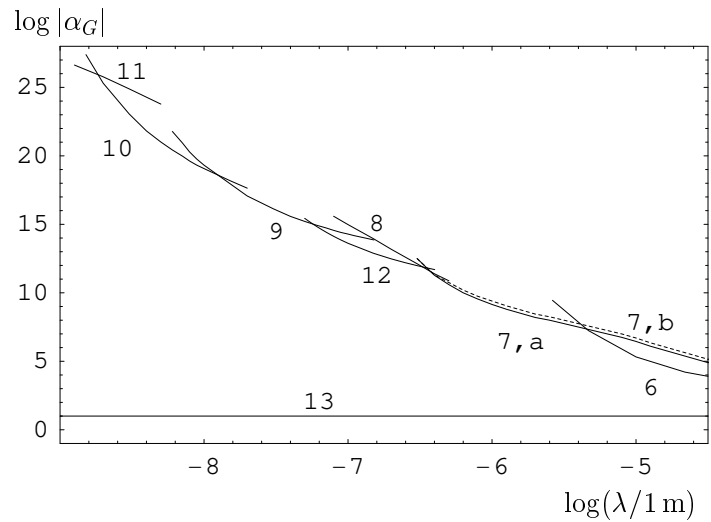}}
 \end{center}\vspace*{-1cm}
  \caption{Plot of the constraints on the Yukawa interaction
parameter $\alpha$ from various ranges of $\lambda$ reproduced
from \cite{Most2003}. Curves 7-10, 12 follow from Casimir force
measurements, Curve 11 from van der Waals force measurements.
Curve 6 is from an experiment that measured deviations from
Newton's law of gravity~\cite{Price2003}. The typical prediction
of extra dimensional physics is shown in Curve 13. The region in
the ($\alpha,\lambda$) plane above each curve is excluded and the
region below each curve is allowed.} \label{constraint}
\end{figure}

\subsection{Constraints on new macroscopic forces}
Keeping in mind that the strengths of the new macroscopic forces
are small, their potentials are scaled with respect to the
gravitational interaction between two point masses as shown below:
\begin{equation}
V(r) = -\frac{GM_1M_2}{r} \left( 1 + \alpha e^{-\frac{r}{\lambda}}\right)%
\end{equation}
where $\alpha$ represents the coupling strength of the interaction
and $\lambda$ the range \cite{Krishnan89-T, Unni92}. The typical
scale of $\lambda$ will vary depending on the source of the
potential. For the extra dimension scenarios it would be the size
of the extra dimension, while for the new string inspired forces,
it would be proportional to the inverse of the mass of the
mediating particle. Thus constraints can be placed on the
parameter space of $\alpha - \lambda$ from experiments that study
long-range interactions \cite{RC1981}-\cite{RC1982-2}.
\nocite{RC1981,RC1982-1,RC1982-2,RC1988-PRL, RC1988-2, RC1989-3,
RC1990} For distances of $\lesssim 0.1$ mm, Casimir force provides
the dominant background and best limits on $\alpha$ for these
$\lambda$ can be obtained from Casimir force measurements. The
available constraints to date on $\alpha-\lambda$ from various
experiments are summarised in Fig.\ref{constraint}.

The torsion balance experiments described in this thesis are
capable of strongly constraining theories with macroscopic extra
dimensions, apart from measuring the Casimir force and its finite
temperature corrections.

\bibliographystyle{plain}
\bibliography{reference}


\clearpage

\chapter{Torsion balance- Design and Fabrication}

\emph{Abstract: This chapter is devoted to the description of the
main components of the experimental set up. An overview of the
experimental scheme will be followed by a detailed explanation of
the various components of the experiment. The procedures followed
during assembly to reduce systematic and environmental noises will
be described.}

\section{General principle of the apparatus}

    Torsion pendulums have been used as transducers for precision
measurements for over two centuries. They are known for their
capability to isolate and measure feeble effects that would
otherwise be difficult if not impossible to observe against the
background gravitational field of the earth.(see for ex.,
\cite{Gillies93, RC1997}) Our experimental set up is aimed at
achieving the sensitivity required to measure the finite
temperature effect in the Casimir force using a torsion pendulum,
optical auto-collimator combination. \cite{MG8,Unni95}. The
experiment was to be performed in the separation range $2~\mu$m -
$10~\mu$m. Casimir force at $10~\mu$m will produce, in static
case, a deflection of $\sim 10^{-6}$ radians on our pendulum. This
is well with in the sensitivity of $\sim 10^{-8}$ radians of our
optical lever. \\

\begin{figure}
\begin{center}
\resizebox{!}{8cm}{\includegraphics*[5.5cm,0.1cm][24.5cm,15.2cm]
{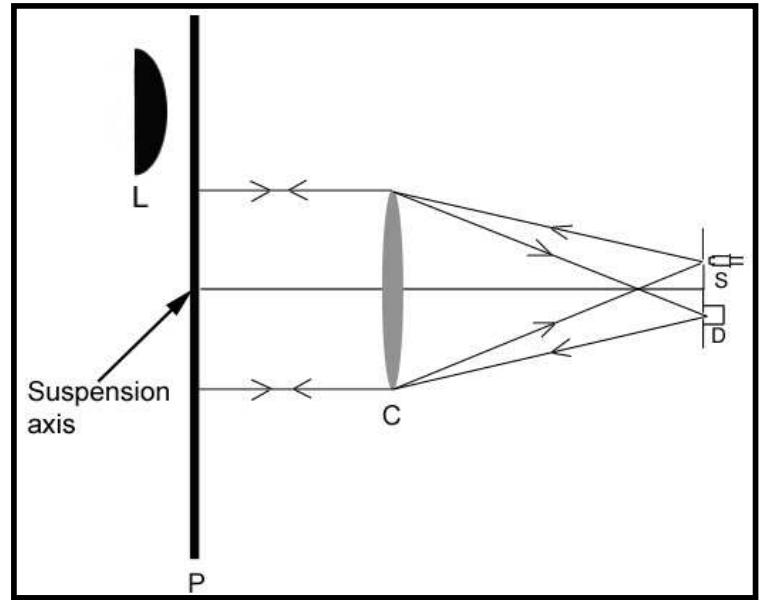}}
\end{center}
\caption{Figure representing the measurement scheme. L = spherical
surface, P~=~pendulum disc, C = collimating lens, S = source of
light, D = Light Detector } \label{scheme}
\end{figure}

   The measurement scheme [Fig.~\ref{scheme}] broadly consists of a
torsional pendulum with a flat circular disc, P , as mass element,
suspended using a thin strip fibre. The source of the force field
we wish to study is provided by a spherical surface, L , located
close to one edge of the suspended disc. The torque on the
pendulum due to this force is inferred from the deflection of the
pendulum measured using an auto-collimating optical lever.

\section{Torsion Pendulum}

\subsection{Mass Element} The mass element of the torsion pendulum
has two parts. A circular disc and a mechanism to attach the disc
to the suspension. The mass element assembly weighs a total of
$52\ \mathrm{grams}$ and has a moment of inertia of $\sim 198$
g.cm$^{2}$

\subsubsection{The Disc:} The active element of the suspension is a
circular disc of thickness $4$ mm and diameter $80$ mm made of
glass. The edges of the disc are chamfered and both the faces are
polished to have a surface finish of $\lambda /2 $ and are coated
with a $1~\mu$m thick layer of gold. One of the faces acts as a
conducting boundary for the force under study and the other face
acts as a mirror viewed by a sensitive optical lever which
measures the angle that the normal to the disc makes with the
optic axis of the optical lever.

\subsubsection{The Disc holder:} 
\begin{figure}
\begin{center}
\includegraphics[width=8cm]{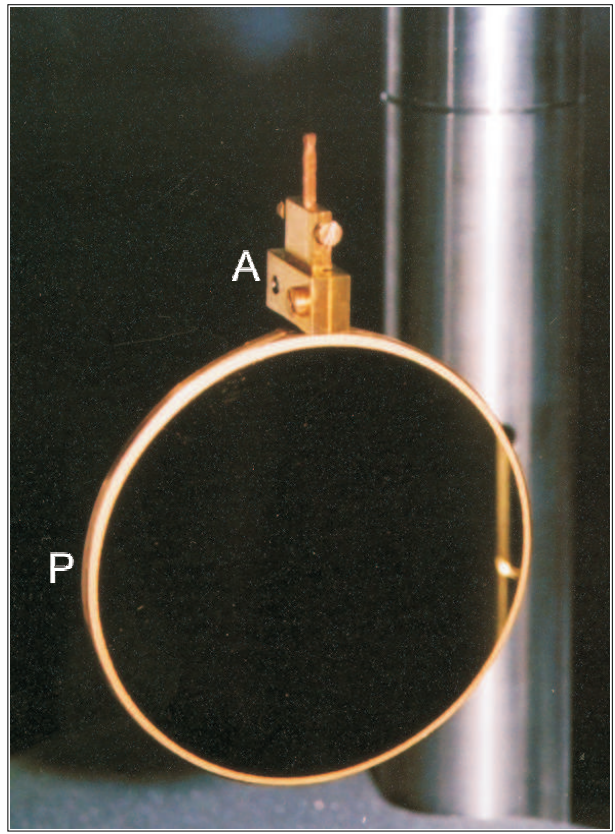}
\end{center}
\caption{Picture of Mass element assembly. A = Aluminium holder,
P~=~Pendulum Disc} \label{Mass-elmt}
\end{figure}

The gold coated glass disc is held
in a frame made of a gold strip that is $80~\mu$m thick and $4$ mm
wide. In order to hold the glass disc firmly, this strip is shaped
to form a groove that matches the chamfered edges of the disc and
forms a circular frame around the disc. The ends of this strip are
held securely between the flat surfaces of an Aluminium holder
[Fig.~\ref{Mass-elmt}]. The flat surfaces that press the strip
ends together have $50~\mu$m deep, $4 $mm wide channels machined
with the central axis of the channel along the axis of the holder.
These locate the strip and hence the pendulum bob along the axis
of the holder. The top of this holder has a $2$ mm diameter hole
through its central axis to hold the torsional fibre. This holder
is also gold coated to avoid aluminium oxide layers that can
accumulate charges.

\subsection{The Pendulum Suspension} The mass element is suspended
in two stages to avoid non-torsional modes of oscillation of the
pendulum. The suspension consists of a pre-suspension, a device to
damp the simple pendular modes of the fibre, and the main
suspension.

\begin{figure}
\begin{center}  
\resizebox{5cm}{!}{
  \includegraphics*[60mm,70mm][155mm,211mm]
  {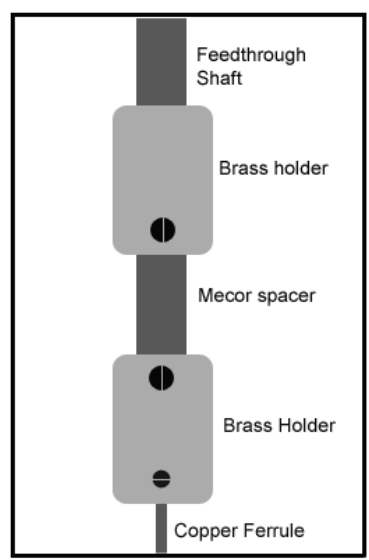}}\\
\end{center}
  \caption{A schematic of the pre-suspension mount}\label{fig-sus1}
\end{figure}

\subsubsection{Pre-suspension:} The first portion of the suspension
is a torsionally stiff copper wire of $100~\mu$m diameter. This
ensures verticality of the main suspension that uses a metal
ribbon. The pre-suspension is essential to avoid spurious
torsional effects due to tilts of the suspensions. The ends of the
stiff wire are passed through a copper ferrule of outer diameter
$2$ mm and bore diameter of about $0.5$~mm. The ferrule is crimped
such that it holds the fibre gently but firmly along its central
axis. The length of the Copper wire is $7$~cm between the
ferrules. The wire is mounted from the shaft of a rotary
feed-through that is attached to the top of the vacuum chamber
housing the experiment. To electrically isolate the suspension
from the chamber, the wire is held through a Macor insulator as
described below [Fig.~\ref{fig-sus1}]. The shaft holds a
cylindrical clamp made of brass, which in turn holds a Macor
cylinder. Another cylindrical brass holder is mounted to this
Macor piece and has a $2$ mm diameter hole along its axis. The
ferrule at one end of the wire is passed through this hole and
held in place by a screw. A copper disc is suspended from the
ferrule at the other end of the wire using a similar mechanism
[Fig.~\ref{fig-damper}]. Kapton insulated copper wires connect the
fibre suspension to an electrical feed-through mounted on the
vacuum chamber.

\subsubsection{Pendular Mode Damper:} 
\begin{figure}
\begin{tabular}{c c}
\includegraphics[height=8cm]{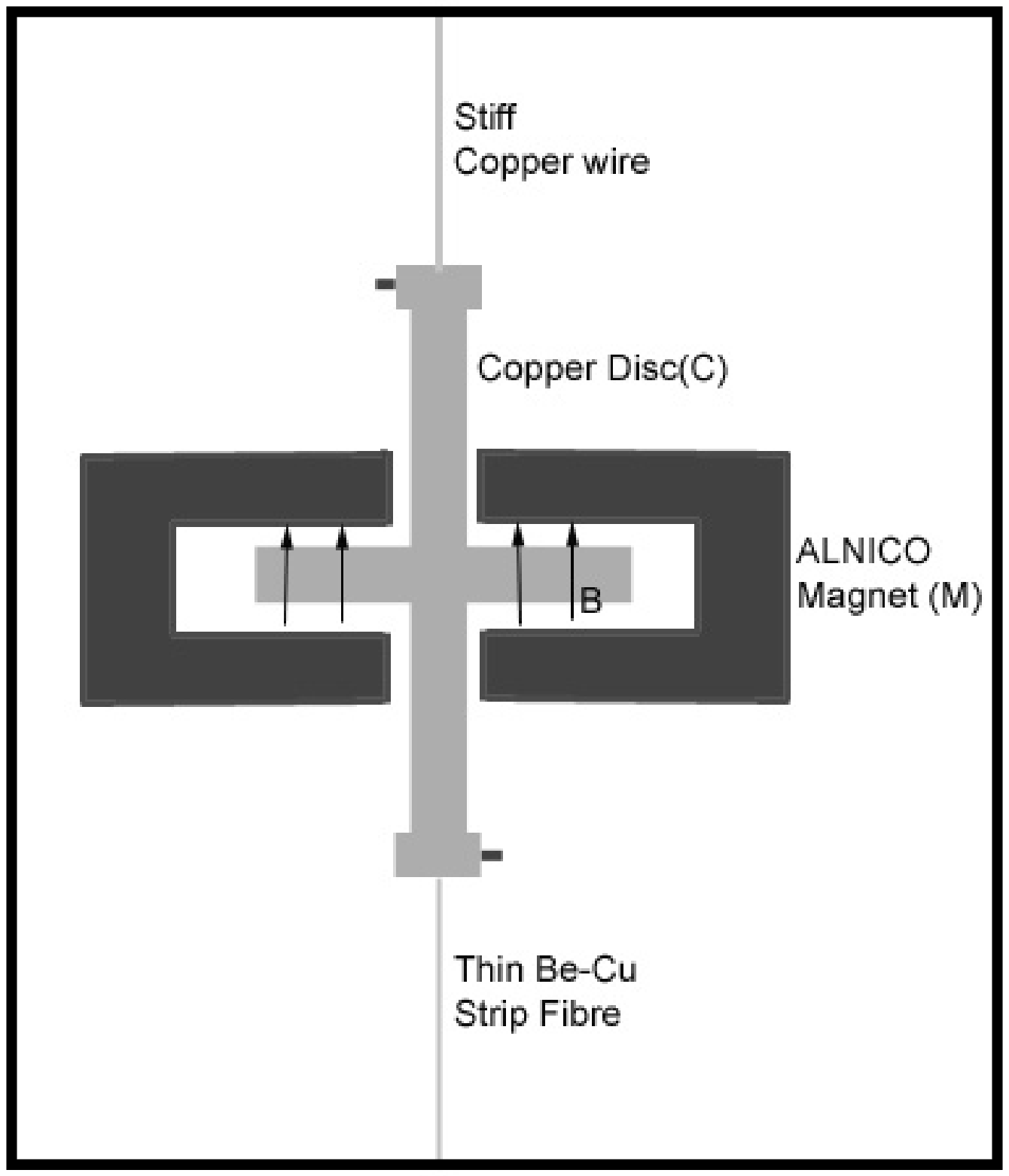} &%
\hspace*{1cm}
\includegraphics[height=8cm]{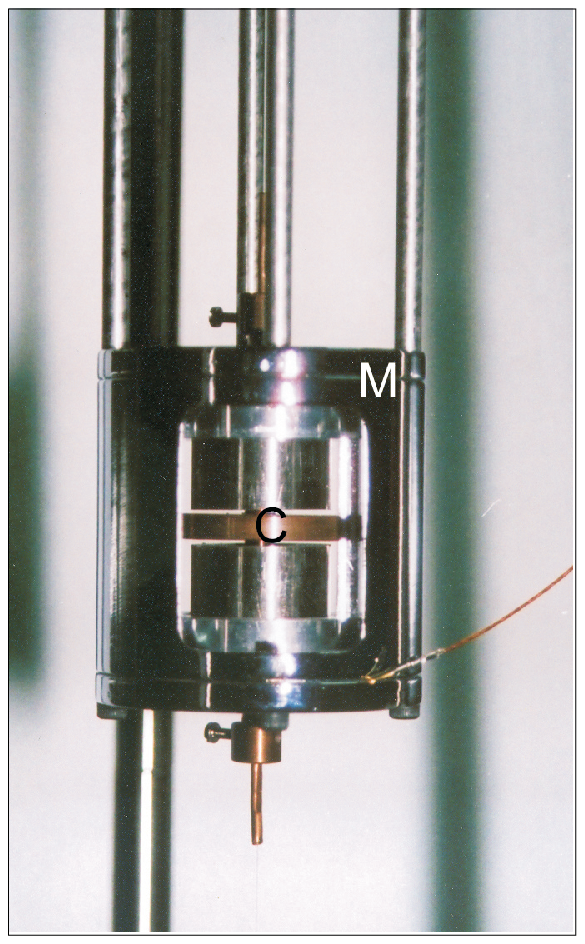} \\
(a) & \hspace*{1cm}(b)\\
\end{tabular}
\caption{The pendular mode damper assembly: (a)Shows the general
scheme and (b) shows a photograph of the assembly}
\label{fig-damper}
\end{figure}

The copper disc (C) passes
between the pole pieces of an aluminium-nickel-cobalt ring
magnets~(M) [Fig.~\ref{fig-damper}] such that it cuts across the
field lines (B) of the magnet. The  fast pendular oscillations of
the fibre and violin modes cause eddy currents to flow in the
copper disc and dissipate these modes. Thus, these modes will be
damped out rapidly. Since the copper disc and the magnet are
axially symmetric, the very slow torsional modes are not damped.

\subsubsection{The Torsional Fibre:} The main torsional element is
a Beryllium-Copper strip fibre of width $=90~\mu$m, thickness
$=9~\mu$m and length $=39$~cm. The breaking strength of the fibre
is approximately $100$ g. Thus our load of $52$ g is well within
its breaking strength. Copper ferrules are crimped at either end
of this fibre as in the case of the torsionally stiff copper wire.
The fibre is anchored to the copper disc on the top and holds the
active mass element at the bottom. The fibre is pre-annealed under
a load of $50$ g at a temperature of about $150^{\circ}$~C for a
day.\\

The torsion constant of the suspension $k_{f} \approx 0.05$ dyne
cm rad$^{-1}$ and the suspension has a time period of $\sim\ 406$
sec. The thermal amplitude of the suspension in the absence of
external disturbances is $\sim 9 \times 10^{-7}$ radians at room
temperature of $\sim 300^{\circ}$~K.

\section{The Capacitor Plates and Torsion Mode Damping} 
\begin{figure}
  \begin{center}
  \includegraphics[width=8.5cm]{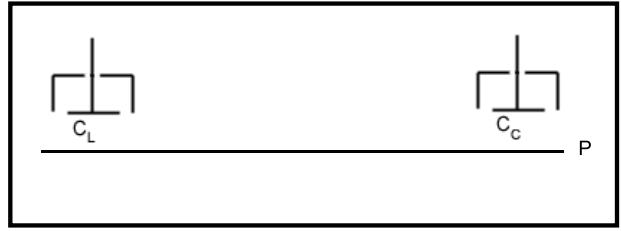}\\
  \end{center}
  \caption{The capacitance arrangement for damping torsional modes}
  \label{capa-scheme}
\end{figure}

The
torsional oscillations of the pendulum can be  damped to an
amplitude of about $10^{-5}$~radians by applying capacitive forces
to the suspended disc.  Two capacitor plates are mounted such that
they produce opposing torque on the disc. The net torque due to
the capacitances is varied by adjusting the individual voltages on
them. A schematic of the arrangement is shown in
Fig.~\ref{capa-scheme}. A voltage applied to $C_L$ rotates the
pendulum, P clockwise as seen from above while a voltage on $C_C$
rotates it anticlockwise. By switching the voltages exactly out of
phase with the torsional oscillations of the pendulum, these
oscillations are damped. \\

       The capacitor plates are made of aluminium and consist
of circular plates of $1.2$ cm diameter that are placed within
grounded guard rings. The guard ring is insulated from the plate
with Macor positioners. The two sets of capacitor plates are
mounted together on another Aluminium fixture such that they are
on diametrically opposite edges of the suspended disc and
positioned at about $2\ \mathrm{mm}$ separation from the disc.
Shielded Kapton insulated copper leads connect the capacitor
plates to electrical feed-throughs on the vacuum chamber. The
capacitors and the mount are gold coated to avoid exposed
Aluminium oxide
surfaces.\\

       These capacitors can also be used to perform a null
experiment in which the deflection of the torsional pendulum is
balanced by capacitive forces. The  position signal from the
optical lever is fed back to control the effective voltage on the
capacitor plates and the torsion pendulum `locked' at a fixed
position. If the torque on the pendulum due to force between the
pendulum and the spherical lens surface is modified, voltage on
the capacitors changes to balance this torque. Thus, the change in
voltage on the capacitor is a direct signal of the
torque acting on the pendulum.\\

   The varying voltages are generated from a $16$ bit DAC in a PCI
interface card. Typically one volt on the capacitor at $1$~mm
separation, gives rise to a torque of $\sim10^{-3}$~dyne.cm. The
voltage on one capacitor is kept fixed at about $4$~V and that on
the other is varied from $0$~V - $10$~V so that both positive and
negative torques may be applied to the pendulum. The information
on the angular position of the suspended disc obtained from the
autocollimator is fed back to a PID loop through software
(Labview) to control the voltage of the capacitor. When the lens
is far away and the only force on the suspension is the restoring
force from the fibre, the PID loop keeps the position of the
pendulum locked to about $1/10$ of a pixel or $5
\times10^{-7}$~radians (1 sec integration). This is below the
thermal amplitude of the pendulum which is of the order of
$10^{-6}$~radians.

\begin{figure}
 \begin{center}
 \resizebox{12cm}{!}{
  \includegraphics[0mm,70mm][214mm,212mm]
  {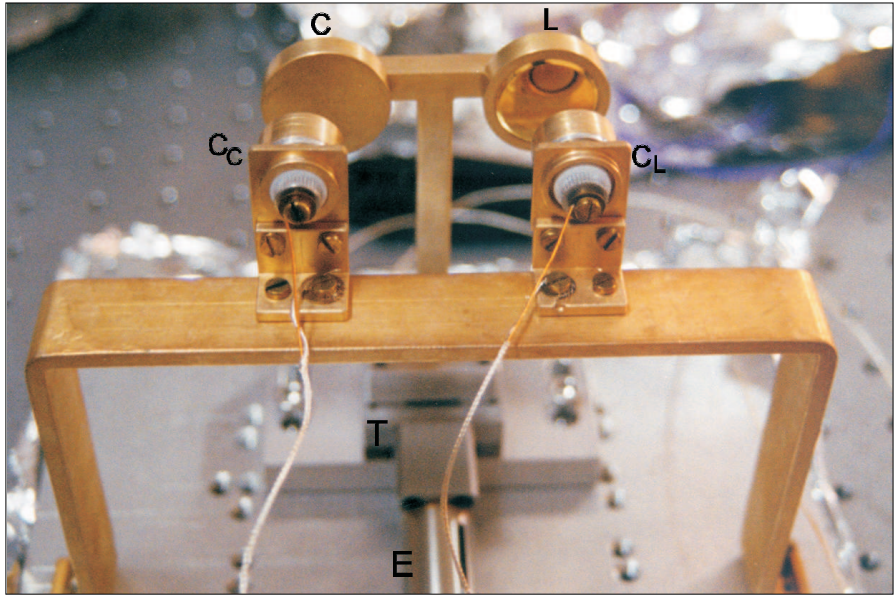}}\\
 \end{center}\vspace*{-1cm}
\caption{The lens (L), compensating plate (C) and the capacitor
plates ($C_L$ and $C_c$) assembled (without the pendulum). T is
the translation stage and E the
EncoderMike$^\circledR$}\label{lensandcap}
\end{figure}
\section{The Spherical Lens and the Compensating Plate} In our
experiment the Casimir force between the suspended disc and the
spherical surface of a lens is measured. This configuration is
simpler to implement as difficulties in holding the 2 plates
parallel to each other while measuring the force are avoided. The
lens is $25$ mm in diameter and has a  radius of curvature of
about $38$ cm. It is coated with $1~\mu$m thick layer of gold. It
is mounted in a cell and held along the diameter of the suspended
disc close to one edge such that the interactions between the lens
and the mass element apply a torque on the pendulum. In this
position, the gravity due to the lens cell assembly will also
apply a torque on the suspension.  This is minimized by using a
compensating plate of Aluminium with mass equal to that of lens
and the cell [Fig.~\ref{lensandcap}]. The outer diameter of the
compensating plate is equal to that of the lens cell, but the
compensating plate has a flat surface facing the suspended disc
and its thickness is adjusted to equalize the masses. The
compensating plate is mounted such that the gravitational torque
due to the plate opposes the gravity due to the lens assembly. The
net gravitational torque on the pendulum is small. More
importantly, changes in the force as the lens is moved through
small distances is negligible compared to the changes in Casimir
force and electrostatics forces. The lens assembly and the
compensating plate are together mounted on a translation stage (T)
(Newport Model- 461 series)(Fig.~\ref{lensandcap}). An
EncoderMike$^{\circledR}$ actuator (E) from Oriel is used to
translate this stage perpendicular to the disc surface.  The
actuator movement is controlled by DC voltages applied to it. The
encoder has a resolution of $0.05~\mu$m and its output is
monitored using a Data acquisition card with a PCI interface
attached to the PC. \\

The lens assembly and the compensating plate are electrically
isolated from each other and from the mount. Kapton insulated
copper wires connect them to separate connectors on an electrical
feed-through attached to the vacuum chamber.

\begin{figure}
\begin{center}
\resizebox{9cm}{!}{\includegraphics[0cm,.5cm][8cm,8.3cm]
 {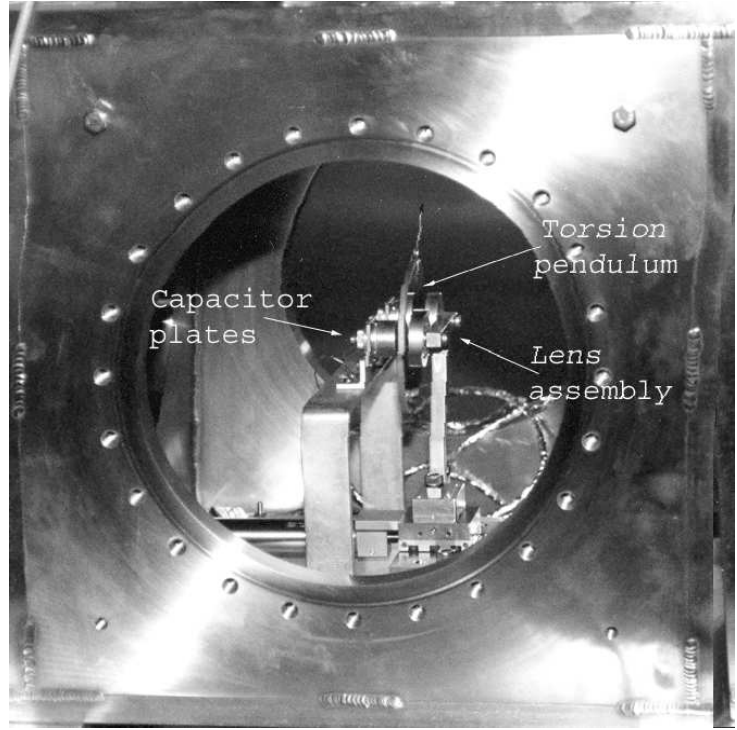}}
\end{center} \caption{Picture of the torsion pendulum, capacitors and the
lens assembled inside the vacuum chamber}
  \label{closeup}
\end{figure}

\section{The Vacuum Chamber}
\begin{figure}
\begin{center}
  \resizebox{12cm}{!}
  {\includegraphics[3.3cm,.3cm][18.4cm,27.1cm]
  {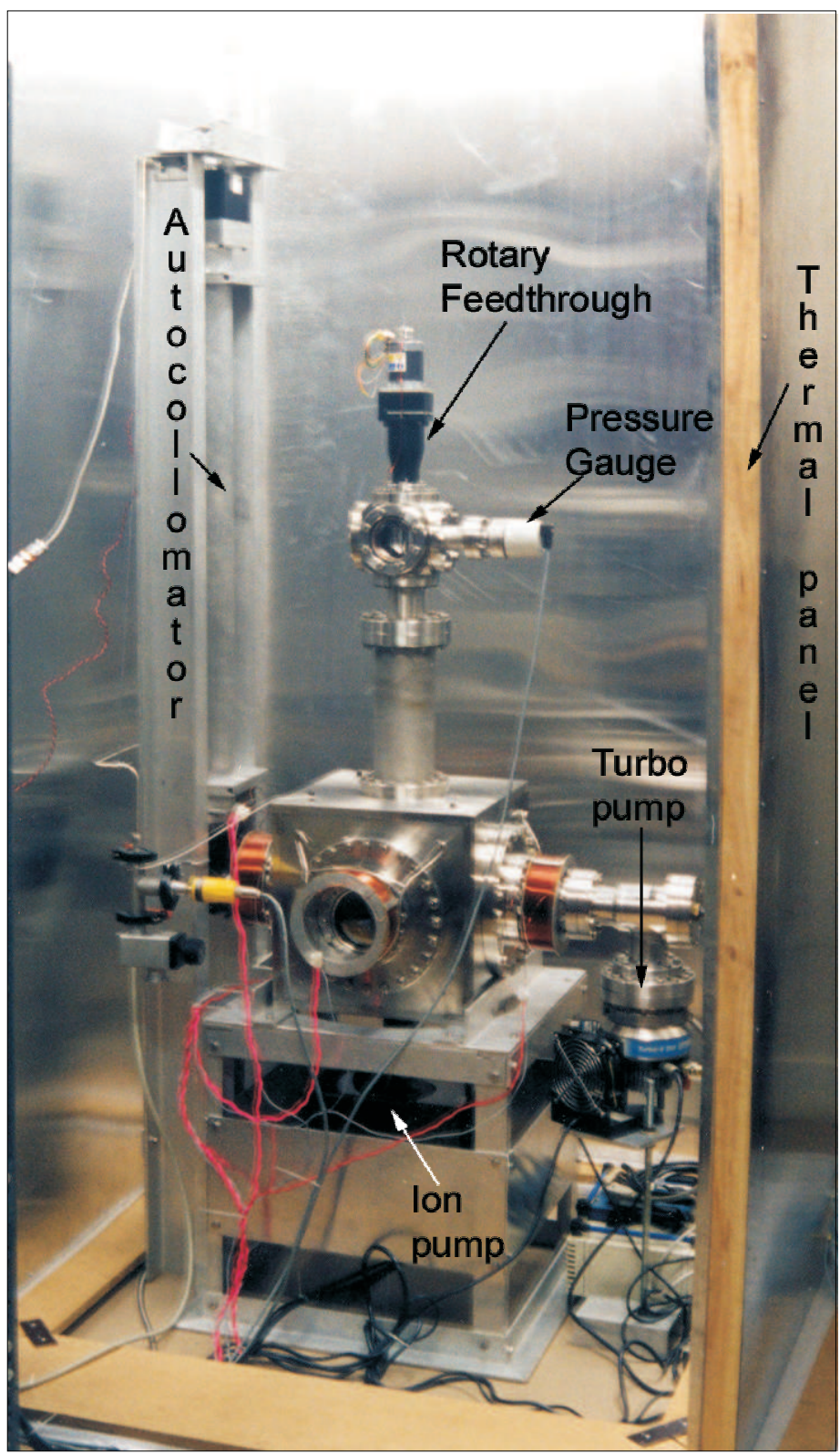}}\\
\end{center}\vspace{-1cm}
  \caption{A photograph of the experimental set up during assembly}
  \label{setup}
\end{figure}

The experiment is conducted in high vacuum ($\sim 4 \times
10^{-8}$ Torr). The presence of gas surrounding the pendulum at
pressures above $10^{-6}$ Torr not only damps its oscillations
very quickly, but also produces pressure gradients which lead to
erratic, anharmonic deflections of the pendulum. A cubical vacuum
chamber with side ports for optical windows with $300$ mm long
extension tube on its top flange was designed and fabricated. The
experimental apparatus are mounted inside this chamber
[Fig.~\ref{closeup} and Fig.~\ref{setup}]. The extension tube is
fitted with a rotary feed through at the top, from which the
pendulum is suspended. This enables us to rotate the pendulum and
change its equilibrium position. The feed through is motorized and
has angular resolution of $0.1$~degree. The
MotorMikes$^\circledR$, translation stages and all other
components were vacuum tested to $\sim10^{-8}$~Torr individually
before assembly into the main vacuum chamber. \\

The vacuum chamber is initially pumped down to $\sim10^{-8}$ Torr
using a turbo molecular pump from Varian with a pumping speed of
$250$ litre/s. An ion pump is switched on at this stage. The
system is baked at $90^\circ$ C for 2 days. The turbo pump is then
valved shut and switched off. The ion pump remains on through out
the experiment and maintains the pressure at $3 \times 10^{-8}$
Torr without any mechanical disturbances. The pressure inside the
vacuum chamber is measured using a compact full range cold cathode
gauge from Pfeiffer Vacuum. The ion current read
out from the ion pump is also used to monitor the pressure.\\

       For the optical lever to view the mirror suspended
inside the vacuum chamber a home made wire sealed optical glass
window is used. The window is made of BK7 glass, $25$ mm thick and
is polished on either side to a surface accuracy of $\lambda /4$.
A commercial glass view port typically does not have the optical
quality required to get sharply focused image of the source slits
of the optical lever on the CCD detector. There are also two other
standard glass view ports to see into the chamber. Glass surface
being dielectric and insulating, accumulates charges. These
charges can create electric fields inside the vacuum chamber that
influences the torsional pendulum. To shield these fields grounded
wire mesh are mounted inside the vacuum chamber in front of all
the glass windows. Without these wire meshes, the pendulum can get
`locked' into position determined by stray electric fields from
the view ports.
\begin{figure}
  \begin{center}
  \fbox{
  \resizebox{15cm}{!}
  {\includegraphics[1mm,0mm][20.9cm,22.8cm]
  {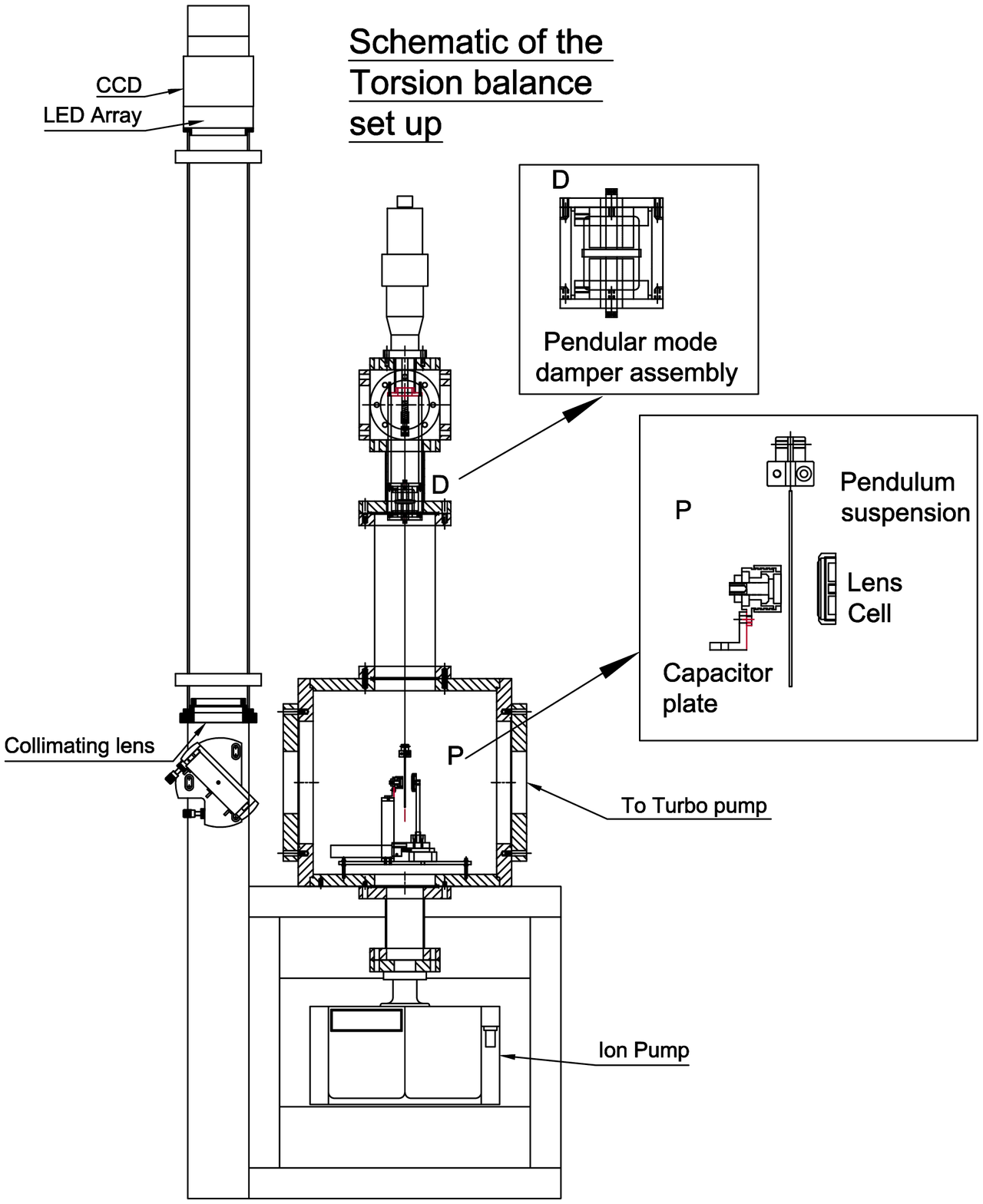}}}
  \end{center}
  \caption{Schematic of the experimental set up}
  \label{setup-scheme}
\end{figure}
\section{The Electrical Wiring and Grounding in the Apparatus}

The vacuum chamber is held at ground potential by a thick copper
cable. This shields the apparatus inside from electrical pick ups.
The pendulum and the compensating plates are externally connected
to the same ground point. The capacitor plates and the lens are
also connected to this ground point when voltages are not applied
on them. \\

A UV lamp is placed inside the chamber and flashed on during pump
down when the pressure in the chamber is dropping from $10$ Torr
to $1$ Torr. This generates lots of electrons (by photoelectric
effect) and some ions; thus allowing a neutralization of the
electrical charges. The usage of the lamp was `empirical' and was
not very systematic. We found small reduction in the residual
electrostatic force when the lamp was operated for short duration
during pump down.

\section{The Thermal Panels} The experimental set up
is shielded from fast temperature variations and temperature
gradients in the environment. The set up is surrounded by four
$1.2 \times 4.2$~meter, insulating panels. Each of these panels is
made of several layers of thermocol and plywood sheets sandwiched
between Aluminium sheets and held together by a wooden frame. The
`walls' formed by these is covered on the top by thermocol layers
attached to Aluminium sheet. Various electrical and signal cables
come out through tightly packed holes in the shroud. The entire
apparatus is placed within a closed room and controlled from
outside. The peak to peak variation in temperature inside the
enclosure is $1$ degree per day (diurnal cycle), while the ambient
temperature changes by as much as $10$ degree per day. However,
fluctuations over time scales of an hour are within
$5$~millidegree.

\bibliographystyle{plain}
\bibliography{reference}

\clearpage

\chapter{The Autocollimating Optical Lever}

\emph{Abstract: An optical lever of novel design built to measure
the deflection of the torsional pendulum will be described in this
chapter. The optical level has a large dynamic range of $10^6$ and
a sensitivity of $\sim 1 \times 10^{-8}$ radians/$\surd$Hz.  The
chapter will begin with the discussion of the principle of the
design and go on to describe its implementation. Finally the tests
and characterization of the autocollimator will be discussed. }

\section{Conceptual aspects of the design}
The angular deflection of the torsion balance contains the signal
in our experiment. The method used for its measurement is the
standard optical lever arrangement [see for example,
\cite{Jones1951,Jones1959}, \cite{Krishnan89} and references
therein] where a beam of light is reflected off a mirror on the
torsional pendulum and the deflection of the light beam is then
proportional to the rotation of the pendulum. This basic scheme
has been modified to give good accuracy and large dynamic range
for angle measurements.\\
\begin{figure}[h]
\begin{center}
\resizebox{10cm}{!}
{\includegraphics*{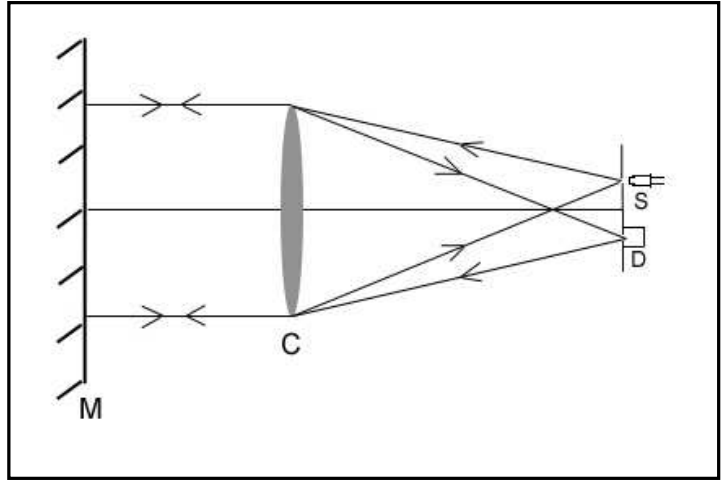}}
\caption{\label{Ray diagram} Classical autocollimating
arrangement}
\end{center}
\end{figure}

The optical lever is arranged in an auto-collimating
configuration. In this configuration, the translation of the
mirror does not change the image on the detector and the optical
lever is sensitive to only rotations of the mirror. An illuminated
array of slits, S~, is placed in the focal plane of an achromatic
lens, C~, with a slight offset with respect to its optical axis.
The collimated beam emerging from the lens falls on the mirror,
M~, whose rotation angle is to be measured. The reflected beam
from the mirror returns  through lens and will form an image of
the slit very close to the slit itself, but with an opposite
offset with respect to the optical axis. This image falls on a
linear array CCD detector, D. The location of the centroid of the
image on the CCD is a measure of the angle between the optic axis
and the normal to the mirror. In the conventional optical lever, a
single slit is imaged on to a photodetector or a position
sensitive photodiode. The resolution of the device improves
roughly as the inverse of the square root of intensity of light
detected and using a multi-slit scheme can increase the
sensitivity significantly compared to the single slit scheme. The
use of the linear CCD array allows measurement of deflections over
much larger range of angles than
what is possible in a conventional optical lever.\\

\begin{figure}
\begin{center}
\resizebox{8cm}{!}
{\includegraphics*{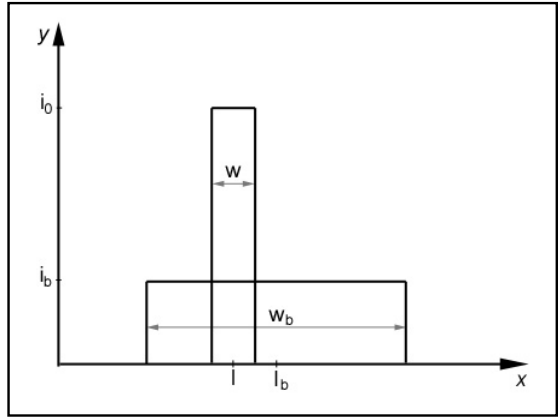}}
\caption{\label{Distribution O} Distribution $i(x)$}
\end{center}
\end{figure}

Let $i(x)$ represent the number density of the photons counted as
a function of their location $x$.  For considerations of the
design let $i(x)$ be a simple box shaped distribution as shown in
Fig.~\ref{Distribution O}.\\

The moment, $m$ of the distribution $i(x)$, say about $x=0$ is
simply given by
\begin{eqnarray}
m &=& \int_{-\infty}^{\infty}i(x)\cdot(x-0) dx; \nonumber\\
  &=& \left.
  i_0\frac{x^2}{2}\right|_{l-\frac{w}{2}}^{l+\frac{w}{2}};\nonumber \\
  &=& i_0lw.
\end{eqnarray}

The fluctuation in $m$ is given by,
\begin{eqnarray}
(\Delta m)^2 &\approx& \int i(x)\ x^2\ dx \nonumber;  \\
             & = & \left. i_0 \frac{x^3}{3}
             \right|_{l-\frac{w}{2}}^{l+\frac{w}{2}}; \nonumber \\
             & = & \frac{i_0}{3}(3l^2w+\frac{w^3}{4}).
\end{eqnarray}

Thus, the centroid of the image is obtained by dividing the moment
by the total number of recorded photons $I_0=i_0w$
\begin{eqnarray}
x_c  =  \frac{m  \pm \Delta m}{i_0w} = %
l \pm \sqrt{\frac{1}{I_0}(l^2+\frac{w^2}{12})}.
\end{eqnarray}

Now consider the presence of a background light which generates
counts $i_b(x)$ spread over a width $w_b > w$ about some location
$l_b$.  This will combine with the image and generate a new
centroid given by
\begin{eqnarray}
\begin{split}\label{centroid-with-background}
 x_b&=\frac{I_0l+I_bl_b}{I_0+I_b}\pm  \\
& \sqrt{\frac{I_0}{{(I_0+I_b)}^2}(l^2+\frac{{w}^2}{12})+
\frac{I_b}{{(I_0+I_b)}^2}(l_b^2+\frac{{w_b}^2}{12})}.
\end{split}
\end{eqnarray}

The presence of such a background induces both a systematic
uncertainty and an additional statistical uncertainty which can be
large.

One element in the design of our optical lever is a strategy to
eliminate the error due to background light. From the total
intensity field $i_t=i(x)+i_b(x)$ we subtract
$i_b+3\frac{\sqrt{i_bp}}{p}$ where p is the width of the
digitizing pixel which is much smaller than $w$ and $w_b$. $\delta
= 3\frac{\sqrt{i_bp}}{p}$ is the statistical fluctuation in the
background light. After such a subtraction we generate a new
intensity field given by
\begin{eqnarray}
i_n &=& \{i(x)+i_b(x)\}-i_b+3\sqrt{\frac{i_b}{p}}\\
& \approx & i(x)-3\delta\ \   for\ i_{n} > 0. \nonumber \\
& = & 0\ \ for\  i_n < 0.
\end{eqnarray}

The new centroid calculated with this $i_n$ is given by
\begin{equation}
x_n = l \pm \sqrt{\frac{(l^2+\frac{w^2}{12})}{I_0-3\delta w}}.
\end{equation}
Notice that $x_n \approx x_c$ when the fluctuation in the
background intensity, $\delta$, are small.

\begin{figure}
\begin{center}
\resizebox{11cm}{!}
{\includegraphics*{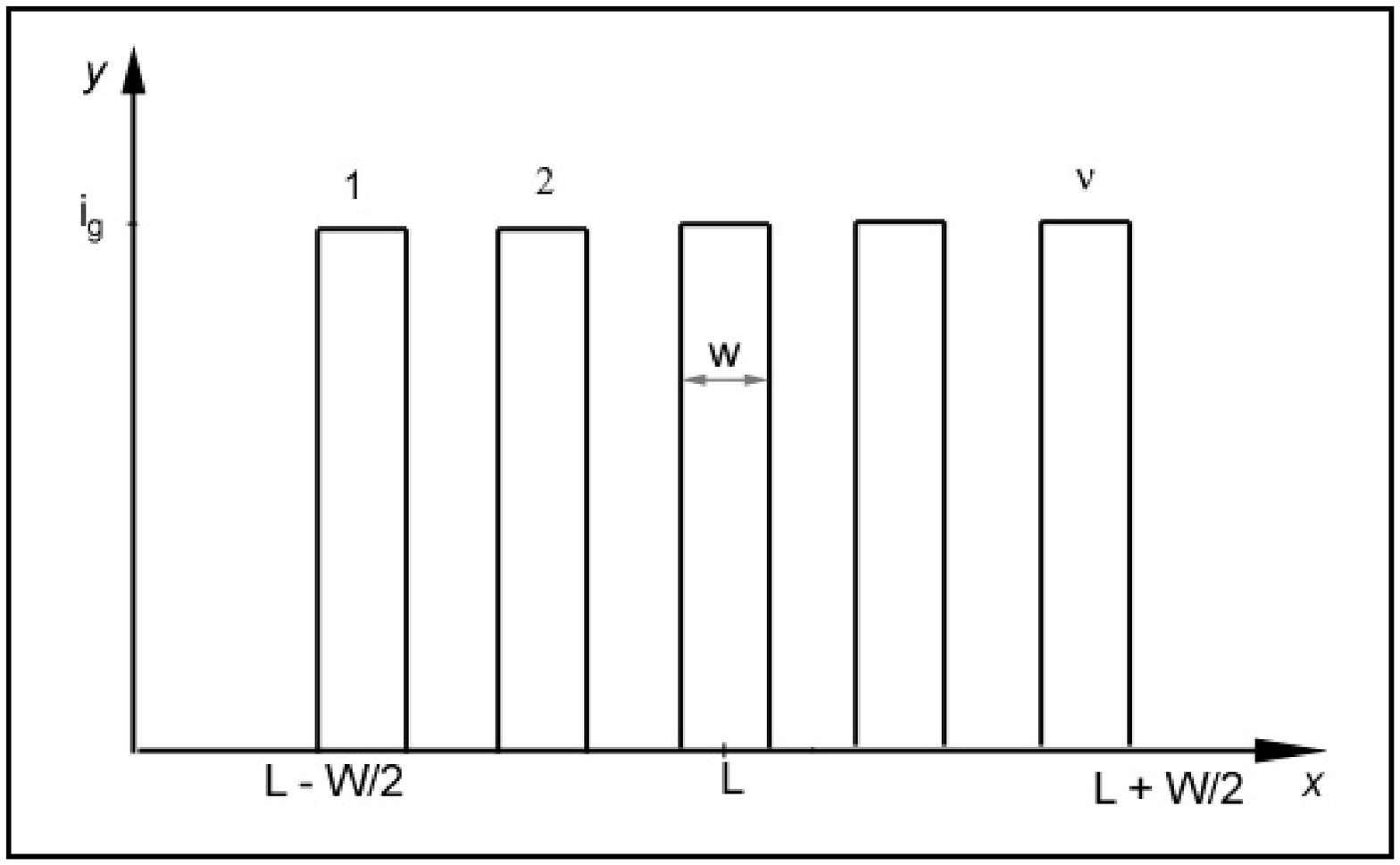}}
\caption{\label{grating} Grating}
\end{center}
\end{figure}

The second element in the design involves having multiple slits
and corresponding multiple peaks in the image.  Let us consider a
grating of $\nu$ elements with a spatial periodicity of $w$, with
$\frac{w}{2}$ opaque and $\frac{w}{2}$ transmitting
[Fig.~\ref{grating}]. The full length of the grating is $W=\nu w$.

Centroiding as before we get
\begin{equation}
x_g = L \pm \sqrt{\frac{2}{\nu I_0}(L^2+\frac{\nu^2
w^2}{12})}.\label{patch}
\end{equation}

The uncertainty in $x_g$ is much larger than that in $x_c$ and is
about $\sqrt{2 \nu} \Delta x_c$. This larger uncertainty is
essentially due to the fact that the width of the light
distribution has increased $\nu$-fold.

To overcome this, consider a set of $\nu$ fiducial points  $x_i$,
$i=1,2,...,\nu$, spaced at intervals of $w$.  The centroid of the
image of the individual grating elements with respect to the
corresponding fiducial elements is given by
\begin{equation}
x_{ci} = x_{ci}^{0} \pm \sqrt{\frac{1}{2
I_0}(\frac{w^2}{4}+\frac{w^2}{12})}.
\end{equation}

Averaging all the $x_{ci}$
\begin{equation}\label{centroid-equation}
\bar{x} = \frac{\sum x_{ci}^0}{\nu} \pm \sqrt{\frac{w^2}{6 \nu
I_0}}.
\end{equation}

The precision in the determination of the centroid is
substantially improved in this case as opposed to
Eqn.~\ref{patch}. It is as though the photon density of an
individual image of the slit has been increased by a factor $\nu$.
Since, the illumination of the slit is limited by the brightness
temperature of the source, the above method of decreasing the
statistical uncertainties proves useful.\\

The considerations related to the spacing of the grating are
straightforward.  The mirror whose deflection angle is to be
measured is smaller than the size of the autocollimator lens and
as such determines the diffraction width of the image of the slit.
Thus, the width $a_0$ of the opaque region between the slits may
be taken as twice the full width of the diffraction width due to
the mirror:
\begin{equation}
a_0 = 2 \frac{1.22 \lambda}{d_m} f .
\end{equation}

Here $d_m$ is the diameter of the mirror and $f$ the focal length
of the lens.  The width $a_t$ of the transparent part of the
grating should be chosen such that the width of its image covers
at least $\sim 5$ pixels, so that the spatial digitization of the
image is adequate and does not lead to inadequate sampling of the
possible asymmetries in the image profile.
\begin{equation}
a_t+\frac{1.22 \lambda}{d_m} f \gtrsim 5 p .
\end{equation}

Although one would like to keep $a_t$ as small as possible in an
attempt to improve the angular resolution of the optical lever, it
is necessary to keep $a_t$ to be at least as large as the
diffraction width itself to achieve a bright enough image, i.e.,
$a_t \sim 2.5 p$.  Further, it may be advantageous to choose its
width large enough so that the light illuminating it does not get
diffracted away even beyond the periphery of the lens.  Thus, the
grating constant $a=a_0+a_t$  and it is of the order of four to
five times the diffraction width of the mirror.

\section{Construction of the Optical Lever}
\begin{figure}
\resizebox{\textwidth}{!}{\includegraphics*[0cm,1.5cm][32.1cm,14.5cm]%
{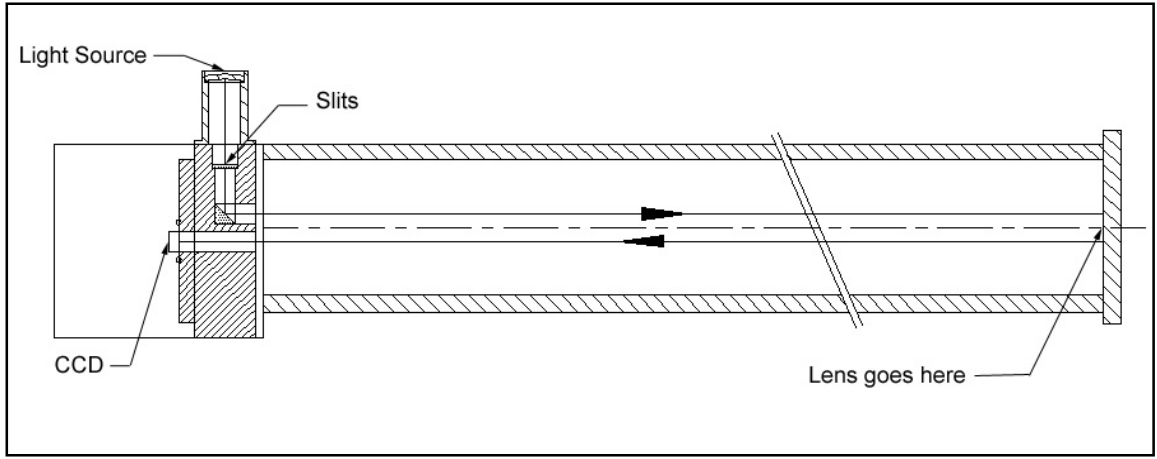}}
 \caption{Layout of optical lever assembly }
\label{collimator}
\end{figure}

A sketch of the optical lever is shown in Fig.~\ref{collimator}.
It consists of an array of slits made with a photographic plate
with transparent slits of $30~\mu$m width, 90 in number, separated
from each other with dark regions of $120~\mu$m width (opacity of
the dark regions is $< 100\%$). This array is illuminated by light
emanating from a brightly illuminated ground glass sheet placed
about five millimeters behind the array.  A bank of red LEDs which
emit in a forward cone angle of $\sim 15$~degree illuminate the
ground glass sheet with overlapping circles of light.  This
increases the brightness and uniformity of illumination of the
slit array. The array is placed in the focal plane of an
achromatic lens of $1000$~mm focal length and diameter $80$~mm.
The light passing through the slits is reflected towards the lens
by a right angled prism, located such that the virtual image of
the array is in the focal plane of the lens.  The collimated light
passing through the lens is reflected by the mirror whose
deflection is to be measured. The reflected light passes back
through the lens and gets imaged on the focal plane, just below
the virtual image of the source.\\

\begin{figure}
  \begin{center}
  \resizebox{10cm}{!}{
  \includegraphics[5mm,10mm][90mm,77mm]
  {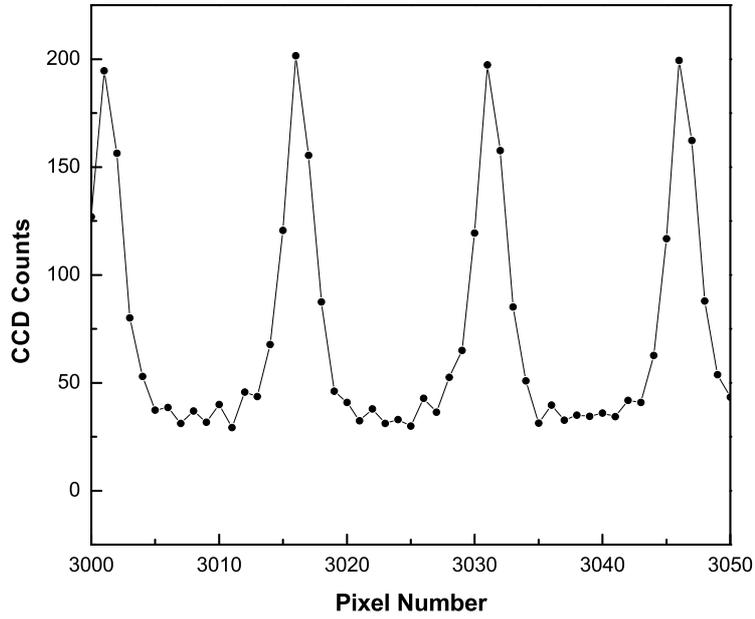}}\\
  \end{center}
  \caption{A sample image of the slits falling on the CCD}\label{Slit-image}
\end{figure}

The image is recorded on a CCD camera with a linear array of
$6000\ Turbosensor^\circledR$ photo-elements with a pixel size of
$10~\mu$m $\times$ $10~\mu$m and center to center spacing of
$10~\mu$m. The well depth is $10^5$~electrons. Two Analog to
Digital Convertors (ADCs) built into the CCD camera digitize the
charge on each pixels. The even pixels are read by one ADC while
the odd pixels are read by the other. A constant bias voltage is
added to the ADC input to ensure that the ADC always has a
non-zero input. The camera is controlled using a National
Instruments image acquisition card PCI 1424 and LabView software.
The card sends the clock signals required to transfer the charges
from the CCD pixels one by one to the ADC and also reads out the
digitized output of the ADC. The clock signals also control the
integration/exposure time of the camera. Thus a digitized image of
the slit array is obtained. This image is diffraction limited by
the size of the mirror attached to the balance, yet clear peaks
corresponding to the slits of the array can be
seen,~[Fig.~\ref{Slit-image}]. The figure shows the counts of the
ADC as a function of the CCD pixels. These images were obtained
with an integration time of $32$~ms. The calculation of the
centroid from these images will be described in \S\ref{Implement}.
When the mirror rotates through an angle $\theta$ the image and
hence its centroid moves through a distance $2F\theta$ on the CCD,
where $F$ is the focal length of the lens.\\

The autocollimator assembly - with the slits, lens and the CCD -
is housed such that the path from the slits to the lens and back
to the CCD can be evacuated down to $10^{-2}$~Torr and stays at
about $1$ Torr for a couple of days when the pump is off. This
helps to reduce image wobble on the CCD due to refractive index
changes in the air path caused by fluctuations of temperature and
by air currents.

\section{Implementation of the Centroiding
Algorithm}\label{Implement}

The  centroid of the image is proportional to the angle the normal
to the mirror subtends with respect to the optic axis. The first
step in the determination of the centroid is to subtract from the
individual pixel values $c_i$ the dark current and the bias. The
average value of the sum of dark and bias, $d_i$ for each pixel is
estimated from 100 frames without any light falling on the CCD.
Accordingly we set the effective counts $m_i$ as,
\begin{alignat}{2}
m_i &= c_i-d_i & \qquad \mathrm{for} \  d_i &< c_i \nonumber \\
m_i &=0        & \qquad \mathrm{for} \  d_i &> c_i
\end{alignat}

\begin{figure}
\begin{center}
\resizebox{\textwidth}{!}
        {\includegraphics*[0cm,3.3cm][32.1cm,14.7cm]
        {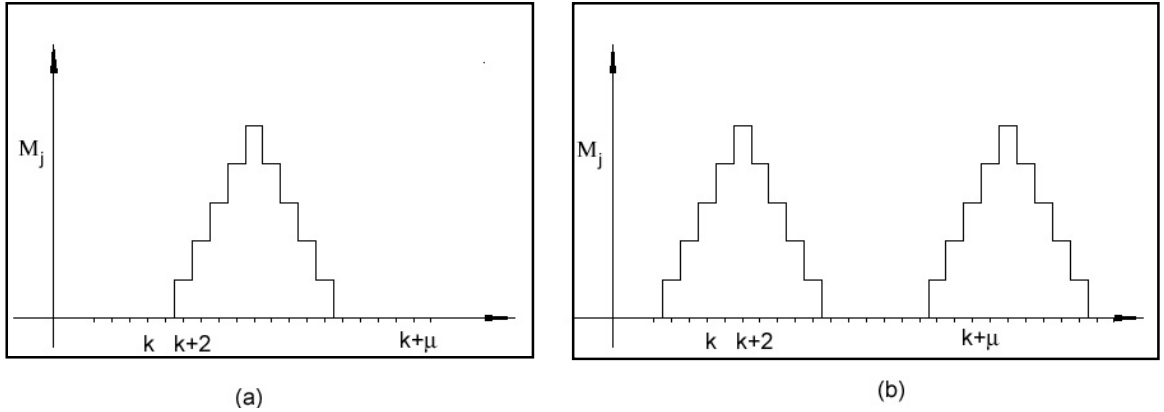}}
\caption{\label{typical image} Two possible image profiles with
respect to the fiducial location, k;  profile in panel (a) will
give an accurate centroid as per Eqn.\ref{centroid}; profile in
panel (b) will have larger errors, since the centroid is
calculated between $k$ and $k+\mu$}
\end{center}
\end{figure}

A typical image profile thus generated would appear as shown in
Fig.~\ref{typical image}. The next step in the analysis is to clip
off the low intensity background regions from the image. The
background counts $M_b$, is estimated as $25\%$ of the average
counts per pixel falling on the CCD,
\begin{eqnarray}
M_b & = 0.25\ \frac{\sum_{i=1}^N m_i}{N} \qquad & \mathrm{where} \
N = \mathrm{Number\ of\ pixels} \\
M_i & = m_i - M_b \qquad & \mathrm{for} \ m_i  > M_b \nonumber \\
M_i & = 0       \qquad & \mathrm{for} \ m_i  \leq M _b
\end{eqnarray}

The intensities $M_i$ represent the image of the sequence of the
grating slits with the background subtracted and the noisy low
intensity regions trimmed-off. \\

The next step in obtaining the centroid of the distribution takes
note of the points made in Eqn.~\ref{patch} and
Eqn.~\ref{centroid-equation}. These requires us to establish a
sequence of fiducial points across the CCD-array so that there is
a fiducial point in close proximity to each of the images of the
grating slits, $\nu$ in number.  The location $x_i$ of the first
set of fiducials is defined by
\begin{equation}
x_i=\left[\mu(i-1)+\frac{1}{2}\right]p\ , \qquad i=1,2,3,...D
\end{equation}
where $\mu$ is the width of the each slit image in pixels ($= 15$
in our case) and D is the total number of fiducial points.

Now computing the centroids of the intensity peaks that lie just
ahead of each of these fiducials and averaging  we get the
effective location of the image
\begin{equation}
{\displaystyle
\bar{x}=\sum_{k=1}^{L}\raisebox{1.5ex}{$\prime$}%
     \left( \frac{\sum_{j=k}^{k+\mu-1}M_j[\{(j-1)+ %
     \frac{1}{2}\}p-x_k]}{\sum_{j=k}^{k+\mu-1}M_j}  \right)}
     \label{bin-centroid}
\end{equation}

Notice that in the averaging process we have to include only those
values of $k$ for which all $M_j^s (j=k\ \mathrm{to}\ k+\mu-1)$
are not zero; this is indicated with a prime on the outer
summation sign. This procedure is similar to folding the image
$\nu$-times over so that all the peaks line up on each other.
$\bar{x}$ thus obtained is the centroid distance modulo $\mu p$
i.e.,  the mantissa.  This may be added to a constant $x_c$
defined by
\begin{equation}
\bar{x}_c={\displaystyle \sum_{i=1}^{L}\raisebox{1.5ex}{$\prime$}\
\frac{x_i}{\nu} }
\end{equation}
to get the complete centroid $X$:
\begin{equation}
X=\bar{x}_c+\bar{x} \label{centroid}
\end{equation}

One final step is needed before we are sure that the best possible
centroid has been obtained.  Consider Fig.~\ref{typical image}
showing two possible image patterns that may occur for the two
different orientations of the mirror.

In order to avoid the enhanced errors that will occur when image
profiles straddle the fiducial points as in Fig~\ref{typical
image}(b), we need to ensure that the fiducials locate the image
well within them. To this end, we introduce $\mu$ set of fiducials
$x_{i,j}$ where $j=1,2\ldots\mu$ such that $x_{i,j+1} = x_{i,j} +
1$. The image counts  at the fiducial locations, $M_{x_{i,j}}$ are
scanned to locate fiducial sets in which,
\begin{equation}
M_{x_{i,j}} = 0 \qquad for\ i = 1,2,\ldots D \label{good-fiducial}
\end{equation}

The centroid, $X_j$ are calculated as given in
Eqns.~\ref{bin-centroid}-\ref{centroid}, for all $x_{i,j}$ that
satisfy Eqn.~\ref{good-fiducial}. The `true' centroid, $C$ is
chosen to be the one for which $\bar{x}_{c,j}$ is closest to $\mu
/2$.

\section{Tests and characteristics of the optical lever}
Several tests were performed to characterize the optical lever.
The first was to study the dark and bias counts of the CCD pixels.
In the CCD the first and last 2 pixels are internally shielded
from light and are used by the CCD electronics to clamp the dark
and bias. The bias voltage applied to the ADC convertor, is
adjusted such that ADC output corresponding to the charges on
these pixels is locked at 4-5 counts. This keeps a check on drifts
due to the internal electronics. Over and above this, the counts
in each pixel of the CCD were monitored for an exposure time of
$32$ ms without any light falling on the CCD. 100 such points were
averaged for each pixel to determine the average sum of dark and
bias of each pixel. This was later subtracted from every frame
acquired, to determine the dark and bias subtracted
counts per pixel.\\

The diffraction limited image of the $30~\mu$m slits falling on
the CCD are 5 pixels wide and are separated by 15 pixels.  For an
integration time of $32$ ms, the peaks are at 200 counts compared
to the saturation counts of $255$ (with 8 bit digitization). The
counts in the region between the peaks is at 40 as against the
dark counts of 5 in the regions of the CCD where no light is
falling. Several such frames are co-added to increase photon
counts. Each frame with about 600 counts per peak spread over 5
pixels and a total of 90 such peaks leads to a photon shot noise
of $2.15 \times 10^{-2}$ pixels/frame. This sets $1.08 \times
10^{-7}$ radians/frame as the shot noise limit to the angular
resolution of
the optical lever.\\

\begin{figure}[h]
\begin{center}
\resizebox{10.0cm}{!}{
\includegraphics*[.5cm,.8cm][9.7cm,8.0cm]
{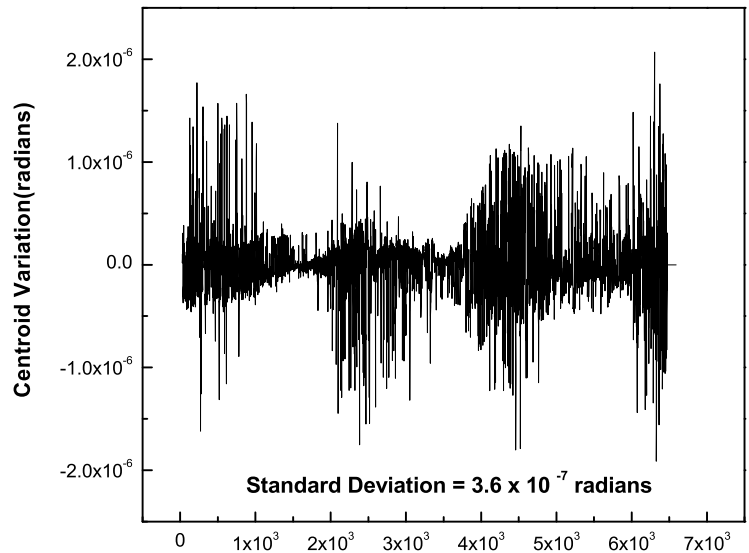}}
\end{center}
\caption{Plot of variation in centroid position of a fixed mirror
with time. The optical lever mount is at atmospheric pressure.
Interval between readings is 1.6 sec.} \label{air}
\end{figure}

In order to characterize the sensitivity of the system, the
optical lever is made to look at a rigidly mounted mirror and the
centroid of the peak intensity is monitored for several hours. The
image is recorded every 1.6 sec after summing over 50 frames of
$32$ ms exposure time. A plot of the mean subtracted value of the
centroid as a function of time is shown in Fig.~\ref{air}. The
fluctuations in the value of the centroid, as defined by the
standard deviation of the data, are found to be at the $3.6 \times
10^{-7}$ radians level, which is worse than the expected shot
noise limit of $1.53 \times 10^{-8}$ radians for 50 frames [see
Fig.~\ref{air}]. The source of this noise could be the air
currents inside the autocollimator tube.\\

\begin{figure}[h]
\begin{center}
\resizebox{10.0cm}{!} {\includegraphics*[.5cm,.5cm][9.7cm,8.0cm]
{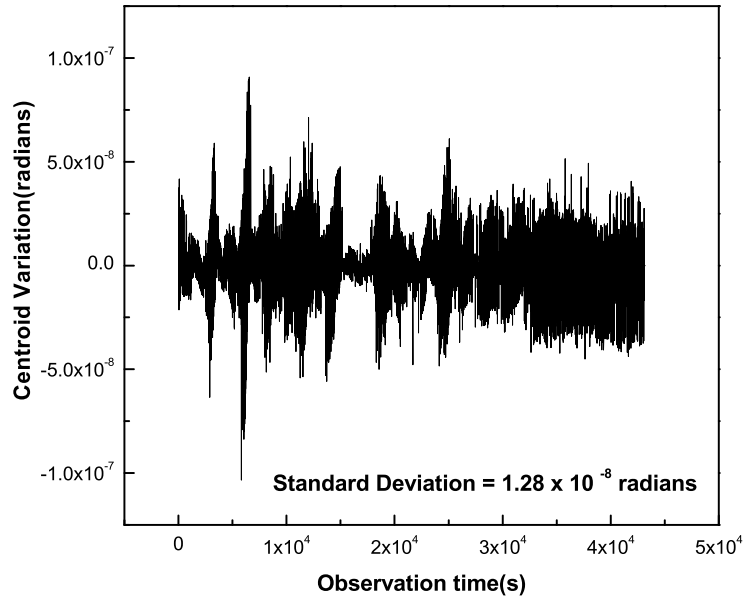}}
\end{center}
\caption{Plot of variation in centroid position of a fixed mirror
with time when the optical lever mount is at about 1 Torr.
Interval between readings is 1.6 sec.} \label{vaccum}
\end{figure}

To reduce the air currents, the path between the slit and the lens
has to be evacuated.  The fixed mirror is monitored as before
after pumping the autocollimator to $10^{-2}$ Torr. The variation
in the centroid in this case is shown in Fig.~\ref{vaccum}.
The fluctuations in the centroid value are reduced to about $1.28
\times 10^{-8}$ radians for 50 frames, which is at the shot noise
limit. Thus we have achieved a sensitivity of $\sim 1.6 \times
10^{-8}$ radians/$\surd$Hz. To improve on this, the total number
of photons incident on the CCD will have to be increased by
improving the mask and by efficiently coupling the light source to
the slits.\\

The dynamic range of the optical lever is $\sim 1.8 \times
10^{6}$. The maximum angle that the instrument can measure is
defined by the total length of the detector. The present mount for
the optical lever masks about 500 pixels on either end of the CCD.
Thus, with a total image width of $1.35$ cm, the largest
displacement in the centroid that we can measure is $3.65$ cm,
which corresponds to $1.8 \times 10^{-2}$ radian deflection of the
mirror.\\

 Even at this stage we can see the salient features of
this design of autocollimating optical lever: it is capable of
measuring angles with a sensitivity of $\sim10^{-8}$
radians/$\surd$Hz and has a dynamic range exceeding a few
millions. These features are quite unique. With a better mask and
light source we hope to reach a sensitivity of $3\times10^{-9}$
radian/$\surd$Hz without sacrificing the dynamic range.

\bibliographystyle{plain}
\bibliography{reference}

\clearpage

\chapter{Characterization of the Apparatus}

\emph{Abstract: This chapter presents the tests and
characterizations performed on the experimental apparatus.}

\section{The Torsional Pendulum}
The force transducer in our experiment is a torsional pendulum.
The sensitivity of the torsional pendulum to any torque acting on
it is related to the moment of inertia, $I$ of the mass element
suspended and torsional constant of the suspension fibre,
$\kappa$. These parameters determine the time period of
oscillation of the pendulum. The period, $T_b$ is given by $$T_b =
2 \pi \sqrt{\frac{I}{\kappa}}\ .$$ Another important parameter of
an oscillator is its quality factor, $Q$. It is a measure of the
damping in the system and is defined as the ratio of  time taken
for the amplitude of the oscillator to fall by $1/e$ to the time
period of the oscillator.\\

\begin{figure}
  \begin{center}
  \resizebox{!}{7cm}{
  \includegraphics[5mm,12mm][90mm,70mm]
  {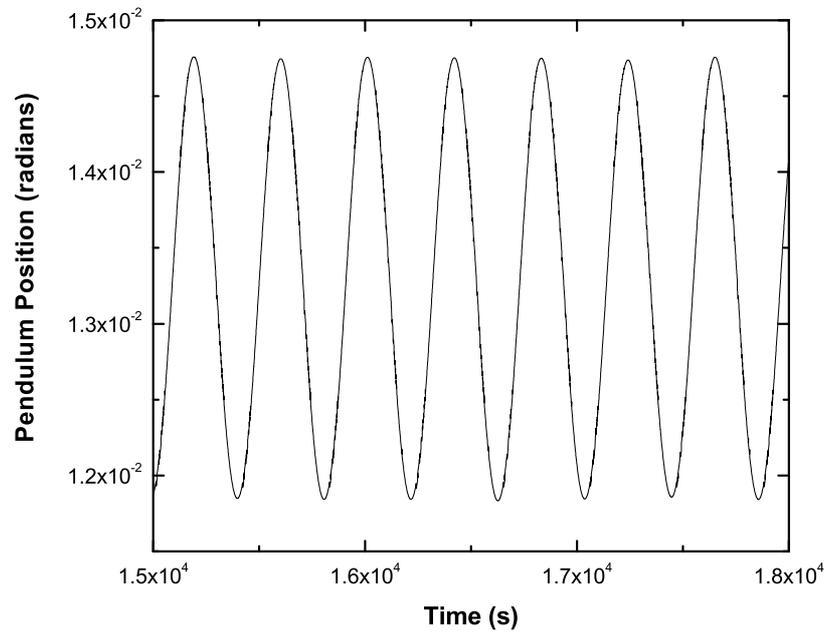}}
  \end{center}
  \caption{Free Oscillations of the pendulum}\label{oscillation}
\end{figure}

In order to characterize our torsional pendulum, it was allowed to
oscillate freely inside the vacuum chamber at a pressure of $4.0
\times 10^{-8}$ Torr with an amplitude of $\sim 3.0 \times
10^{-3}$ radians. The oscillations are monitored continuously for
about a day. Fig.~\ref{oscillation} shows a few cycles of the
oscillations.
\begin{figure}
  \begin{center}
  \resizebox{!}{7cm}{
  \includegraphics[58mm,12mm][140mm,70mm]
  {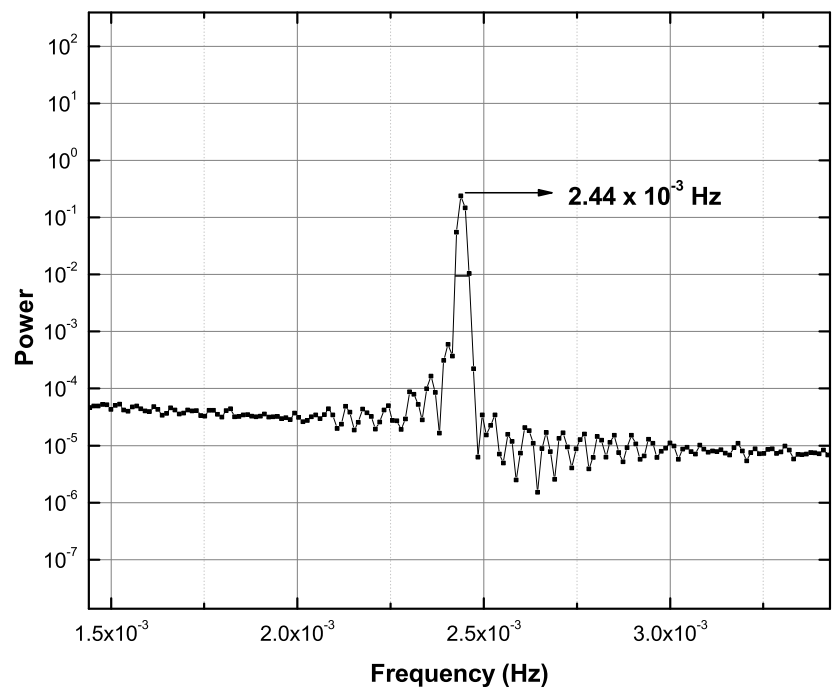}}
  \end{center}
  \caption{Power spectrum of the torsional oscillations}\label{FFT}
\end{figure}
The time period and Q are obtained from by Fourier power spectrum
of this time series data. The power spectrum calculated from the
Fast Fourier Transform(FFT) of the data is shown in
Fig.~\ref{FFT}. The peak in the spectrum occurs at a frequency of
$2.44 \times 10^{-3}$ Hz corresponding to a period of $409$ sec.
The quality factor as determined from the width of this peak is
$\sim 56$. But this is limited by the resolution of the FFT fixed
by the total observing time of about $20$~hrs. The true $Q$ is
$\sim 100$ as we can see from several such one day stretches of
oscillations. The equilibrium position of the pendulum drifts by
$\sim 5.0 \times 10^{-5}$~radians, peak to peak, over a day
(diurnal cycle). The signal to noise ratio achievable in the
system depends on the Nyquist noise which puts a limit on the
smallest torque that we can measure. The Nyquist noise is given
by,
\begin{eqnarray}
\label{t-ny} \tau_{Nyq} & = &\left( \frac{4k_B T
\kappa}{3Qt\omega_b} \right)^{1/2}
\end{eqnarray}
Here, $\kappa$ is the torsion constant of the suspension fibre,
$T$ the laboratory temperature, $Q$ is the quality factor of the
fibre, $t$ is  the total time of observation and  $\omega_b$ is
natural oscillation frequency of the balance. Taking typical
values $T = 300^\circ$ K, $\kappa = 0.05$ dyne.cm.rad$^{-1}$, $Q =
100$, $t = 50$ s, and $\omega_b = 1.5 \times 10^{-2}$
rad.s$^{-1}$, $\tau_{Nyq} \approx 6.1 \times 10^{-9}$ dyne.cm.
This value is two orders of magnitude smaller than the force
sought for in the present set of experiments.

\section{Pendulum behaviour during evacuation}
It is repeatedly noticed that the level of vacuum in the chamber
plays a crucial role in determining the trajectory of the pendulum
within the assembly. During assembly the lens is typically
positioned at a separation $> 5$ mm from the pendulum while the
capacitors are fixed at about $2$ mm from it. When pump down
starts, the pendulum tends to remain `stuck' to either of the
capacitors until a pressure of $\sim 10^{-3}$ Torr is reached. At
this stage the pendulum starts to oscillate with very short
periods of about $20$ s. As the pressure drops the period
progressively increases and reaches $\sim 200$ s at $10^{-5}$
Torr. The period stabilizes to the `free oscillation period' of
about $400$ s when the pressure reaches $\sim 10^{-7}$ Torr.
Further drop in the pressure does not change the period of the
pendulum. We have also tested the behaviour of the pendulum when
the pressure is allowed to rise due to degassing, with pumps shut
off. The period begins to drop as the pressure increases to about
$10^{-6}$ Torr. This behaviour of the pendulum was not
systematically characterized. But, it seems clear that even the
low pressure 'cushions' sandwiching the pendulum disc in the small
gaps between the capacitors and the disc can severely alter the
effective torsional spring constant of the oscillator. Our system
is operated at a vacuum of $\sim 4 \times 10^{-8}$~Torr to avoid
these effects.

\section{Calibration of the force arm}
\begin{figure}
  \begin{center}
  \resizebox{7cm}{!}{
  \includegraphics*[9mm,7mm][95mm,68mm]
  {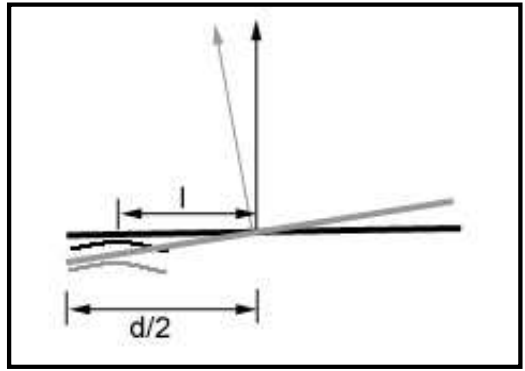}}\end{center}
  \caption{Schematic representing the `touch position' and its
   relation to the force arm}
  \label{lens-calib}
\end{figure}

\begin{figure}
  \resizebox{\textwidth}{!}{
  \includegraphics[5mm,5mm][147mm,78mm]
  {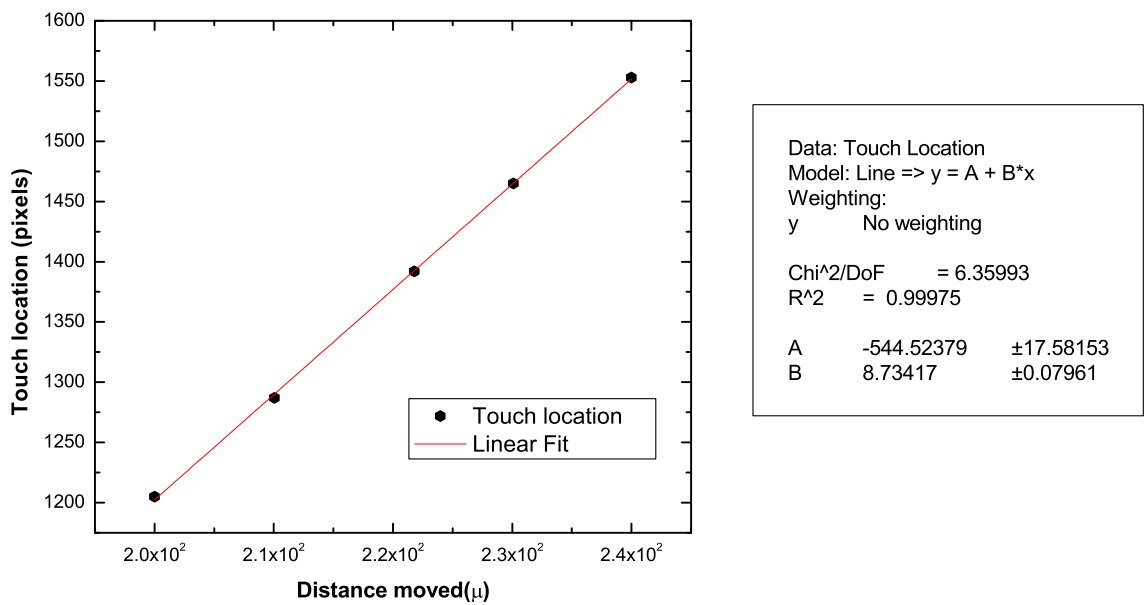}}
  \caption{Plot of the touch location as a function of lens position}
  \label{calibrate}
\end{figure}

The experiment attempts to measure the Casimir force between a
spherical lens surface and the flat surface of the torsion
pendulum. The torque due to this force on the pendulum, depends on
the lateral distance (force arm) of the point at which the force
acts on the pendulum as measured from the suspension axis. The
force arm is determined from the 'point of contact' between the
lens and the pendulum, which is estimated as follows: at some
initial position of the lens, the pendulum is pulled towards the
lens by applying voltages on the capacitor plates. The pendulum
position is monitored using the optical lever as it approaches the
lens. The position where the pendulum touches the lens is
deciphered from the centroid location on the optical lever at
which the pendulum suddenly turns around. The average point of
contact of the lens is measured from several touches at each
position of the lens. The lens is then translated by known amount
using the EncoderMike$^{\circledR}$ actuator [see \S2.4]. The
point of contact of the lens on the pendulum and the force arm
change negligibly, but the angular position of the pendulum at
which the contact occurs changes significantly
[Fig.~\ref{lens-calib}].
This new angular position at which the contact occurs is measured
for several displacements of the lens and a plot of it is shown in
Fig.~\ref{calibrate}.
Thus, the relation between the linear displacement of the lens and
the angular displacement of the centroid of the image at the point
of contact is obtained from the slope of the data. The slope tells
us that $0.114~\mu$m translation of the lens corresponds to $1$
pixel movement of the image centroid. Hence when the point of
contact moves transverse to the plane of the pendulum (towards or
away from the lens) by $0.114~\mu$m, image shifts by  $1$~pixel.
From the optical lever parameters, we know that a image
displacement of $1$ pixel corresponds to a transverse movement of
$0.2~\mu$m at the edge of the pendulum disc which is at $d/2 = 40$
mm from the suspension axis [Fig.~\ref{lens-calib}]. Thus the
force arm $l$ or point of contact is at $22.8$ mm from the axis of
the pendulum.


\clearpage

\chapter{Casimir Force Measurement - Strategy, Data Acquisition
and Analysis}

\emph{Abstract: The basic procedures followed in our experiment
will be described in this chapter. The chapter begins with an
overview of the forces influencing our torsion pendulum transducer
and goes on to describe how the data is acquired in order to
comprehend these. The analysis of this data to extract information
about Casimir force  is then discussed in detail. There is clear
evidence for the finite temperature Casimir force, detected for
the first time, in the analyzed data.}

\section{Strategy of the experiment}
The main objective of the experiment is to measure the force of
attraction between a fixed lens and a flat plate suspended as a
torsional pendulum. In the absence of the lens, the only torque
acting on the pendulum is the restoring torque from the fibre. As
the lens is brought close to the fibre, various forces begin to
act on the pendulum. The basic torques acting on the pendulum when
the pendulum oscillates near the lens are the fibre restoring
torque and the torques due to electrostatic and Casimir force.\\

The presence of all these forces will introduce changes on the
torsional oscillations of the pendulum. Since these forces in
general arise from potentials that differ from the quadratic
harmonic potential due to the torsion in the fibre, they make the
motion of the pendulum anharmonic. As described in the previous
chapters, we record the angular position, $\theta$ of the fibre as
a function of time, $t$. Thus the strategy adopted is to extract
information about the forces from the $t$ versus $\theta$ data.\\

The free oscillations of a pendulum will be simple harmonic. The
existence of other potentials, besides the harmonic potential due
to the fibre, will contribute to the acceleration $\ddot\theta$,
thereby changing the simple linear relationship of $\ddot\theta$
with the amplitude of the oscillations. If $f(\theta)$ is the net
force on the pendulum then,
\begin{eqnarray}
\hspace*{-5cm}\mathrm{The\ torque\ on\ the\ pendulum,\ \ }&& \nonumber\\
\tau(\theta) & =& f(\theta) l \mathrm{\ \ and}\\
I \ddot{\theta} & =& \tau(\theta)
\end{eqnarray}

where $l$ is the lateral distance of the point at which the force
acts on the pendulum as measured from the suspension axis,
$\ddot{\theta}$ is the angular acceleration of the pendulum.
Similarly the angular velocity $\dot{\theta}$ is related to the
potential $U(\theta)$,
\begin{eqnarray}
\frac{1}{2}I \dot{\theta}^2  + U(\theta)& = & E = const. \label{vel-pot}\\
\mathrm{where}\; \; \tau(\theta) & = & - \frac{\partial
U(\theta)}{\partial \theta} \label{accle-force}
\end{eqnarray}

 Thus, by measuring $\dot{\theta}$ or $\ddot{\theta}$ the
potential and the forces that affect the pendulum can be
characterized.

\subsection{Summary of Forces acting on the Pendulum}
Several forces between the lens and the pendulum begin to become
sizeable as they approach each other to within about $100~\mu$m.
We now discuss the forces that play a major role in the motion of
the pendulum.

\subsubsection{Fibre restoring torque:}
\begin{figure}
\begin{center}
  \resizebox{10cm}{!}{\includegraphics[10mm,10mm][223mm,173mm]
  {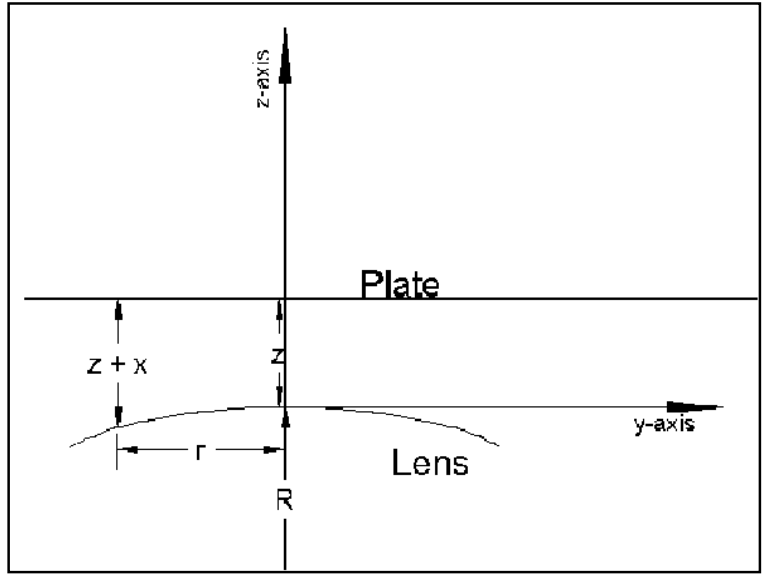}}
\end{center}\vspace*{-1cm}
  \caption{Reference axes for the system.}\label{axis-c}
\end{figure}

The restoring torque due to the fibre makes the pendulum execute
simple harmonic motion about a mean position, hereafter  referred
to as the `null' position $\theta_0$. As is well known, this
torque always acts towards $\theta_0$ and in proportion to the
deflection of the torsion balance from this position. In the
following descriptions, the distances and the angles are measured
with respect to the position of the lens as
origin~[Fig.~\ref{axis-c}]. During the experiment, the presence of
other forces will generate new torques which will shift the
balance to a new equilibrium position $\theta_e$. By appropriate
rotation of the shaft from which the balance is suspended, the
equilibrium position $\theta_e$ is made to lie between the lens at
$\theta_l =0$, and $\theta_0$. The restoring torque, $\tau_{fib}$
and the acceleration, $\ddot{\theta}_{fib}$ on the pendulum due to
this torque are given by:
\begin{eqnarray}
  \tau_{fib}(\theta)  &=& \kappa (\theta_0 - \theta) \\
  \ddot{\theta}_{fib} &=& \frac{\kappa}{I}(\theta_0 - \theta)\\
 \ddot{\theta}_{fib} &=& 2.46 \times 10^{-4} (\theta_0 - \theta)
 \label{fibref}
\end{eqnarray}
where $\kappa $ is the torsion constant of the fibre, $I$ the
moment of inertia of the pendulum and $\theta_0$ is the
equilibrium position of the pendulum. Similarly the potential,
$U_{fib}$ and the angular velocity, $\dot{\theta}_{fib}$ are given
by,
\begin{eqnarray}
  U_{fib} & = & \frac{\kappa}{2} (\theta_0 - \theta)^2 + const.
\end{eqnarray}
It is convenient to write in general,
\begin{eqnarray}
 \dot{\theta}_{fib}^2 & =& -\frac{2 U }{I}+ const. \\
                     & \equiv& - U_N + const.
 \label{fibreU}
\end{eqnarray}
With this we may think of $\dot{\theta}^2$ as just the negative of
the normalized potential $U_N$.

\subsubsection{Casimir Force}
The Casimir force per unit area between two parallel conducting
plates separated by a distance $z$, at $T=0$ K is given by,
\begin{eqnarray}
f_{c0}(z)& = &-\frac{\pi^2 \hbar c}{240z^4} \label{f-cas0}\\
      & = & -\frac{0.013}{z_\mu^4} \quad \mathrm{dyn.cm}^{-2}\ \
          \mathrm{where} \ z_\mu \equiv z\ \mathrm{in\ microns}
\end{eqnarray}
For a finite temperature $T$ we approximate the distance
dependence as
\begin{eqnarray}
f_{cT}(z)  & = & f_{c0}(z) \hspace*{1.25cm}\mathrm{for}\ \ z < \lambda_T\\
 f_{cT}(z) & = & f_{c0}(z)\frac{z}{\lambda_T }
  \hspace*{0.7cm} \mathrm{for}\ \ z > \lambda_T \label{f-casT}
\end{eqnarray}
where $\lambda_T$ is the distance at which the finite temperature
effects become important. Comparing Eqn.~\ref{f-casT} with the
expression for Casimir force at `high temperature' described in
Chapter 1 (Eqn.1.5),
\begin{equation}
F_c^T(d)  \simeq -\frac{1.2k_BT}{4\pi z^3}\quad\text{at\;high\;}T
\label{fcT}%
\end{equation}
we can estimate that for $ T=303^{\circ}\ \mathrm{K,\;}\lambda_T\sim%
3.26~\mu$m.

 For convenience, we write Eqns.~\ref{f-cas0} and
\ref{f-casT} as,
\begin{eqnarray}
f_{c0}(z) & \equiv & -\frac{A_c}{z^4} \ \ \ \ dyn.cm^{-2}\\
f_{cT}(z) & \equiv & -\frac{A_c}{\lambda_T z^3}  \ \ \ \
dyn.cm^{-2}
\end{eqnarray}

\begin{figure}
\begin{center}
  \resizebox{9cm}{!}{
  \includegraphics*[11.6cm,14.5cm][19.3cm,20.3cm]
  {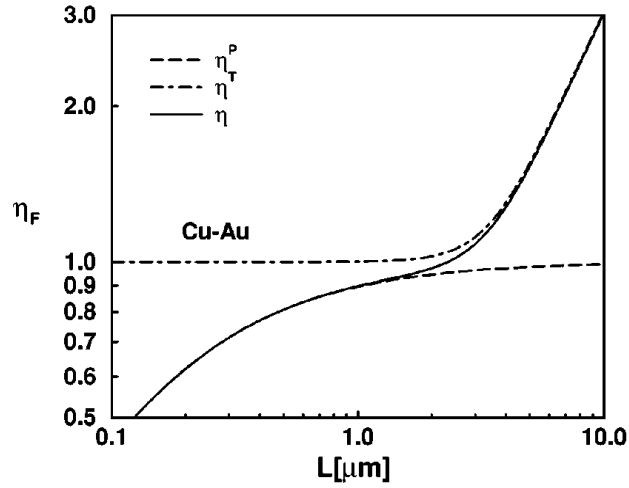}}
\end{center}
   \caption{Ratio of the Casimir force calculated taking into account
the finite conductivity and finite temperature effects to the
Casimir force at zero temperature between ideal conductors is
plotted as a function of spacing between plates
(from,~\cite{Reynaud2000}). Our experiment is in the distance
range of $2~\mu$m - $10~\mu$m }
  \label{Rey}
\end{figure}
 The Casimir force between the lens and the pendulum can now
be estimated using the ``Derjaguin (proximity) approximation",
often called ``proximity theorem"~\cite{Proximity}. For our
geometry the expression has two parts to it depending on the
separation between the lens and the plate. At small separations,
the zero temperature Casimir force dominates. As the separation
increases, finite temperature effects become important and the
force law changes. Fig.~\ref{Rey} shows this change in the law
(see chapter 1 for the definition of the ratios). The Casimir
force for the lens-plate configuration is given by (see Appendix~B
for details),
\begin{eqnarray}
f_{cl}(z) & =&- \frac{2 \pi R A_c}{3}
          \left [ \frac{1}{z^3}-\frac{1}{\left(z +\lambda_T\right)^3}%
         + \frac{3}{2 \lambda_T}\ \left \{ \frac{1}{(z + \lambda_T)^2}- %
          \frac{1}{(z + x_{max})^2} \right \} \right ], \label{Casimir-l-1} \\
&  & \hspace*{75mm} \mathrm{for\ } z < \lambda_T   \nonumber \\
f_{cl}(z) & = & - \frac{\pi R A_c}{\lambda_T}
        \left\{ \frac{1}{z^2} - \frac{1}{(z + x_{max})^2}\right\},
        \hspace*{20mm} \mathrm{for\ } z >  \lambda_T
        \label{Casimir-1-2}
\nonumber
\end{eqnarray}

where $x_{max} = \frac{r_{max}^2}{2R}$ and $z = l \theta$ is the
separation between the lens and the plate, $R$ is the radius of
curvature of the lens and $r_{max}$ is the aperture radius of the
lens (Fig.~\ref{axis-c}). The potentials for the same
configuration are,
\begin{eqnarray}
U_{cl}(z) & = & \frac{ \pi R A_c}{4}
          \left [ \frac{1}{z^2}-\frac{1}{\left(z +\lambda_T\right)^2}%
         + \frac{8}{3 \lambda_T}\ \left \{ \frac{1}{(z + \lambda_T)}- %
          \frac{1}{(z + x_{max})} \right \} \right ], \label{U-Casimir-l} \\
&  & \hspace*{84mm} \mathrm{for\ } z < \lambda_T   \nonumber \\
U_{cl}(z) & = &  \frac{2 \pi R A_c}{3 \lambda_T}
        \left\{ \frac{1}{z} - \frac{1}{(z + x_{max})}\right\}
        \hspace*{33mm} \mathrm{for\ } z >  \lambda_T
        \label{U-Casimir2}
\end{eqnarray}
The angular velocity and acceleration due to Casimir force can be
calculated using Eqns.~\ref{vel-pot} and \ref{accle-force}.\\

It may be noted that the roughness of the surface is not a major
correction term to the Casimir force in our experiment, since it
focuses on Casimir force at relatively large separations between
the metal surfaces. However, the finite conductivity correction, a
reduction in force amounting to approximately $10\%$ needs to be
applied throughout the measurement range.

\subsubsection{Electrostatic Forces}
There are various sources of electrostatic force between the lens
and the plate. Several effects lead to potential differences
between metal coated surfaces even when they are externally
connected to ground \cite{Speake1996,Speake2003,Adel2003}. Small
difference in the contact potentials in these connections can lead
to potential differences between these surfaces. With care, such
potentials can be minimized to a few mV level. Another source of
electrostatic interaction between the bodies are patch fields.
These are microscopic electric potentials found on surfaces of
metals. The different crystalline planes of a given material can
have work functions that can differ by as much as $1$ V . If the
coated metal surface is a mosaic of random microscopic crystal
planes, local potential differences will occur between each of
these micro-size patches. Surface contamination in the metal
coatings can also lead to local variations in the work function
and also provide sites for trapping electrical charges.

\subsubsection{Force due to Capacitative coupling:} The conductive
surfaces of the lens and the plate, form a capacitance which is
given by ``proximity theorem" to be,
\begin{eqnarray}
C_l & = & \epsilon_0 \langle \frac{A}{d} \rangle\\
    & =& \epsilon_0 \int_{r=0}^{r=r_{max}}
              \frac{2 \pi r dr}{[z + \frac{r^2}{2R}]} \\
    & =& 2 \pi R \epsilon_0 \ln \left\{\frac{z + \frac{r_{max}^2}{2R}}{z}\right\}
\end{eqnarray}
where $\epsilon_0$ is the dielectric constant in free space, $R$
is the radius of curvature of the lens, $z$ the separation between
the lens and the pendulum in metres and $r_{max}$ is the aperture
radius of the lens. If the voltage difference, $V$ between the
lens and the plate were zero, the net force due to this
capacitance will be zero. But potentials can be present due to any
of the reason mentioned above. The electrostatic potential energy
due to these between the lens and the plate is given by,
\begin{eqnarray}
  U_{el} &=&  \frac{1}{2}C_l V^2 \\
         &= & \pi R \epsilon_0 V^2 \ln \left\{\frac{z + \frac{r_{max}^2}{2R}}{z}%
         \right\} \label{U-el}
\end{eqnarray}

The resulting capacitative force is given by,
\begin{eqnarray}
  f_{el} &=& - \frac{d}{dz} \left[ U_{el} \right],\\
         &\approx& - \frac{\pi R \epsilon_0 V^2}{z}\hspace*{1cm} %
          \mathrm{for}\ \  r_{max}^2 \gg 2Rz. \\
\end{eqnarray}
We approximate the corrections to the above formula, due to
surface roughness and a variety of other causes and write for the
torque,
\begin{eqnarray}
  \tau_{el} &=& - \frac{\pi R \epsilon_0 V^2}{(\theta - \theta_r)} \\
\end{eqnarray}
  where the roughness parameter $\theta_r$ will be determined
self-consistently from the data. The voltage $V$ comprises of the
applied voltage $V_{ap}$, the D.C voltage $V_{(0,dc)}$ due to
contact potentials and other causes and $V_{(0,ac)}$ due to the
stray pick up fields and leakage, etc. This contributes to the
normalized force a term,
\begin{eqnarray}
  \ddot{\theta}_{el} &=& - \frac{\pi R \epsilon_0 V^2}{I (\theta-\theta_r)}. \label{el}
\end{eqnarray}

\subsubsection{Force due to charges on the lens/pendulum:}
Stray electric charges on the surfaces of the lens and the plate
also lead to electrostatic forces. The potential between these
charges depends in detail on their distribution on the surfaces
and cannot be modeled easily [see for example, \cite{Smythe39}].
An upper limit to the strength of this force may be obtained by
assuming that all the charge is concentrated at the point of
contact between the lens and the plate and that a simple inverse
square law applies for the force with an equidistance image
charge. The potential due this is given by,
\begin{eqnarray}
  U_{qq} &=& \frac{q^2}{4 \pi \epsilon_0 z_m} \label{U_qq},\\
  z_m &=& 2z ,\mathrm{the\ separation\ between\ the\ lens\ and\ the\ pendulum.}
\end{eqnarray}

The force and the acceleration on the pendulum due to it are,
\begin{eqnarray}
  f_{qq} &=& - \frac{q^2}{4 \pi \epsilon_0 z_m^2}. \\
  \mathrm{Torque,\ \ }
  \tau_{qq} &=& - \frac{q^2}{16 \pi \epsilon_0 l \theta^2}, \\
  \ddot{\theta}_{qq} &=& - \frac{q^2}{16 \pi \epsilon_0 l I \theta^2}, \\
   &=& - 2.94 \times 10^{15} \frac{q^2}{\theta^2}.\label{qq}
\end{eqnarray}
   where $q$ is effective charge on the lens/pendulum. For a
   distributed charge the $\sim 1/\theta^2$ dependence will change over
   to a slower dependence of $\sim 1/\theta$. This will be the case
   with metallic surfaces when there are some residual charges on
   them.

\subsubsection{Force due to charge and voltage on the lens:} One
more source of electrostatic force is the combined effect of the
charges and voltage differences present. The electric field,
$\mathcal{E}$, due to these voltages, $V$ interacts with the
charge, $q$ to apply an additional force on the pendulum of the
form,
\begin{eqnarray}
  f_{vq} &=& - q \mathcal{E} = - \frac{q V}{z}, \\
  \mathrm{Torque,\ \ }
  \tau_{vq} &=& - \frac{q V l}{z},\\
  \mathrm{Acceleration,\ \ }
  \ddot{\theta} &=& - \frac{q V}{I \theta}, \\
   &=& - 5.2 \times 10^{4}\frac{q V}{ \theta}.\label{vq}
\end{eqnarray}
$V_{0} = V_{(0,dc)} + V_{(0,ac)}$; however, since the acceleration
is linear in V as shown in Eqn~\ref{vq} and the pick up voltages
are at frequencies much higher than the frequency of oscillation
of the balance at a few milliHertz, the A.C torques will average
to zero and will not contribute.

\subsubsection{Gravitational force:} The gravitational attraction
of the lens on the pendulum gives rise to a torque of $\sim 4
\times 10^{-5}$ dyne.cm on it. This corresponds to the attraction
between disc shaped masses of approximately $10$ g each separated
by an effective distance of about $5$ mm. The `compensating plate'
(refer \S2.5) is used to balance this gravitational torque of the
lens. The mass of the compensating plate is equal to that of the
lens but its position with respect to the torsion pendulum disc is
not exactly the same as the lens. Therefore, the cancellation of
the gravitational torque is only approximate and the residual
gravitational torque on the pendulum is estimated to be $\sim
10^{-6}$ dyne.cm. This changes only by $\sim 10^{-8}$ dyne.cm when
the lens is moved over the distance of $50~\mu$m, since the
fractional change in the effective separation between the centers
of gravity of the plate and the lens assembly is only $1/100$. The
change in angular acceleration of the pendulum due to this is
below the present sensitivity of the experiment.
\begin{figure}[ht]
  \resizebox{14cm}{!}{\includegraphics
  {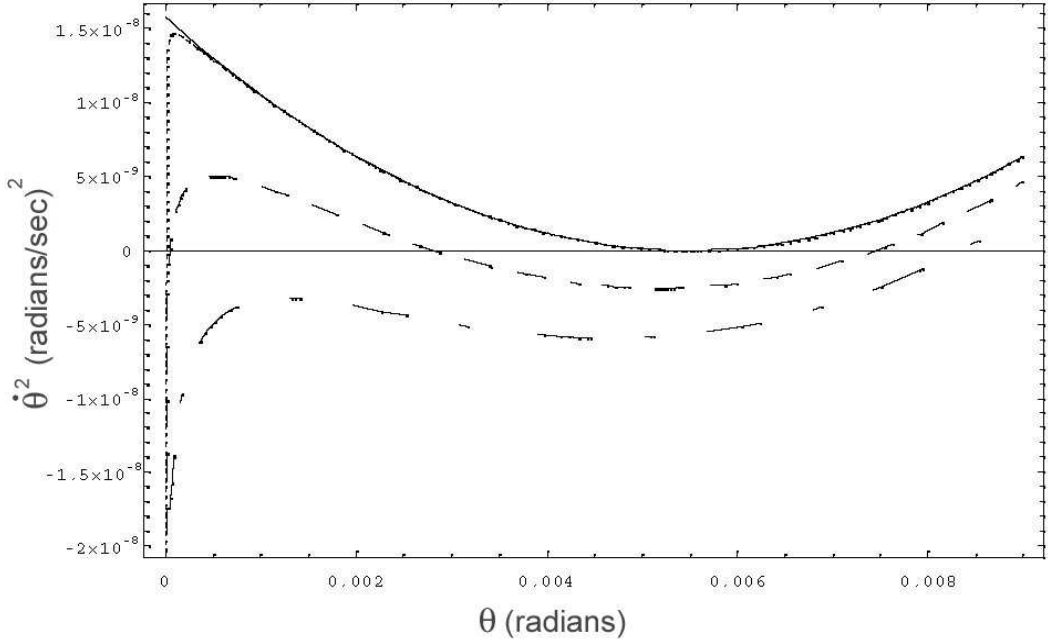}}
  \caption{A plot of normalized potential $U_N = 2U/I$ is shown as a
function of $\theta$. The solid line maps the potential due to the
fibre restoring force when the equilibrium position of the
pendulum, $\theta_0=5.3\times 10^{-3}$ radians. The dotted line is
a plot of the net potential due to the lens and the fibre for
$V=5$ mV, $q=1\ \times \ 10^{-15}$ C and $\lambda=3.26~\mu$m. The
dashed line is a plot for $V=50$ mV and the dashed dot line for
$V=75$~mV on the lens}
   \label{theta-V}
\end{figure}

\subsection{The Effect of these Forces on the Pendulum}
A plot of the square of the angular velocity, $\dot{\theta}^2=
-U_N$ of the pendulum as a function of $\theta$ is shown in
Fig.~\ref{theta-V}. This maps the potential in the region of
oscillation of the pendulum in the absence of the lens (solid
line) and when the lens is at a distance of about $120~\mu$m from
the null position $\theta_0$ (various dashed lines). For
calculating the potential due to the lens, the net voltage
difference between the lens and the pendulum $V$ is set to $5$ mV
(dotted line), $50$ mV (dashed line) and $75$ mV (dashed dot
line). The net charge, $q$ is taken to be $1\times10^{-15}$ C and
the distance at which the Casimir force changes from zero to
finite
temperature form, $\lambda$ is taken to be $3.26~\mu$m.\\

 From the plot it can be seen that the presence of the
electrostatic and Casimir forces, considerably changes the
potential between the lens and the plate. The potential is highly
attractive towards the lens for small angular separations. The
angular distance at which this change over to attractive potential
happens depends on the net voltage on the lens.  Thus, when the
voltage on the lens is $5$ mV, the change over happens at a
distances of about $5\times10^{-5}$ radians. As the voltage on the
lens increases, the change over point extends away from the lens.
The equilibrium position of oscillation of the pendulum is also
shifted for voltages higher than $20$ mV. The potential beyond
$7\times10^{-3}$ radians is shifted for higher voltages but is
parallel to the pontetial due to free fibre. The potential below
$7\times10^{-3}$ radians is flatter than the potential due to free
fibre. Thus the oscillation period of the pendulum is
modified even for small amplitude oscillations of the pendulum.\\

 As the lens is brought close to the pendulum \emph{the
signatures of the electrostatic forces can be seen by looking at
changes in:
\begin{description}
    \item[(a)] the equilibrium position of the pendulum,
    \item[(b)] the time period of the pendulum, and
    \item[(c)] the angular acceleration $\ddot{\theta}$ of the
    pendulum as a function of its angular position  $\theta$.\\
\end{description}
}

 Since the Casimir force falls off more rapidly than the
electrostatic force, to study the effects of Casimir force, we
need to go closer to the lens, \emph{i.e.}, $<1\times10^{-3}$
radians. At those angular distances, the pendulum will no longer
oscillate but will start to fall towards the lens. The force in
that region then can be elucidated only from the $\theta$ versus
$\ddot{\theta}$ of the pendulum. \emph{Hence, to estimate Casimir
force, we need to look at the $\theta$ versus $\ddot{\theta}$ of
the pendulum at angular distances of $<1\times10^{-3}$ radians.}\\

 To completely fathom the electrostatic forces
though, in addition to measurements of the changes on the pendulum
as a function of lens position, knowledge about these changes as a
function of voltage on the lens are also essential. The force due
to the capacitive coupling increases as the square of the net
voltage on the lens, while the charge effect is linear in voltage.
Hence, by studying the force dependance on the voltage, an
estimate of the effect of both the residual voltage and the
charge on the pendulum can be obtained.\\

\section{Data Acquisition:}
Despite considerable care taken during the assembly to minimize
the voltages and charges on the lens and the pendulum, certain
residuals are unavoidably present in the apparatus and  it is
essential to characterize these electrostatic forces. All the
earlier experiments that looked for Casimir forces have seen the
effect of these forces
\cite{Derj54,Derj60,Sparnaay57,Overbeek78,Lamor97,Mohi98,Chan01,Ruoso02}.
Since the distance range in our experiment is larger than the
earlier experiments on Casimir force, electrostatic forces are a
strong background. Hence, the data acquisition was tuned to be
able to estimate the electrostatic forces and subtract them in
order to look at the Casimir force at separations between $2~\mu$m
- $10~\mu$m.\\

 The Casimir force falls off more rapidly with distance
than the electrostatic forces. Thus, at distances $>25~\mu$m the
electrostatics forces will dominate by more than two orders of
magnitude. At these separations corresponding to a angular
distance of $> 1\times10^{-3}$ radians, even for a net
voltage,~$V$ of $75$~mV between the lens and the pendulum, it will
execute anharmonic oscillations.  Hence, the presence of the
electrostatic force can be studied by looking at the oscillations
of the pendulum in the potentials summarized in
the previous section. \\

 Data was acquired in three main modes:
\begin{itemize}
    \item Small amplitude oscillations that sampled the region
    within $ 5\times10^{-4}$ radians of the equilibrium position.
    \item Large amplitude oscillations that sampled regions up to
    an angular separation of $1\times10^{-3}$ radians from the lens.
     \item Accelerated fall of the pendulum onto the lens starting
     from an angular separation of $< 5\times10^{-4}$ radians.
\end{itemize}

\subsubsection{Small amplitude oscillations}
\begin{figure}
\begin{center}
  \includegraphics[5mm,15mm][95mm,80mm]
  {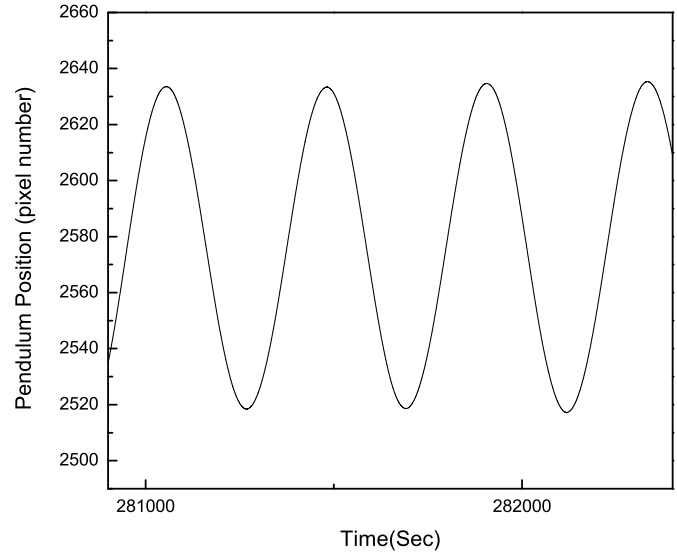}\\
\end{center}
  \caption{The raw data showing a few cycles of small amplitude
   oscillations of the pendulum }\label{smalloscil}
\end{figure}

To start with, the lens is positioned at a distance of about
$0.35$ mm from the equilibrium position of the pendulum. The
oscillations of the pendulum for amplitude of $\sim
2.5\times10^{-4}$ radians, i.e, $5.7~\mu$m are monitored using the
autocollimator. The lens is then moved stepwise slowly towards the
pendulum  until the separation between the two is reduced to
$100~\mu$m. At each separation, the pendulum is allowed to
oscillate with an amplitude of $ \sim 2.5\times10^{-4}$ radians
for several ($>10$) cycles. The position of the pendulum in units
of CCD pixel numbers is recorded every $0.16$~sec.
Fig.~\ref{smalloscil} shows
the raw data of few cycles of oscillations of the pendulum. \\

 From this data the change in the period and the
equilibrium position of the pendulum as a function of lens
position is derived.  As the lens is brought closer the
equilibrium position, the period of the pendulum remained
unchanged until the separation is $> 250~\mu$m. For lower
separations, the equilibrium position of the pendulum shifts
closer to the lens. The period changes from  about $406$ sec at
$0.35$ mm separation to about $503$ sec at $100~\mu$m separation.\\
\begin{figure}
\begin{center}
  \includegraphics*[5mm,5mm][90mm,80mm]
  {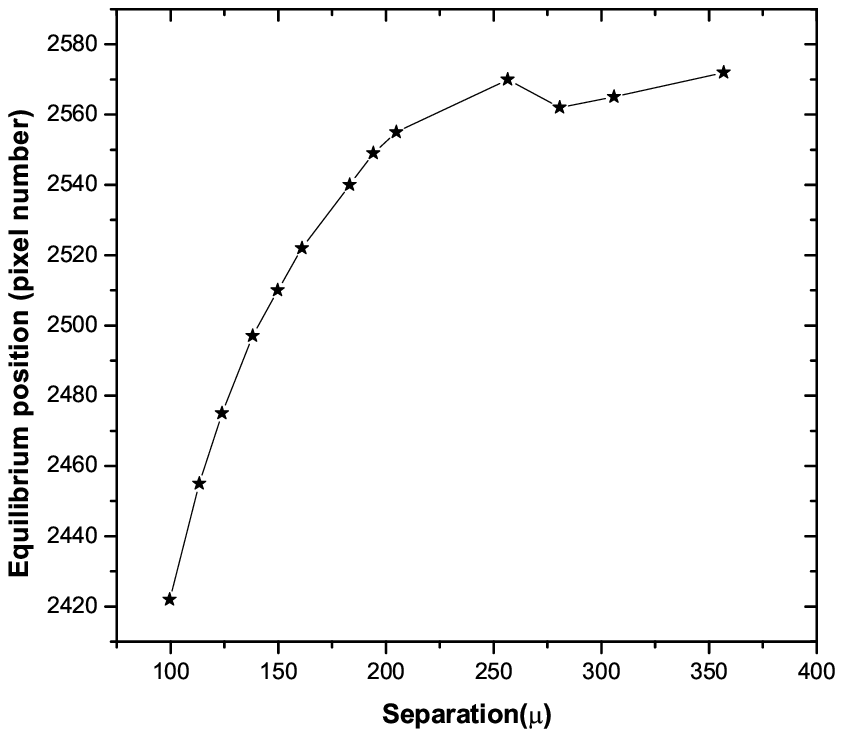}\\
\caption{The change in the equilibrium position of the pendulum as
the lens is moved closer to it.}
  \label{sep-equil}
\end{center}
\end{figure}

\begin{figure}
\begin{center}
  \includegraphics*[5mm,5mm][90mm,80mm]
  {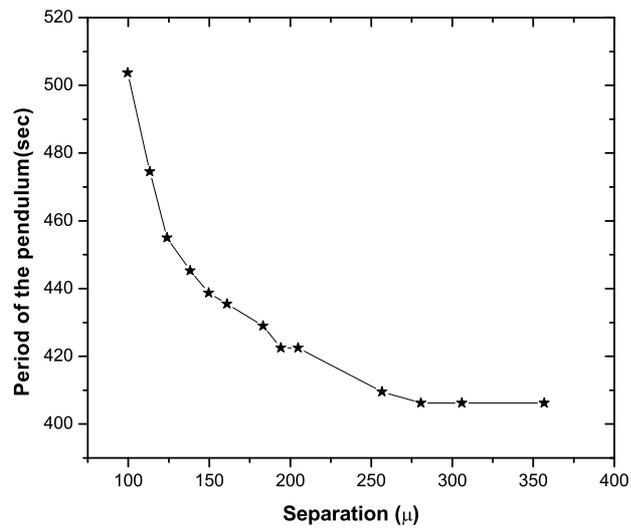}\\
\caption{The change in the time period of the pendulum as the lens
is moved closer to it.}
  \label{sep-period}
\end{center}
\end{figure}

 These changes are shown in Figs.~\ref{sep-equil} and
~\ref{sep-period}. Fig.~\ref{sep-equil} shows the change in
equilibrium position as the lens is moved closer to the pendulum.
The pixel numbers on the CCD are used as the reference to see this
change. The equilibrium position is not measured with the lens as
origin as the position of lens itself changes. The separation
marked is the effective separation calculated taking into account
both, the change in equilibrium  position and the change in the
lens position. Fig.~\ref{sep-period} shows the variation in the
time period of the pendulum for the same separations.\\

Thus, at \emph{separations $> 250~\mu$m} the interaction of the
pendulum with the lens is negligible and the motion of the
pendulum is defined by properties of the suspension fibre. Hence,
this region \emph{can be used to characterize the fibre
properties}. To study the effects of the interaction with the
lens, separations $< 120~\mu$m are ideal as the change in the
equilibrium position and time period are steeper in this
region.\\

\begin{figure}[h]
\begin{center}
\resizebox{\textwidth}{!}{\includegraphics
{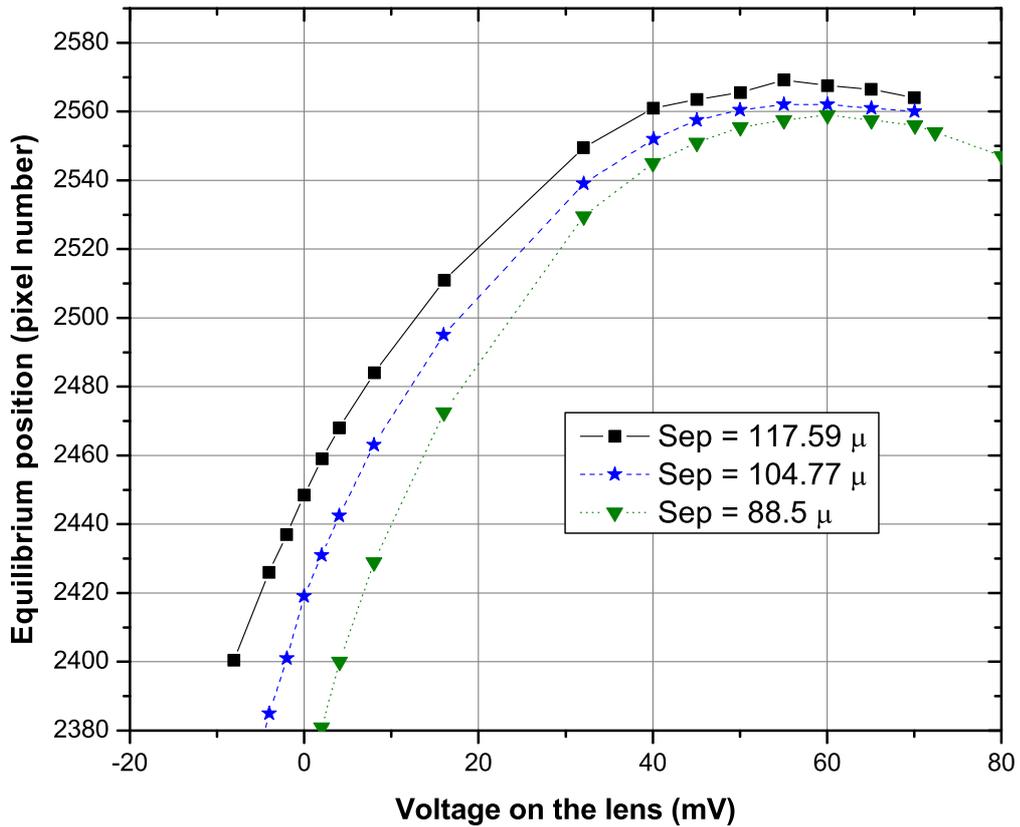}}\\
\caption{The equilibrium position of the pendulum as a function of
voltage applied on the lens at various lens positions}
  \label{volt-shift}
\end{center}
\end{figure}

 In addition to looking at the changes in the equilibrium
position of the pendulum as a function of separation, the changes
in the equilibrium position are also studied as a function of
voltages applied on to the lens. The lens is kept at a fixed
position and the voltage on the lens is scanned until the pendulum
sees the least force from the lens.  This would be the voltage at
which the applied voltage $V_{ap}$ balances the residual voltage
on the lens $V_{0,dc}$. The data is taken for different null
positions of the lens. The observations are summarized
in Fig.~\ref{volt-shift}\\

 In Fig.~\ref{volt-shift}, the separations marked are the
distance between the lens and the pendulum equilibrium when the
applied voltage on the lens is zero. The separation of
$117.59~\mu$m corresponds to the data taken when the lens is at
pixel number $1417$, $104.77~\mu$m to lens at pixel number $1500$
and $88.5~\mu$m to lens at pixel number $1585$ respectively. As
the voltage on the lens is increased, the equilibrium position
shifts away from the lens, implying a reduction in the force of
attraction from the lens until the voltage reaches between $55$ to
$60$ mV beyond this voltage the attraction increases again. Hence,
the electrostatic
forces are minimum around this voltage range.\\

 Thus, from the small amplitude data, we have an idea of
the magnitude of the electrostatic forces acting on the pendulum.
The distance and the voltage ranges over which observations are to
be carried out in order to fully characterize the electrostatic
and other forces are now clear.\\

\subsubsection{Large amplitude oscillations}
Another mode in which data is obtained is by sampling the
potential in the regions that are very close to the point where
the potential becomes completely attractive.  As we observed in
the small amplitude data, the electrostatic force on the pendulum
is the least when the voltage is about $57$ mV on the lens. This
then means that at angular distance of about $1 \times 10^{-3}$
radians, the potential in the region between the lens and the
pendulum would be considerably different from the harmonic
potential of the fibre, but will still have a minimum. Thus, if
the pendulum is made to execute oscillations of very large
amplitude about $2 \times 10^{-3}$ radians, the oscillations of
the pendulum will show marked anharmonicity especially during one
half of the cycle.\\

 The period and the equilibrium position of the pendulum
in this case will be different from those in the small amplitude
case. The $\dot\theta$ and $\ddot\theta$ of the pendulum will show
marked change between the two halves of the cycle. Since in this
mode we are sampling the potentials in regions that are reasonably
far away from the lens - separations $ > 20~\mu$m, the dominant
force on the pendulum will be the electrostatic force. Thus,
quantitative information about the electrostatic forces can be
obtained from this data.\\
\begin{figure}
\begin{center}
  \includegraphics[5mm,15mm][95mm,80mm]
  {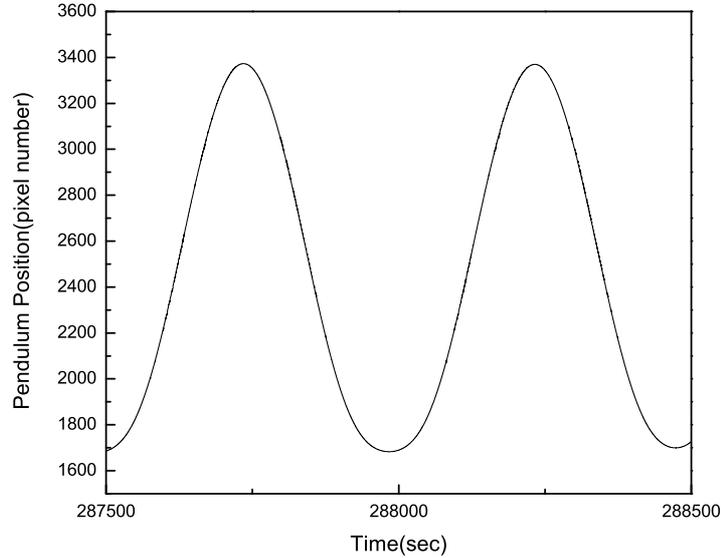}\\
\end{center}
  \caption{The raw data showing large amplitude
  oscillations of the pendulum when an voltage of 75~mV was applied
  to the lens}\label{largeoscil}
\end{figure}

 The amplitude of the pendulum is increased using the
capacitor plates that were described in Chapter 2. The amplitude
of the torsional oscillations of the pendulum can be controlled by
applying appropriate voltage on these plates. The pendulum is made
to oscillate such that it approaches to within  $1\times 10^{-3}$
radians from the lens for about $10$ cycles. Data is taken at
$0.16$ sec time intervals. The time period of these oscillations
varies from $400$~sec to $600$ sec depending on the voltage
present on the lens. The angular position of the pendulum is
recorded in units of pixel numbers of the CCD.
Fig.~\ref{largeoscil} displays a portion of the large amplitude
data taken and shows the asymmetry in the oscillations on the side
closer to the lens (lens is at about pixel number 1580 for the
data shown). To fully understand the electrostatic forces, the
large amplitude data is obtained by applying voltages on the lens.
Since we expect the minimum in the force to be at about
$V_{ap}=57$~mV, voltages starting from $50$~mV to $100$~mV are
applied on the lens in $5$~mV steps.

 Further analysis of these oscillations to comprehend the
electrostatic forces will be presented later in this chapter.

\subsubsection{Pendulum `falling' onto the lens}
The Casimir force falls very rapidly with the distance from the
lens. In order to measure Casimir force, the potential very close
to the lens has to be studied. But in this region, the potential
is highly attractive and the pendulum cannot oscillate. Instead,
it is pulled steadily towards the lens until it hits the lens and
bounces away from it. Thus, to quantify Casimir force, data is
taken as the pendulum falls on to the lens.\\

\begin{figure}
\begin{center}
  \includegraphics[5mm,15mm][95mm,80mm]
  {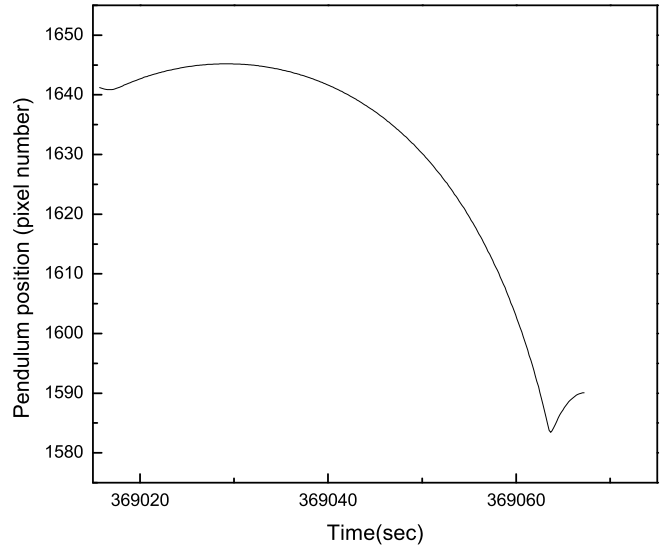}\\
\end{center}
  \caption{The raw data showing the pendulum `falling' onto the lens
after being pulled away from it using the capacitor plates. For
this data the lens is at pixel number of about 1580 and a voltage
of $80$ mV is applied to it.} \label{fall}
\end{figure}

 The initial velocity of the fall is controlled by using
the capacitor plates that damp the angular oscillations. By
applying appropriate voltages to the capacitor plates, the
pendulum is first pulled over the potential hill onto the
attractive region of the potential. As the pendulum falls towards
the lens, it is stopped and mildly pulled away from the lens. The
velocity away from the lens is not sufficient for the pendulum to
go over the potential hill again, but it just falls back onto the
lens after moving away from it for a short distance. Thus, the
fall data is available with zero initial velocity. [Fig.\ref{fall}]\\

 Data is taken by making the pendulum fall onto the lens
several times. The angular position of the pendulum $\theta$ is
recorded every $0.16$ sec as the pendulum falls. The voltage on
the lens is also varied as in the case of the large amplitude
oscillations. Fig.~\ref{fall} shows the raw data of a single
`fall' of the lens on to the pendulum when a voltage of $80$ mV is
applied to the lens. The analysis of this data to extract
information about the Casimir force is presented in the next
section.

\section{Analysis:}
    The angular position data obtained from the autocollimator, has to
be analyzed to gain knowledge about the forces acting on the
pendulum. The strategy followed involves the use of $\ddot\theta$
as a function of $\theta$ to extract information about the forces.
Thus, the first step in the analysis is to convert the
$(t-\theta)$ data to $(\theta-\ddot\theta)$ map. This is then
further analyzed to extract information about the force.
      The analysis of the data acquired as described in the
previous sections is presented in this section.

\subsubsection{Large amplitude  oscillation:}
     The large amplitude oscillations of the pendulum give us
information about the forces acting on the pendulum at separations
$>20~\mu$m from the lens. The electrostatic forces are the
dominant forces at these separations.\\

   An angular position plot of large amplitude oscillation is
shown in Fig.~\ref{largeoscil}. As a first step in the analysis of
the data, the angular position is converted to angular separation,
$\theta$, taking into account the lens position during the run of
the experiment. This data is filtered to remove $\theta$ values
that have $>3\sigma$ errors in them. These errors arise due to the
non-uniformities in the CCD pixels.  The $\theta$ values with
errors $>3\sigma$ are replaced by the mean $\theta$ of the
neighbouring $10$ points. Since the time period of the pendulum is
much larger than the sampling time between points, the change in
$\theta$ within this $10$ point interval of $1.6$ sec can
be considered linear.\\
\begin{figure}
\begin{center}
  \includegraphics[15mm,15mm][95mm,80mm]
  {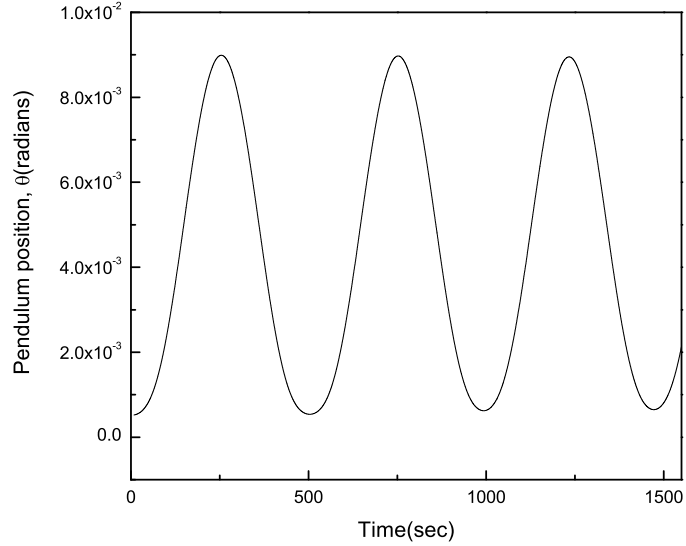}\\
\end{center}
  \caption{Plot of smoothened $t-\theta$ Data taken with a voltage
   of $75$ mV on the lens}
   \label{smooththeta}
\end{figure}

 The $(t-\theta)$ data thus generated is further smoothened
using a third order polynomial fit for every $51$ points. This is
a small fraction of the total of about $2500$ points available per
cycle of the pendulum. The central, $26^{th}$ point of this set is
replaced by the smoothened $\theta$ obtained from the fit. A plot
of the smoothened data is shown in Fig.~\ref{smooththeta}. The
next step is to obtain the angular velocities and accelerations.
For this, in every $51$ points of the smoothened $(t-\theta)$
data, the time at the $26^{th}$ point $t_{26}$, is subtracted from
all $t$s. This is then fitted with a second order polynomial in
$t$. Let this be $u + v t + w t^2$. Here $u$ would be $\theta$ at
$t=0$ which is the $\theta$ of the $26^{th}$ point. The angular
acceleration and velocity are then calculated as follows,
\begin{eqnarray}
  \theta(t) & = &  u + v t + w t^2, \\
  \dot\theta &=&  v  + 2 w t,\ \ \mathrm{at\ }\theta = u,\\
  \ddot\theta &=& 2 v ,\ \ \mathrm{at\ }\theta = u.
  \label{eqn-thetadbldot}
\end{eqnarray}

This is done for entire $(t-\theta)$ data by sliding over every
consecutive $51$ points. Thus, we get $\dot\theta, \ddot\theta$ at
every $\theta$. Since the $(t-\theta)$ is cyclic, the $(\theta -
\dot\theta)$ and $(\theta -\ddot\theta)$ data will also be cyclic.
The cycles are overlapped by ordering according to  increasing
$\theta$. These are then  averaged over a $\theta$ range of $1
\times 10^{-5}$ radians. Thus, for every set of $(t-\theta)$ data,
a set of $\theta, \dot\theta, \ddot\theta$ data is generated.\\
\begin{figure}
  \begin{center}
  \resizebox{\textwidth}{!}{
  \includegraphics[8mm,15mm][140mm,80mm]
  {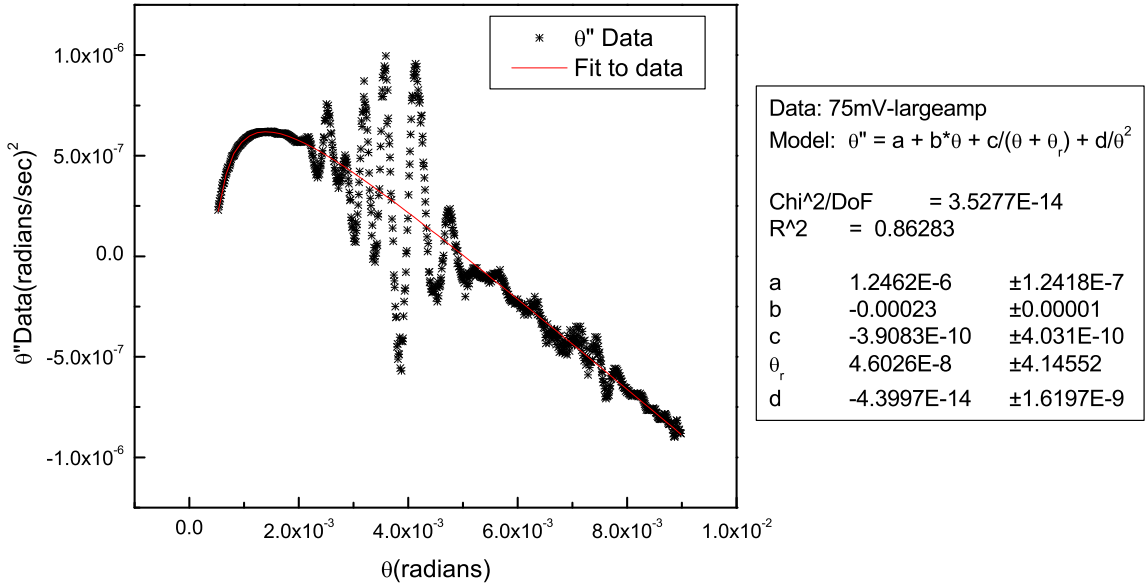}}\\
  \end{center}
  \caption{$(\theta-\ddot\theta)$ for large amplitude oscillations
    data taken with a voltage of 75 mV on the lens and a fit to the
    data}
   \label{thetadbldot-large}
\end{figure}

Fig.~\ref{thetadbldot-large} shows a plot of
$(\theta-\ddot\theta)$ data. The `asterisks' mark the data points
obtained as described above. Almost periodic errors  are present
even after the averaging as can be seen. The error is large when
the velocity of the pendulum is large and is perhaps caused by
systematic errors in the timing generated by the PC. The PC runs
Windows operating system and hence interrupts from the operating
system could lead to non-uniformity in the time difference between
2 successive data
acquisitions.\\

 The linear portion of the plot, for $\theta > 4 \times 10^{-3}$
radians represents the force due to the fibre. The sharp curvature
in the plot for $\theta < 2 \times 10^{-3}$ radians is
predominantly caused by the electrostatic forces. To study these
forces, described in \S1 of this chapter, we fit an equation of
the form,
\begin{eqnarray}
  \ddot\theta &=& a + b\ \theta + \frac{c}{\theta+\theta_r}
              + \frac{d}{\theta^2}. \label{fit-eqn}
\end{eqnarray}

The coefficient $b$ represents the force due to the fibre and $c$
the voltage dependant electrostatic forces. The coefficient $d$ is
an estimate of the Casimir force at these separations. It was seen
that the fitting coefficients $\theta_r$ and $d$, for this data at
large separations, do not play a major role in determining the
best fit parameters. The values of coefficients $a$, $b$ and $c$
do not change much even if $\theta_r$ and $d$ are made zero.
Parameters $a$ and $c$ change by about $0.001\%$ while parameter
$b$ does not change at all. A fit of this form to the
$(\theta-\ddot\theta)$ data when a voltage of $75$ mV
is applied to the lens is shown in Fig.~\ref{thetadbldot-large}.\\
\begin{table}[ht]
\footnotesize{
  \centering
  \caption{Table of coefficients of the fit to $\theta-\ddot\theta$
for large amplitude oscillation data obtained with various
voltages applied to the lens}\label{tab-lcoef} \vspace*{24pt}
\begin{tabular}{|c|c|c|c|c|c|}
  \hline
  Voltage on lens & \multicolumn{5}{c|}{ Fit
  parameters}\\ \cline{2-6}
(volts) & a & b & c & $\theta_r$ & d \\ \hline
 0.05  & $ 1.278\times10^{-6}$ &$ -2.3\times10^{-4}$ & $-5.0136\times10^{-10}$
 &$ -6.3578 \times10^{-7}$  & $ -1.198 \times10^{-13}$ \\ \hline

 0.055 & $ 1.2433 \times10^{-6}$  & $ -2.3 \times10^{-4}$ & $ -4.2532\times10^{-10}$
 & $ -6.7919 \times10^{-8}$& $-1.5841\times10^{-13}$ \\ \hline

 0.06 & $ 1.2636 \times10^{-6}$  & $-2.3 \times10^{-4}$ & $-4.4659 \times10^{-10}$
 & $1.3246 \times10^{-8}$ &$ -9.2951\times10^{-14}$ \\ \hline

 0.065 & $1.2612 \times10^{-6}$  & $-2.3 \times10^{-4}$ & $-4.0365 \times10^{-10}$
 & $-5.4028 \times10^{-8}$ & $-9.0942 \times10^{-14}$ \\  \hline

 0.07 & $1.2657 \times10^{-6}$   & $ -2.3 \times10^{-4}$ & $-4.2719 \times10^{-10}$
 & $9.3035 \times10^{-8}$ & $-3.79\times10^{-14}$ \\ \hline

 0.075& $1.2462 \times10^{-6}$ & $ -2.3 \times10^{-4}$ & $-3.9083 \times10^{-10}$
 & $4.6026\times10^{-8}$ &$ -4.3997 \times10^{-14}$ \\\hline

 0.08 & $1.2495 \times10^{-6}$ & $ -2.3 \times10^{-4}$ & $-4.0554\times10^{-10}$
 & $-3.0652 \times10^{-8}$ & $-3.1951 \times10^{-14}$ \\ \hline

 0.085 & $1.2525 \times10^{-6}$  & $ -2.3 \times10^{-4}$ &  $-4.1011 \times10^{-10}$
 & $1.0153 \times10^{-7}$ & $ -1.8389 \times10^{-14}$ \\\hline

 0.09 & $1.2543 \times10^{-6}$ &  $-2.3 \times10^{-4}$ & $-4.3337\times10^{-10}$
 & $5.0976\times10^{-7}$ & $-1.1825 \times10^{-14}$ \\ \hline

 0.095& $ 1.256 \times10^{-6}$  &  $-2.3 \times10^{-4}$ & $-4.7088 \times10^{-10}$
 &  $ -1.1626 \times10^{-6}$ & $-6.8344 \times10^{-15}$ \\\hline

 0.1& $ 1.246 \times10^{-6}$   & $-2.3 \times10^{-4}$ & $-5.1666 \times10^{-10}$
 &$ 2 \times10^{-5}$& $-1 \times10^{-16}$\\ \hline
\end{tabular}
}
\end{table}

\begin{figure}[h]
  \resizebox{\textwidth}{!}{
  \includegraphics[8mm,15mm][147mm,90mm]
  {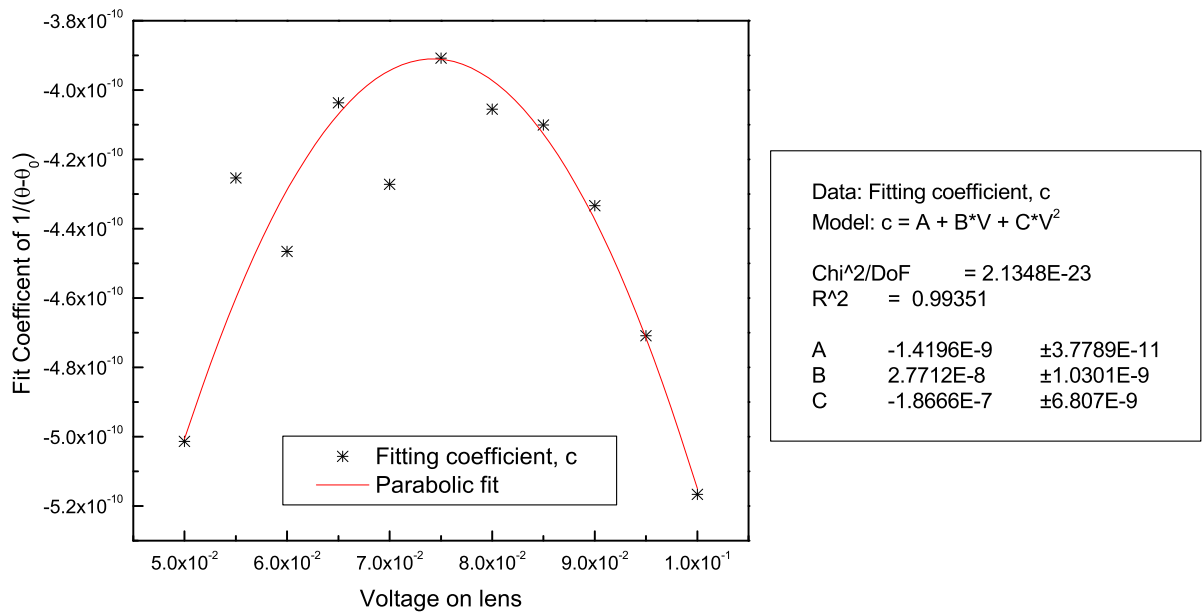}}\\
  \caption{A plot of the fitting parameter $c$, as a function of
 voltage on the lens calculated for the large amplitude oscillations
 of the pendulum.}
 \label{fitcoeff-large}
\end{figure}

Data obtained with voltages in the range of $50-100$ mV are
analyzed following the same procedure. The fit parameters obtained
for these data are tabulated in Table~\ref{tab-lcoef}. As seen
from the table, the coefficient $b$ is a constant independent of
the voltage on the lens as is to be expected. Its value of $-2.3
\times10^{-4}$ compares well with the value of $-2.46
\times10^{-4}$ estimated from the time period of the pendulum in
the absence of the lens. A plot of the coefficient, $c$, which
describes the voltage and charge dependent forces as a function of
voltage applied to the lens is shown in Fig.~\ref{fitcoeff-large}.
The figure also contains a parabola fitted to the data. The
minimum in the electrostatic force is determined by the balance
between the charge dependent and the charge independent forces.
The figure shows that $c$ is minimum for an applied voltage of
$75$ mV.  Thus, the background electrostatic forces acting on the
pendulum at these separations are balanced when a voltage of $75$
mV is present on the lens.

\subsubsection{Fall Data:}
The other kind of data that was acquired is the data as the
pendulum fall on the lens from separations of $<15~\mu$m. The
electrostatic forces and the Casimir force together determine the
motion of the pendulum in this region.\\

\begin{figure}[h]
 \resizebox{\textwidth}{!}{
  \includegraphics[8mm,15mm][150mm,80mm]
  {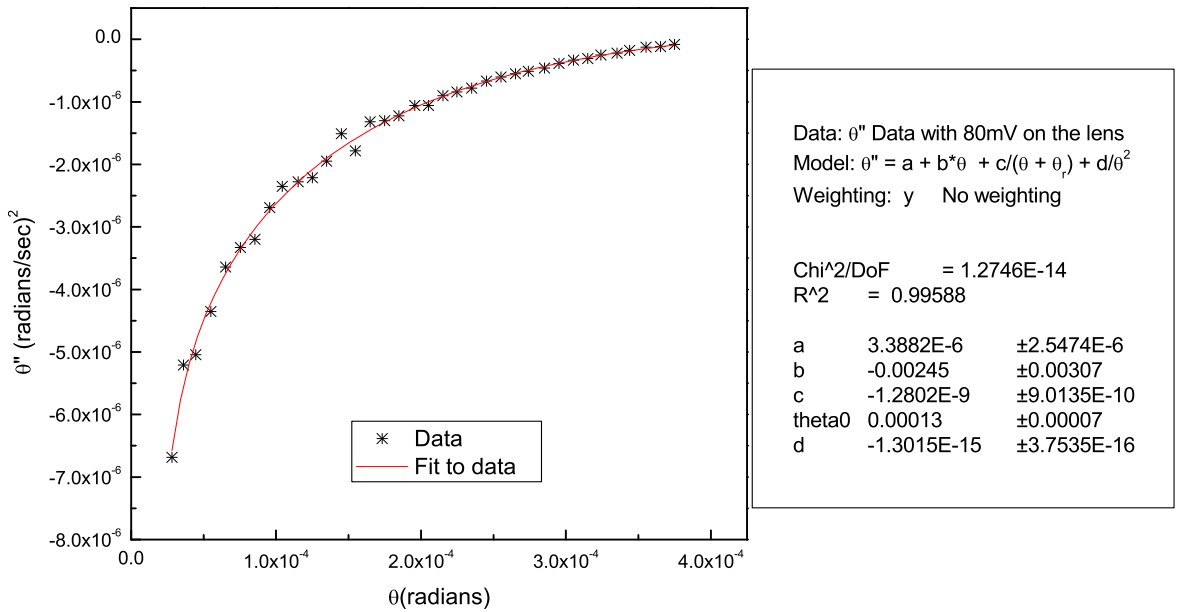}}\\
  \caption{$(\theta-\ddot\theta)$ for fall data taken with a voltage of 80 mV on
   the lens and a fit to the data.}\label{thetadbldot-fall}
\end{figure}

A sample plot of this data is shown in Fig.~\ref{fall}. The
angular acceleration is estimated by fitting a second order
polynomial as before to every 10 points in the $(t-\theta)$ data.
The $(\theta-\ddot\theta)$ are given by
Eqn.~\ref{eqn-thetadbldot}. Here too the data from multiple falls
is combined by ordering them increasing in $\theta$. Data is
averaged over a $\theta$ range of $1 \times 10^{-5}$ radians.
Fig.~\ref{thetadbldot-fall} shows a plot of $(\theta-\ddot\theta)$
for the fall data with a voltage of $80$ mV on the lens. The
`asterisks' mark the data points obtained as described above. This
data is fitted with Eqn.~\ref{fit-eqn}. The coefficients $a$, $b$,
$c$ and $d$ correspond to the various forces acting on the
pendulum as mentioned earlier. The presence of parameters
$\theta_r$ and $d$ is essential for this fit. In the absence of
$d$, the curvature in the data is not matched by the form of the
fitting curve. Thus this data has signature of the Casimir
force.\\

\begin{figure}
\begin{center}
  \includegraphics[5mm,15mm][95mm,80mm]
  {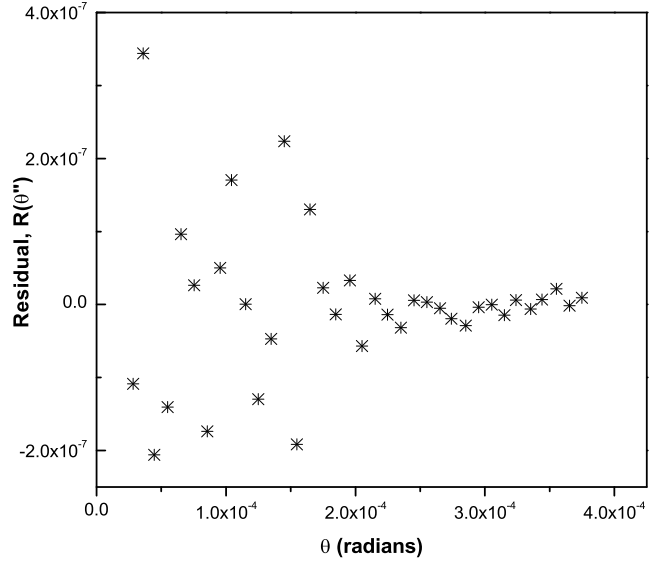}\\
\end{center}
  \caption{Plot of the residual $R(\ddot{\theta})_i$, for the data
  with a voltage of 80 mV on the lens.}
   \label{residual}
\end{figure}

The electrostatic and fibre forces have to be subtracted from this
data in order to study Casimir force. For this, the angular
accelerations obtained using the fitting parameters
$\ddot{\theta}(fit)$, are subtracted from the observed
accelerations, $\ddot{\theta}(obs)$ to get the
residuals,$R(\ddot{\theta})$  as a function of $\theta$. Thus, for
each $\theta_i$ in the data set, $R(\ddot{\theta})_i =
\ddot{\theta}(obs)_i - \ddot{\theta}(fit)_i $. A plot of these
residuals is shown in Fig.~\ref{residual} . To minimize errors,
$R(\ddot{\theta})$ are averaged over a $\theta$ range of $5 \times
10^{-5}$ radians to get average residuals and error in the
residuals as function of average $\theta$ as follows,
\begin{eqnarray}
  \langle \theta \rangle_i  &=& \frac{\sum_{j=1}^{n}{\theta_{ij}}}{\sum_{j=1}^{n}{1}} \\
  \langle R(\ddot{\theta})\rangle_i  &=&  %
          \frac{\sum_{j=1}^{n}{R(\ddot{\theta})_{ij}}}{\sum_{j=1}^{n}{1}}\\
  \vartriangle \hspace*{-1.5mm}R_i &=& \frac %
        { \left\{ \sum_{j=1}^{n}{ \left( \langle R(\ddot{\theta})\rangle_i -%
        R(\ddot{\theta})_{ij} \right)^2}
        \right\}^{\frac{1}{2}}}{n},
\end{eqnarray}
 where $n$ is the number of points with the $\theta$ range of $5 \times
10^{-5}$ radians.\\
\begin{figure}
\begin{center}
  \includegraphics[5mm,15mm][95mm,80mm]
  {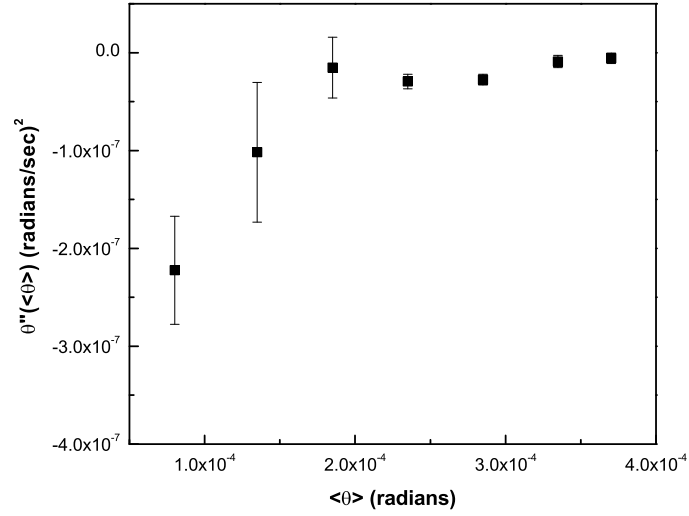}\\
\end{center}
  \caption{Plot of $ \ddot{\theta}_i(\langle \theta \rangle)$, for the data
  with a voltage of 80 mV on the lens.}
   \label{Avgresidual}
\end{figure}

The average $1/\theta^2$ term of the fit has to be added to the
$\langle R(\ddot{\theta})\rangle_i $ to get the remanent angular
accelerations with only the electrostatic and fibre background
forces removed, $\ddot{\theta}_i(\langle \theta \rangle_i)$. Thus,
$ \ddot{\theta}_i(\langle \theta \rangle_i)= \langle R(\ddot{\theta})\rangle_i  + %
\langle \frac{d}{\theta^2} \rangle = \langle R(\ddot{\theta})\rangle_i  + %
\frac{d}{\theta_{min,i}*\theta_{max,i}}$, where $\theta_{min,i}$
and $\theta_{max,i}$ are the minimum and maximum value
respectively of $\theta$ in the range $i$ that was averaged.
Fig~\ref{Avgresidual} shows a plot of remanent angular
acceleration, $\ddot{\theta}_i(\langle \theta \rangle_i)$ as a
function of $\langle \theta \rangle$ with a voltage of $80$ mV on
the lens. $\vartriangle \hspace*{-1.5mm}R_i$, represents the error
bars on the data.\\

\begin{figure}
\begin{center}
 \resizebox{\textwidth}{!}{
  \includegraphics[8mm,8mm][102mm,80mm]
  {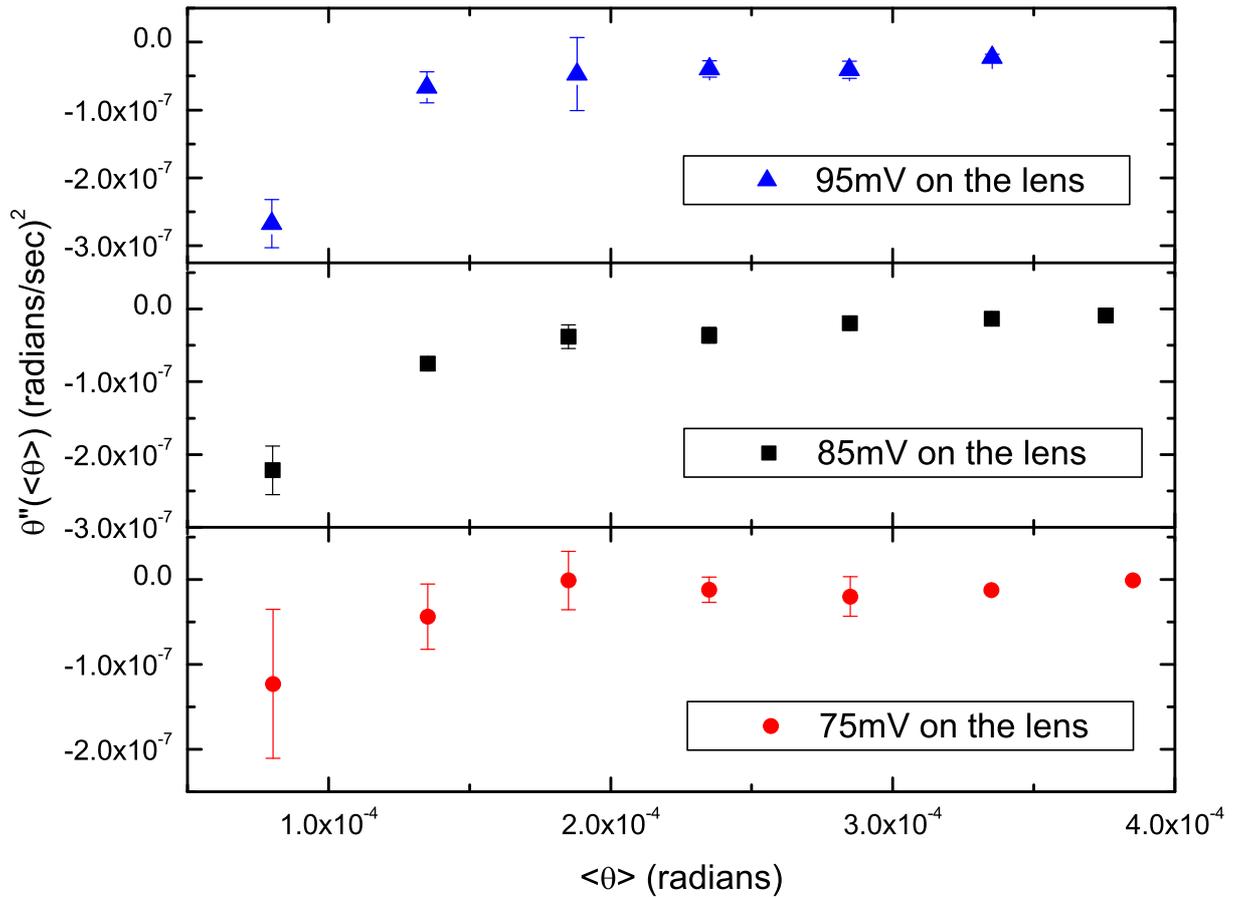}}\\
\end{center}
  \caption{Plot of $ \ddot{\theta}_i(\langle \theta \rangle)$, for the data
  with a voltages of 75 mV, 85 mV and 95 mV on the lens.}
  \label{Avgresidual-3set}
\end{figure}

The estimated $(\theta,\ddot{\theta})$ data, obtained with
voltages $V_{\alpha}$ in the range of $70-100$ mV on the lens were
analyzed following the same procedure to get the background
subtracted, $\ddot{\theta}_i(\langle \theta \rangle, V_{\alpha})$
and $\vartriangle \hspace*{-1.5mm}R_i(V_{\alpha})$. The fit
parameters obtained are tabulated in Table~\ref{tab-fcoef}. The
average value of the coefficients obtained in the fit and the
errors in them are tabulated in Table~\ref{tab-fcoef-avg}. A plot
of $\ddot{\theta}_i(\langle \theta_i \rangle)$ as a function of
$\langle \theta \rangle_i$ for three such data sets is shown in
Fig.~\ref{Avgresidual-3set}.
\begin{table}
\footnotesize{
  \centering
  \caption{Table of coefficients of the fit to $\theta-\ddot\theta$
for fall data obtained with various voltages applied to the
lens}\label{tab-fcoef} \vspace*{24pt}
\begin{tabular}{|c|c|c|c|c|c|}
  \hline
  Voltage on lens & \multicolumn{5}{c|}{ Fit
  parameters}\\ \cline{2-6}
(volts) & a & b & c & $\theta_r$ & d \\ \hline

 0.07 & $2.5782 \times10^{-6}$   & $ -1.13\times10^{-4}$ & $-1.1153 \times10^{-9}$
 & $1.0 \times10^{-4}$ & $-8\times10^{-16}$ \\ \hline

 0.075& $2.7099\times10^{-6}$ & $ -1.59 \times10^{-4}$ & $-1.0454 \times10^{-9}$
 & $8.0\times10^{-5}$ &$ -8\times10^{-16}$ \\\hline

 0.08 & $3.3882 \times10^{-6}$ & $ -2.45 \times10^{-4}$ & $-1.2802\times10^{-9}$
 & $1.3 \times10^{-4}$ & $-1.3015 \times10^{-15}$ \\ \hline

 0.085 & $2.1666 \times10^{-6}$  & $ -1.09 \times10^{-4}$ &  $-8.2853 \times10^{-10}$
 & $8.0 \times10^{-5}$ & $ -1.0614\times10^{-15}$ \\\hline

 0.09 & $2.9664 \times10^{-6}$ &  $-1.9\times10^{-4}$ & $-1.3116\times10^{-9}$
 & $1.2\times10^{-4}$ & $-8 \times10^{-16}$ \\ \hline

 0.095& $ 2.962 \times10^{-6}$  &  $-2.1 \times10^{-4}$ & $-1.1186\times10^{-9}$
 &  $ 1.3 \times10^{-4}$ & $-1.0701 \times10^{-15}$ \\\hline

 0.1& $ 1.3246 \times10^{-6}$   & $ -2.6 \times10^{-4}$ & $-5.7839 \times10^{-10}$
 & $7.35 \times10^{-5}$ & $-9\times10^{-16}$ \\ \hline
\end{tabular}
}
\end{table}
\begin{table}
  \centering
  \caption{Table of average coefficients of the fit to $\theta-\ddot\theta$
for fall data obtained with various voltages applied to the
lens}\label{tab-fcoef-avg} \vspace*{24pt}
\begin{tabular}{|c|c|c|c|c|c|}
  \hline
Fit parameter & Average value & Error in estimation \\
\hline
 $a$         & $ 2.585 \times 10^{-6}$      & $2.541\times10^{-7}$  \\ \hline
 $b$         & $-0.00184$                   & $2.2625\times10^{-4}$\\\hline
 $c$         & $-1.014  \times 10^{-9}$     & $8.87796\times10^{-11}$\\\hline
 $\theta_r$  & $1.01929  \times 10^{-4}$    & $9.35569\times10^{-6}$\\\hline
 $d$         & $-9.61857 \times 10^{-16}$   & $7.22763\times10^{-17}$\\\hline
\end{tabular}
\end{table}
A weighted average of the remanent acceleration from various
voltages is obtained using,
\begin{eqnarray}
\mathrm{Average\ Remanent\ Acceleration,\ \ }%
\langle \ddot{\theta}(\langle \theta \rangle_i)\rangle & = & \frac{%
\sum_{\alpha}{ \ddot{\theta}_i(\langle \theta \rangle_i, V_{\alpha})  %
\frac{1}{\vartriangle~R_i^2(alpha)}}} %
{\sum_{\alpha}{\frac{1}{\vartriangle~R_i^2(alpha)}}} \\
\triangle(\langle \theta \rangle_i) & = & \frac{ %
\left \{ \vartriangle~R_i^2(alpha \right \}^{1/2}}%
{\sum_{\alpha}{1}}
\end{eqnarray}

\begin{figure}
\begin{center}
 \resizebox{10cm}{!}{
  \includegraphics[8mm,8mm][107mm,84mm]
  {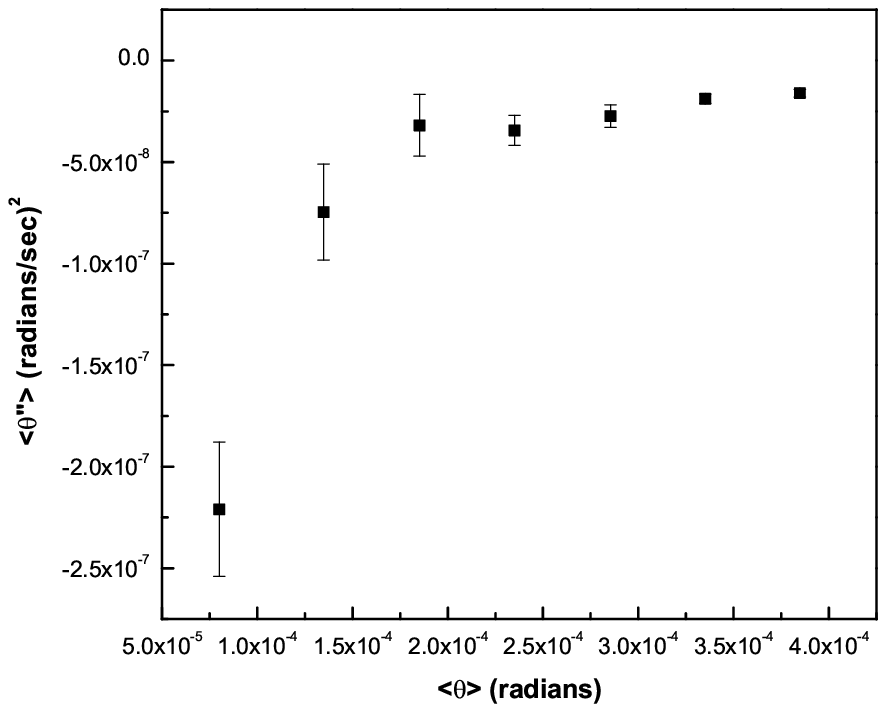}}\\
\end{center}
  \caption{Plot $\langle \ddot{\theta}(\langle \theta
  \rangle_i)\rangle$ estimated from the fall data, is shown.}
  \label{Avgthetadbldot-fall}
\end{figure}

A plot of this is shown in Fig.~\ref{Avgthetadbldot-fall}.

\section{Discussion of results}
\begin{figure}
\begin{center}
 \resizebox{10cm}{!}{
  \includegraphics[18mm,12mm][100mm,80mm]
  {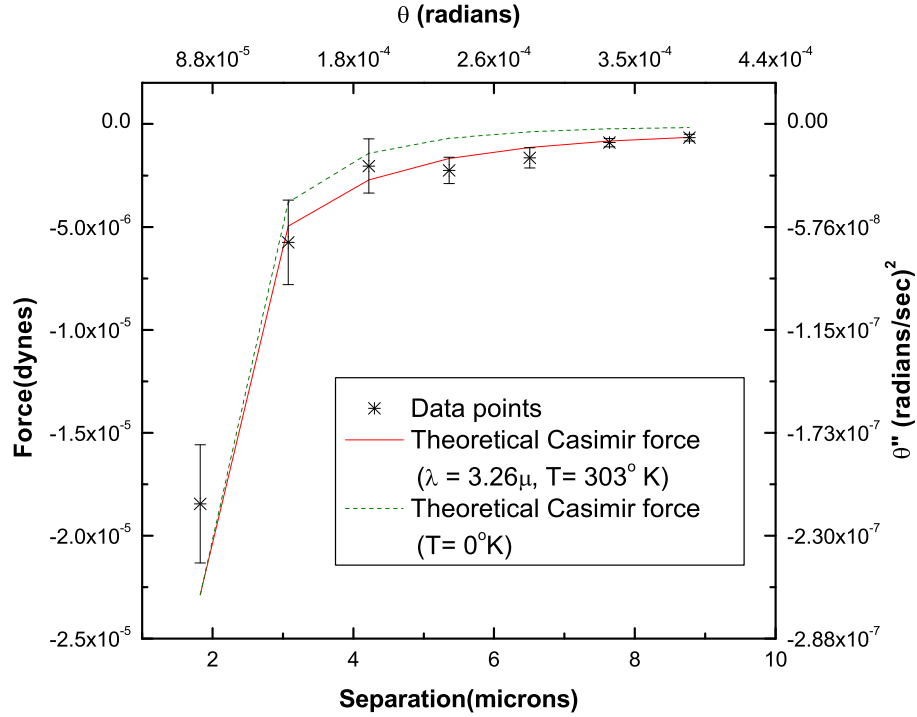}}\\
\end{center}
 \caption{Plot of force between the lens and the pendulum disc
  estimated from the fall data(stars) is shown
  along with theoretically estimated Casimir force. The solid lines
  represents the Casimir force estimated assuming $\lambda=3.26~\mu$m and
  the dashed line represents the zero temperature Casimir
  force.}
  \label{Plot-Force-theta2}
\end{figure}

The reduced data in Fig.~\ref{Avgthetadbldot-fall} is the force of
interaction between the spherical surface and the flat disc after
subtracting the electrostatic forces, fibre force and constant
(distance independent) forces and offset. This measured force
plotted along with theoretically expected Casimir force is shown
in Fig.~\ref{Plot-Force-theta2}. The `asterisks' represent the
remanent force obtained after the subtraction of the electrostatic
and the fibre background forces as described in the previous
section. The Casimir force is estimated using
Eqns.~\ref{Casimir-l-1},~~\ref{Casimir-1-2} with $\lambda =
3.26~\mu$m consistent with the theoretical estimates of Genet,
Lambrecht and Reynaud~\cite{Reynaud2000}. This is averaged over a
range of $5 \times 10^{-5}$ radians in separation $\theta$ and
overploted (solid line). The Casimir force between the lens and
the parallel plate without including finite temperature
corrections is also averaged and plotted (dashed line).
The error in the measurements at separations $> 4~\mu$m are less
as compared to that at closer separations. This is because the
pendulum spends less time in the region close to lens due to
larger `fall' velocity, and hence there are less number of data
points in this region when $\theta$ is recorded as a function of
$t$. There is an offset of the measured force from the
theoretically expected force even in the region where the errors
are small. This offset is a small fraction of the error on the
constant term `$a$' in the fit, Eqn.\ref{fit-eqn}, to the entire
data. If the data is normalized to the finite temperature Casimir
force at $9~\mu$m by adding a constant offset to all the data
points, the measured force agrees with the theoretically
calculated finite temperature Casimir force within the statistical
errors of the data. The required offset of $8.4 \times 10^{-9}$
(radians/sec)$^{2}$ or $3.61 \times 10^{-7}$ dynes is $\sim
1/30^{th}$ of the error in the estimation of the constant
coefficient from the fit, which is $2.541\times10^{-7}$
(radians/sec)$^{2}$ [Table~\ref{tab-lcoef}]. Also this offset is
of the order of the expected residual gravitational force on the
pendulum due to the lens [see \S5.1.1]. The deviation of the data
from the finite temperature Casimir force around separations of
$4~\mu$m is due to the fact that we have imposed a strict cut-off
for the change over from zero temperature to finite temperature
theory at $\lambda = 3.26~\mu$m. The actual force will gradually
change from the dependance of $1/z^2$ to $1/z^3$ around $z =
3.26~\mu$m. Also there will be about $10\%$ reduction in the
estimated Casimir force through out the range due
to finite conductivity corrections.\\

Thus, the measured force is consonant with the presence of a force
that is $\sim 1/z^2$, viz., finite temperature Casimir force.

\bibliographystyle{plain}
\bibliography{reference}

\clearpage

\chapter{Comparison of experiments on Casimir force}

\emph{Abstract:  A brief review of various experiments performed
to study the Casimir effect will be presented. We then compare the
results from the recent experiments with our results. The first
ever detection and measurement of finite temperature corrections
to Casimir force in our experiment is highlighted.}

\section{Experiments to Study Casimir force}
Experimental studies on Casimir force started half a century ago,
a few years after Casimir predicted the existence of the force in
1948~\cite{Cas48}. A historical overview of the experiments was
presented in the first chapter. Extensive reviews may be found
in~\cite{Bordag01, Miloni94}. We now discuss the details about
some of the pioneering experiments measuring Casimir force.\\

Overbeek and Sparnaay~\cite{Sparnaay52} were the first to attempt
to measure the Casimir force. They tried to measure the force
between two parallel polished flat glass plates with a surface
area of $1$~cm$^2$, in the distance range of $0.6~\mu$m to
$1.5~\mu$m. The distance was sometimes made $0.2~\mu$m. The force
was measured using a spring balance. The displacement of the
spring was measured by means of a capacitive method. The
measurements at $1.2~\mu$m, `pointed to the existence of a force
which were of the expected order of magnitude'~\cite{Sparnaay89}.
The hygroscopic behaviour of the glass (quartz) surfaces and
presence of dust particles
presented difficulties.\\

Results in the distance range $0.1~\mu$m to $1.0~\mu$m with
several different materials were first obtained by Derjaguin and
Abrikossova~\cite{Derj54,Derj60}. They measured the force between
plano-convex fused silica lens and a flat attached to a knife-edge
balance. The deflection of the balance was observed optically. A
current in proportion to the deflection of the balance from its
mean position was generated and passed though a coil. The coil was
pivoted between the pole pieces of a magnet. The magnet and the
coil were mounted such that the current flowing through the coil
produced a turning force on the balance opposite to that due to
attractive molecular forces and kept the balance from turning.
This current was, thus, a measure of the force. The separation
between the lens and a flat was measured using ``Newton's rings".
They could show that Casimir's formula provided a better fit to
the experimental data in contrast to London's theory for the van
der Waals forces. They also did experiments with two quartz
plates, quartz and chromium plates and with crystals made of
thallium-bromide and thallium iodide. It was these experiments
that, in a way, motivated Lifshitz to derive a macroscopic theory
of molecular forces~\cite{Lift56}. \\

Sparnaay~\cite{Sparnaay57} repeated his measurements with metal
plates in 1957. He measured the force between chromium plates and
chromium steel plates. The measured scheme was similar to the
previous experiment with a spring balance of sensitivity between
$(0.1-1)\times10^{-3}$~dynes. The distance was varied between
$0.3~\mu$m and $2~\mu$m. The measurements did not `contradict' the
expected force per unit area from Casimir's relation. But large
systematic errors and  electrostatic forces prevented a detailed
quantitative study.\\

The van der Waals force between two curved mica surfaces were
measured in the distance range of $1.5$ nm to $130$ nm by
Israelachvili and Tabor~\cite{Tabor1972a,Tabor1972b}. They
measured the force between two crossed cylindrical sheets of mica
with the effective contact resembling that between a sphere and a
plate. The surfaces were silvered and  the distance between them
was measured by observing interference patterns with white light.
The force was measured by two methods. In the short distance range
of $1.5$ nm -$20$ nm a `jump method' was used. One surface was
rigidly fixed to a moveable base and the other was mounted on a
cantilever spring. When the fixed surface was moved towards the
cantilever, at some point - depending on the stiffness of the
spring - the two surfaces jumped into contact. The measurement of
this jump distance as a function of spring distance was used to
measure the force of attraction. For the distance range of $10$ nm
- $130$ nm a resonant method was used. One surface was supported
on a rigid piezo-electric crystal and was set vibrating at very
small amplitudes over a convenient range of frequencies. The other
was supported facing it on a stiff spring. The natural frequency
of the latter depended both on the spring constant and on the van
der Waals forces exerted on it by the other surface. By
determining the resonant frequency as a function of separation,
the force law was deduced. They showed that there is a gradual
transition between the non-retarded and retarded van der Waals as
the separation is increased from $12$ nm to $50$ nm.\\

The next major set of improved measurements with metallic surfaces
were performed by van Blokland and Overbeek~\cite{Overbeek78} in
1978. They measured the forces between  a lens and a flat plate
coated with chromium using a spring balance at distances between
$0.13~\mu$m and $0.67~\mu$m. Precautions were taken to
characterize and eliminate electrostatic forces by careful
experimentation. Precise measurement of the separation was made by
measuring the lens-plate capacitance. A detailed comparison of the
data with Lifshitz theory, taking into account finite conductivity
effects was done before concluding that ``the measured force and
the calculated force are in excellent agreement''. It can be
considered as the first unambiguous demonstration of the Casimir
force between metallic surfaces.\\

After couple of dormant decades, interest in the measurement of
forces in the sub-millimeter range was aroused by theories that
predicted the existence of new forces in this range. The earliest
of these experiments was by Lamoreaux in 1997~\cite{Lamor97} and
around the same time we started the design of our
experiment~\cite{Mg8}. The apparatus built by Lamoreaux consists
of a  glass plate coated with gold over copper suspended as a
torsional pendulum. Its attraction towards a spherical lens also
coated with copper and gold was measured for distances between
$0.6~\mu$m - $6~\mu$m. The plate was held fixed with respect to
the spherical surface by a voltage applied to two compensating
capacitor plates held parallel to the `Casimir' plate. By
measuring the change in the voltage required to hold the plates
parallel when the  lens was moved towards the `Casimir' plate, the
Casimir force was measured. He detected and corrected for a
contact potential between the lens and the plate of $250$ mV. The
measured Casimir force agreed with theory at the level of 5\% up
to separation of about $2~\mu$m. The sensitivity of the experiment
was not adequate to measure the force beyond this
range.\\

In a later experiment, Mohideen and Roy~\cite{Mohi98} measured the
Casimir force  between a metallized sphere of diameter $196~\mu$m
and a flat plate for separations from  0.1 to $0.9~\mu$m. They
used the high sensitivity of an Atomic force microscope to report
a statistical precision of 1\% at the smallest separations. The
electrostatic force due a residual voltage of $29$ mV was measured
and subtracted from the
data.\\

A micromechanical actuator based on Casimir force was developed by
researchers at Bell Labs~\cite{Chan01} in 2001. The device,
fabricated using standard nanofabrication techniques, consists of
a $3.5~\mu$m thick, $500~\mu$m square heavily doped polysilicon
plate freely suspended on two opposite sides by  thin torsional
rods. The other ends of the torsional rods were anchored to the
substrate. There was a $2~\mu$m gap between the top plate and the
underlying substrate. A gold coated ball was suspended above one
side of the plate, which caused the plate to tilt due to Casimir
force when the plate was brought close to the ball. This tilt was
determined by measuring the capacitance between the plate and the
substrate accurately. A residual voltage of $30$ mV was determined
and compensated for. The measurements matched with Casimir force
theory at the 1\% level. The errors were mainly due to the
corrections for finite conductivity and surface roughness that
have to be applied to the
theory.\\

Recently, Bressi \emph{et al.}~\cite{Ruoso02} have measured the
Casimir force between parallel metallic surfaces. This is the
first and only precision measurement of the Casimir force in the
`parallel plate' configuration. The force was exerted between a
silicon cantilever coated with chromium and a similar rigid
surface and was detected by looking at the  shifts induced in the
cantilever frequency when the latter was approached. The motion of
the cantilever was monitored by means of a fiber optic
interferometer. Electrostatic force due a residual voltage of
$\sim -68$ mV is systematically determined and corrected for. The
scaling of the Casimir force with the distance between the
surfaces was tested in the $0.5~\mu$m - $3.0~\mu$m range, and the
related force coefficient was determined at the $15\%$ precision
level.\\

Very recently, the Casimir force between two dissimilar metals was
measured by Decca \emph{et al.} \cite{Decca03E} in the separation
range of $0.52~\mu$m - $2.0~\mu$m. They measure the force of
attraction between a copper layer evaporated on a micro-mechanical
torsional pendulum and a gold layer deposited on an Aluminium
oxide sphere with $600~\mu$m nominal diameter. The force is
inferred from the deflection angle of the torsional oscillator
determined by measuring the capacitance between the oscillator and
two fixed electrodes. The sensitivity of the force measurement was
further improved by measuring the change in the resonant frequency
of the oscillator due to the presence of the sphere. Residual
voltage difference of $632.5$ mV was determined by experimentation
and corresponds to the difference in work functions of the gold
and copper layers. The experiment highlights the need for
simultaneous measurement of the dielectric constant of the
metallic films used for a better understanding of the Casimir
force between non-ideal bodies at these separations.

\section{Comparison of results}

The Casimir force has been measured over a wide range of
separation from $1.5$ nm to $3~\mu$m by various techniques. Most
experiments have been performed in the sphere-plate geometry
rather than in the plate-plate geometry due to stringent
conditions of parallelism required in the latter. All the
experiments encountered a background of strong electrostatic
forces due to residual voltages present on the interacting bodies
even when these are well grounded. The Casimir force for the
sphere-plate geometry, depends on the radius of the sphere used
and is typically much larger than the electrostatic force at
sub-micron separations. But at larger separations, the
measurements are dominated by the electrostatic forces, making it
difficult to perform experiments in the several microns range. For
example, an electrostatic contribution comparable to the Casimir
force at $1~\mu$m becomes $10$ times or more stronger than the
finite temperature Casimir force at $10~\mu$m. The experiments
performed thus far, have been able to detect effects due to finite
conductivity and surface roughness of the metals on the Casimir
force but the finite temperature correction has not been observed.

\begin{figure}
\begin{center}
  \resizebox{\textwidth}{!}
  {\includegraphics[5mm,5mm][98mm,80mm]
  {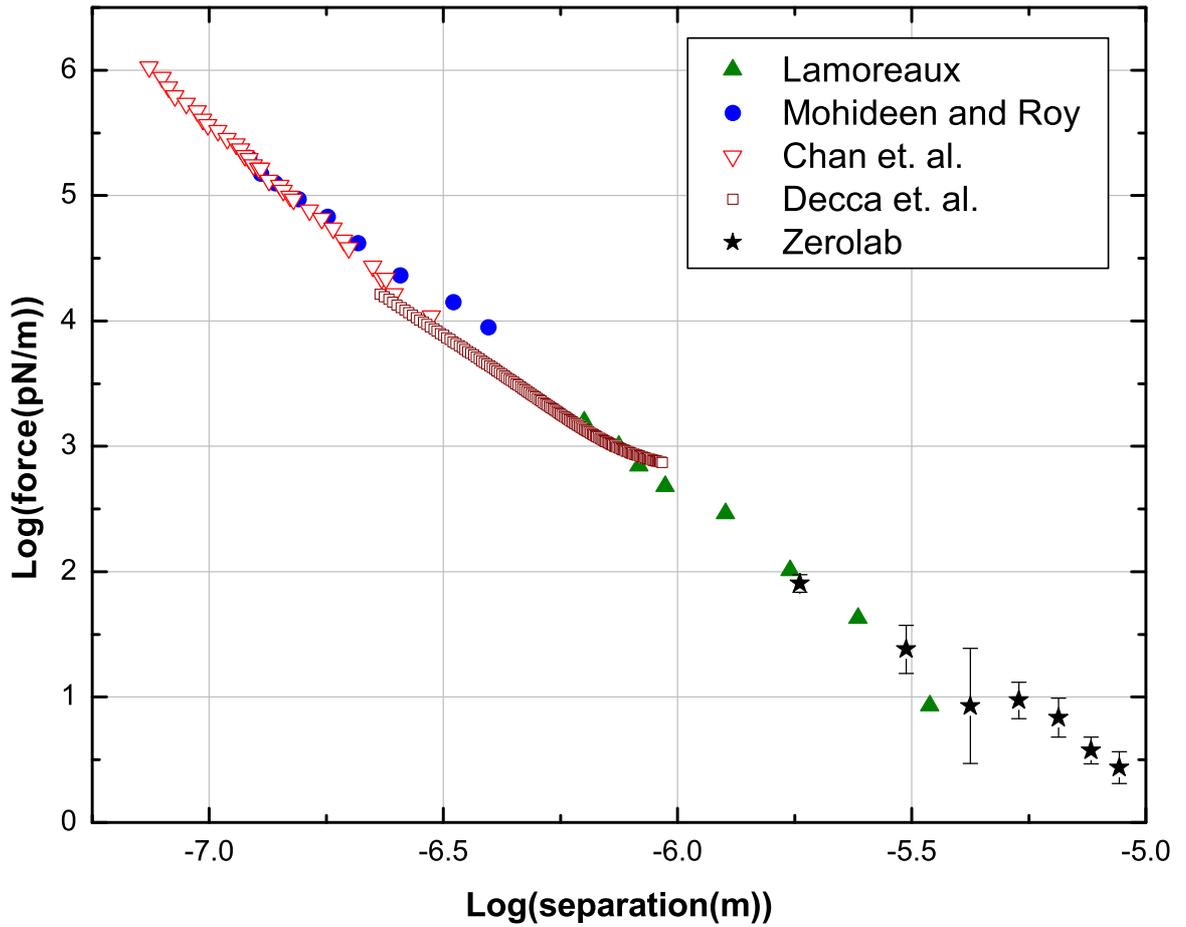}}\\
  \end{center}
\caption{Data from recent Casimir force experiments using the
sphere-plate geometry by Lamoreaux~\cite{Lamor97}, Mohideen and
Roy~\cite{Mohi98}, Chan \emph{et. al.}~\cite{Chan01}, Decca
\emph{et. al.}~\cite{Decca03E} and data from our experiment.}
  \label{Worlddata}
\end{figure}

Our experiment measures the Casimir force in the separation range
of $2~\mu$m to $9~\mu$m. This region extends well into spacings
beyond $\sim 3.3~\mu$m where corrections to the Casimir force due
to the $\sim 300^{\circ}$ K environment become significant. As
expected, the data clearly indicate the presence of the effects of
finite temperature. Data from recent experiments that measured the
force in the sphere-plate configuration are summarized in
Fig.~\ref{Worlddata}.
The figure shows the force measured in these experiments divided
by $2 \pi R$, where $R$ is the radius of curvature of sphere used
in the respective experiments. The results presented in this
thesis are represented by the `stars'. The data from various
experiments overlap in the region where the errors are small. Our
data, below $3.3~\mu$m overlaps well with the measurements by
Lamoreaux. The slope of the data represents the power of the force
law obeyed by the data. Around separation of $\sim 4~\mu$m the
plot distinctly shows the change in slope from $\sim -3$ to $\sim
-2$ within the errors of the experiments. There is a clear
indication of the change-over in the Casimir force law from the
zero temperature theory to that due to finite temperature around
this separation.\\

In summary, our current study extends the observation of the
Casimir effect up to separations of $9~\mu$m, i.e., well into the
region where finite temperature effects become important. The
experiment confirms the existence of these effects and agrees with
the standard theory at the $20\%$ level.

Keeping in mind the wide-ranging importance of Casimir forces it
would be useful to conduct such studies with greater accuracy and
also extend them to longer separations of $\sim 50~\mu$m or more.

\bibliographystyle{plain}
\bibliography{reference}

\clearpage

\chapter{Bounds on the Strength of New Macroscopic Forces
          and Future Directions}

\emph{Abstract: We briefly review the experimental bounds on the
strength of inverse square law violating short-range interactions.
The contribution of our experiment in this direction will be
presented. The future scope of our experiment to test and
constrain the predictions of fundamental theories will be
discussed.}

\section{Constraints on new macroscopic forces}
Extensions of the Standard Model predict the existence of a
variety of neutral light bosons. The exchange of these will
mediate forces that lead to deviations from the inverse square law
of gravity at distances related to the mass of the bosons. String
and M-theories that attempt to unify the fundamental forces close
to the Planck scale or the theories with large extra dimensions
that attempt to overcome the ``hierarchy problem" by invoking
unification of the fundamental forces at the TeV scale of
electro-weak symmetry breaking, predict variations from the
inverse square law of gravity at sub-millimeter distances. An
overview of these effects was presented in Chapter 1. In general,
the new macroscopic forces can be described by an addition of a
Yukawa type potential term to the gravitational interaction
between two point masses as given by,
\begin{equation}
V(r)  = -\frac{GM_1M_2}{r} \left( 1 +
                  \alpha e^{-\frac{r}{\lambda}}\right)
\label{Pot-Yukawa}
\end{equation}
where $\alpha$ represents the coupling strength of the interaction
and $\lambda$ the range. A recent review of the experimental and
the theoretical status of the inverse square law tests can be
found in ~\cite{Adel2003}. Given this form for the potential,
constraints can be placed on the parameter space of $\alpha-
\lambda$ from experiments that study long-range interactions

\subsection{Astrophysical Bounds}
\begin{figure}[h]
\begin{center}
  \resizebox{9cm}{!}{
  \includegraphics{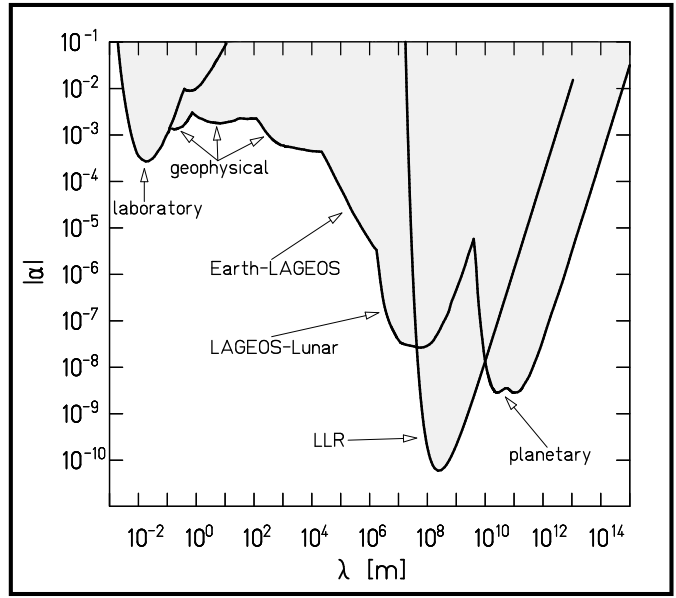}}
  \end{center}\vspace*{-8mm}
  \caption{Constraints on inverse square
law-violating Yukawa interactions with $\lambda > 1$cm. The Lunar
Laser Ranging (LLR) constraint is based on the anomalous
precession lunar orbit; the remaining constraints are based on
Keplerian tests. This plot is reproduced from ~\cite{Adel2003}.
The region in the ($\alpha,\lambda$) plane above each curve is
excluded, and below each curve is allowed.}
  \label{Astro}
\end{figure}

Astronomical tests provide the best constraint on the parameter
$\alpha$ at distance scales $\lambda > 1$~km. These are typically
obtained from observation of the Keplerian orbits of the planets
and the satellites and looking for deviations from normal gravity.
The measurements of precession of the lunar orbit using Lunar
Laser Ranging studies provides very good constraints in the
$10^8$~m scale. Precession is expected due the quadrupole field of
the Earth, gravitational perturbations from the other planets in
the solar system and general relativistic effects. Presence of
Yukawa type interactions, will add to the precession and hence
limits can be placed on $\alpha$ after accounting for the expected
sources of the precession. A summary of the constraints from
astrophysical data is reproduced from ~\cite{Adel2003,
Fischbach1999} in Fig.~\ref{Astro}. In the range $\lambda < 1$~m,
these limits have been surpassed by laboratory experiments
\cite{Adel1997, Hoyle2001, Price2003}. Details on geophysical and
laboratory constraints in the range $\lambda > 1$~m can be found
in \cite{Krishnan89-T, Unni92} and \cite{RC1981}-\cite{RC1982-2}.
\nocite{RC1981,RC1982-1,RC1982-2,RC1988-PRL, RC1988-2, RC1989-3, RC1990}\\
Astronomical data can also provide evidence for the presence of
extra dimensions \cite{ADD1999}. The cooling rate of the
supernovae will increase by the presence of the extra dimensions
because energy from the explosion will be radiated into these
dimensions as well. This would lead to the reduction in the number
of neutrinos emitted from the supernovae. SN 1987A data constrains
the size of the extra dimensions to  $<0.7~\mu$m when the number
of extra dimensions is 2 ~\cite{ADD1999,Cullen1999,
Hanhart2001-1,Hanhart2001-2}. This does not imply that experiments
at sub-millimeter scales will not observe effects due to the new
dimensions. A single large dimension of size $1$~mm with several
much smaller extra dimensions is still allowed.

\subsection{Bounds from Laboratory tests of Inverse square law}
\begin{figure}[h]
\begin{center}
  \resizebox{9.0cm}{!}{
  \includegraphics{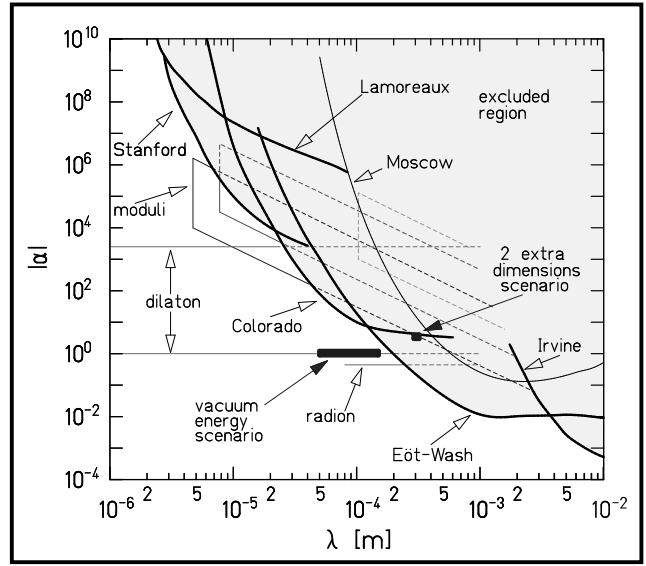}}
  \end{center}\vspace*{-8mm}
\caption{Constraints on inverse square law-violating Yukawa
interactions with $1~\mu$m $< \lambda < 1$ cm. This plot is from
~\cite{Adel2003}. The heavy curves give experimental limits. The
curve marked Irvine is a E\"{o}tvos type experiment. `Lamoreaux'
is a Casimir force measurement. The other curves are from
Canvendish-type experiments. The theoretical expectations from
various scenarios are also shown.}
  \label{ISL}
\end{figure}

Constraints on $\alpha$ in the sub-millimeter to centimeter scales
can be obtained from laboratory experiments of the E\"{o}tvos and
Cavendish type. The E\"{o}tvos-type experiments test the
equivalence principle by measuring the acceleration imparted by
Sun, Earth or some laboratory attractor to various materials of
the same mass. The presence of any additional force that couples
to the material properties other than the mass will show up as a
difference in the acceleration experienced by the various
materials. The Cavendish-type experiments are direct tests of the
Newton's inverse square law. The tightest constraints in the $1$
mm to $200~\mu$m region to date are provided by such experiments.
These are shown in Fig.~\ref{ISL} (reproduced from
\cite{Adel2003}). The shaded region in the ($\alpha,\lambda$)
plane above the continuous curves is excluded.

\subsection{Bounds from Casimir Force Measurements}
\begin{figure}[h]
 \begin{center}
 \resizebox{10cm}{!}{
 \includegraphics*[3.2cm,13.7cm][16.2cm,23.8cm]
 {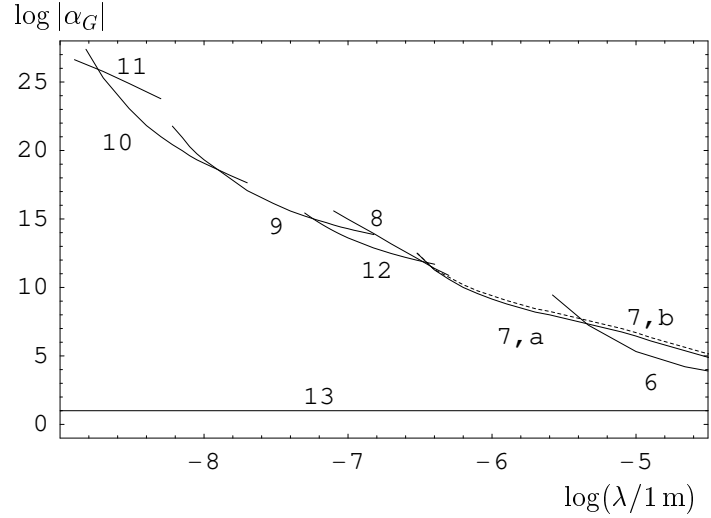}}
 \end{center} \vspace*{-1cm}
  \caption{Plot of the constraints on the Yukawa interaction
parameter $\alpha$ from various ranges of $\lambda$ reproduced
from \cite{Most2003}. Curves 7-10, 12 follow from Casimir force
measurements, Curve 11 from van der Waals force measurements.
Curve 6 is from an experiment that measured deviations from
Newton's law of gravity~\cite{Price2003}. The typical prediction
of extra dimensional physics is shown in Curve 13. The region, in
the ($\alpha,\lambda$) plane, above each curve is excluded and the
region below each curve is allowed.} \label{constraint-Moste}
\end{figure}

For distances of $\lesssim 0.1$~mm, Casimir force measurements
provide the tightest constraints. Casimir force has generated
extensive experimental and theoretical interest in the last
decade. Precision experiments have been performed in the distance
range of about $0.1~\mu$m to $3~\mu$m and the theoretical
correction to the Casimir force due to finite temperature, finite
conductivity and surface roughness have been calculated
accurately. This has enabled comparison between theory and
experiment at the $1\%$ level for the smallest separations. If the
inverse square law violating interactions are present due to any
of the scenarios mentioned in Chapter 1, they will show up as
additional forces in the measurement. Thus, constraints on the
strength of these interactions can be obtained by looking at the
deviation of the measured Casimir force from the theoretically
expected value and attributing it to Yukawa type
interactions~\cite{Bordag1998,Bordag01,Fischbach01,Decca03C}. A
plot of the constraints to date is summarized in
Fig.~\ref{constraint-Moste}.

\subsection{Estimation of bound from our experiment} Our experiment
measures the Casimir force between a spherical lens and a flat
disc made of `BK7' glass (density, $\rho_v = 2.51$~g.cm$^{-3}$),
coated with gold layer of thickness, $\delta = 1~\mu$m (with
density, $\rho_s = 19.32$~g.cm$^{-3}$). The radius of curvature of
the lens, $R = 38$~cm and its total height $x_{max}= 0.07$~cm. The
radius of the flat disc, $l = 4$~cm and its thickness, $g =
0.4$~cm. The force due to Yukawa type potential for this
configuration is given by (see Appendix C for details),
\begin{eqnarray}
\mathcal{F}_a(z) & = &  4 \pi^2 G \alpha R \lambda^2%
                     \left[ \rho_s - (\rho_s - \rho_v)%
                       e^{-\delta/\lambda}\right]\ %
                       \rho_s\ \delta\ e^{-z/\lambda},\quad %
                       \mathrm{for}\ \lambda  << \delta,
                       \label{f-yuk-a}\\
\mathcal{F}_{b}(z)& = & 4 \pi^2 G \alpha R \lambda^2%
                     \left[ \rho_s - (\rho_s - \rho_v)%
                       e^{-\delta/\lambda}\right]%
       \left(\ \rho_s\ \delta + \rho_v\ \lambda\ e^{-\delta/\lambda}%
       \  \right)\ e^{-z/\lambda}, \label{f-yuk-b}\\
&& \hspace*{8.5cm} \mathrm{for}\ \lambda >> \delta. \nonumber
\end{eqnarray}
where $\alpha$ and $\lambda$ are parameters of the theory as
defined by Eqn.~\ref{Pot-Yukawa}.

\begin{figure}
  \begin{center}
  \includegraphics{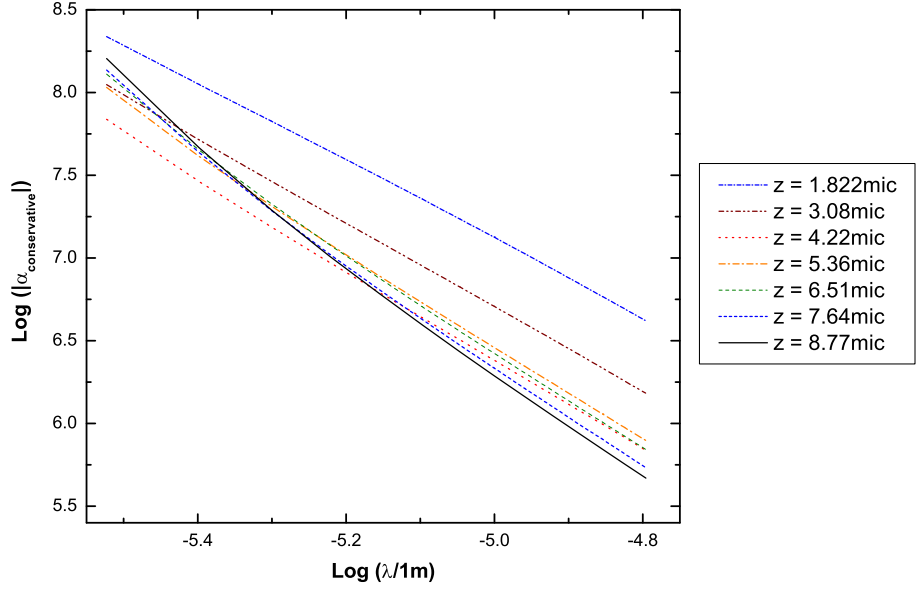}
  \end{center}
  \vspace*{-1cm}
 \caption{$\alpha_{conservative}(\lambda)$ estimated from the
 total force measured in the experiment.}\label{alpha-force}
\end{figure}

The upper limit of constraints on the Yukawa-type interactions
from our experiments can be estimated by two methods. \underline{A
conservative estimate} of the parameters of the Yukawa-type
interactions can be obtained by assuming that all of the force,
$f_{obs}(z)$ measured in our experiment is due to the Yukawa-type
potential. Thus,
\begin{eqnarray}
f_{obs}(z) & = & \mathcal{F}(z) \label{cond-conserv}
\end{eqnarray}
We solve Eqn.~\ref{cond-conserv} for the observed values of the
force at $z = 1.82~\mu$m, $3.08~\mu$m, $4.22~\mu$m, $5.36~\mu$m,
$6.51~\mu$m, $7.641~\mu$m, $8.77~\mu$m to get $\alpha$ as a
function of $\lambda$, we call this
$\alpha_{conservative}(\lambda)$. Since the thickness of the gold
coatings, $\delta = 1~\mu$m, we derive
$\alpha_{conservative}(\lambda)$ in the range $\lambda = 2~\mu$m
to $16~\mu$m using Eqn.~\ref{f-yuk-b}. This gives us seven
possible functions for $\alpha_{conservative}(\lambda)$
corresponding to the seven separations at which we have
measurements. A plot of these is shown in Fig.~\ref{alpha-force}.
The lowest values of $\alpha_{conservative}(\lambda)$ are obtained
from data at $z = 8.77~\mu$m in the range $7~\mu$m $\leq \lambda
\leq 16~\mu$m and from data at $z = 4.22~\mu$m in the range
$3~\mu$m $ \leq \lambda < 7~\mu$m.

\underline{An optimistic estimate} of the parameters of the
Yukawa-type interactions can be obtained from the error in the
measurements of the force. Our experimental data matches with the
finite temperature Casimir force between the lens and the plate
within the errors of force measurement at separation $z > 4~\mu$m.
The dominant contribution to the finite temperature Casimir force
for our configurations [see Appendix A], at these separations is
given by,
\begin{eqnarray}
f_{cl}(z) & = & - \frac{\pi R A_c}{\lambda_T}
        \left\{ \frac{1}{z^2} - \frac{1}{(z + x_{max})^2}\right\},
        \hspace*{5mm} \mathrm{for\ } %
        z >  \lambda_T = 3.26~\mu\mathrm{m};
        \label{Casimir2}\\
A_c &  = & \frac{\pi^2 \hbar c}{240} = 0.013 \times 10^{-18}\
\mathrm{dyn.cm}^{2}.
\end{eqnarray}

\begin{figure}
  \begin{center}
  \includegraphics{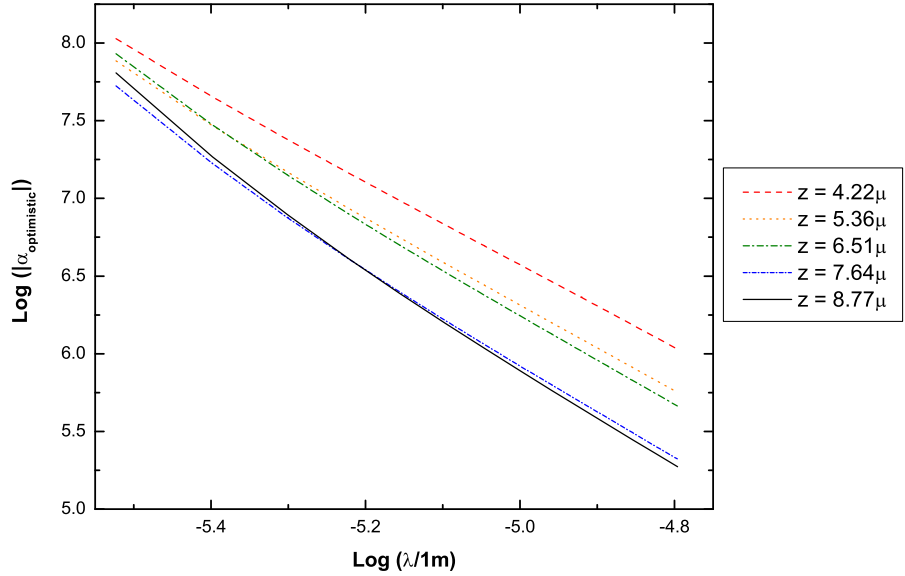}
  \end{center}
  \vspace*{-1cm}
 \caption{$\alpha_{optimistic}(\lambda)$ estimated from the errors
 in the measurement.}\label{alpha-error}
\end{figure}

The corrections to Casimir force due to surface roughness are $<
0.6\%$ in the $4.5~\mu$m to $9~\mu$m range \cite{Bordag1998} and
can be ignored. The finite conductivity of gold surface gives rise
to a reduction in the Casimir force by $10\%$ over the entire
range of the experiment \cite{Reynaud2000, Genet-T}. Thus, the net
theoretical force between the lens and the plate is given by,
\begin{eqnarray}
f_{th}& = & f_{cl}(z) + \Delta_p f_{cl}(z) + \mathcal{F}(z);
\label{ftheory}
\end{eqnarray}
where $\Delta_p f_{cl}(z)$ represents the corrections due to
finite conductivity and $ \mathcal{F}$ is the force due to
hypothetical Yukawa-type interactions. This is derived in Appendix
C for our geometry and is given by Eqns.~\ref{f-yuk-a} and
\ref{f-yuk-b}. Thus, the Yukawa parameters can be estimated using,
\begin{eqnarray}
\mid f_{th}(z) - f_{cl}(z) \mid & \leqslant \Delta F(z);
\label{deltaf}
\end{eqnarray}
where $\Delta F(z)$ is the error in the measurement. If absolute
error in our force measurement at a distance $z$ is $e(z)$, we can
solve Eqn.~\ref{deltaf} assuming $\Delta F(z) =  3 e(z)$. The
corrections due to surface roughness become less important for
larger separations and for $z < 4~\mu$m, the theoretical Casimir
force is not strictly defined by Eqn.~\ref{Casimir2}. We use the
data at $z = 4.22~\mu$m, $5.36~\mu$m, $6.51~\mu$m, $7.64~\mu$m,
$8.77~\mu$m and solve for $\alpha$ from Eqns.~\ref{ftheory} and
\ref{deltaf} in the range $3~\mu$m $\leq \lambda \leq 16~\mu$m.
The values of $\alpha$ obtained by this method are represented by
$\alpha_{optimistic}$ and are plotted in Fig.~\ref{alpha-error}.
The lowest values of $\alpha_{optimistic}(\lambda)$ are obtained
from data at $z = 8.77~\mu$m in the range $6~\mu$m $\leq \lambda
\leq 16~\mu$m and from data at $z = 7.64~\mu$m in the range
$3~\mu$m $\leq \lambda <
6~\mu$m.\\

\begin{figure}[h]
  \begin{center}
  \resizebox{\textwidth}{!}{
  \includegraphics{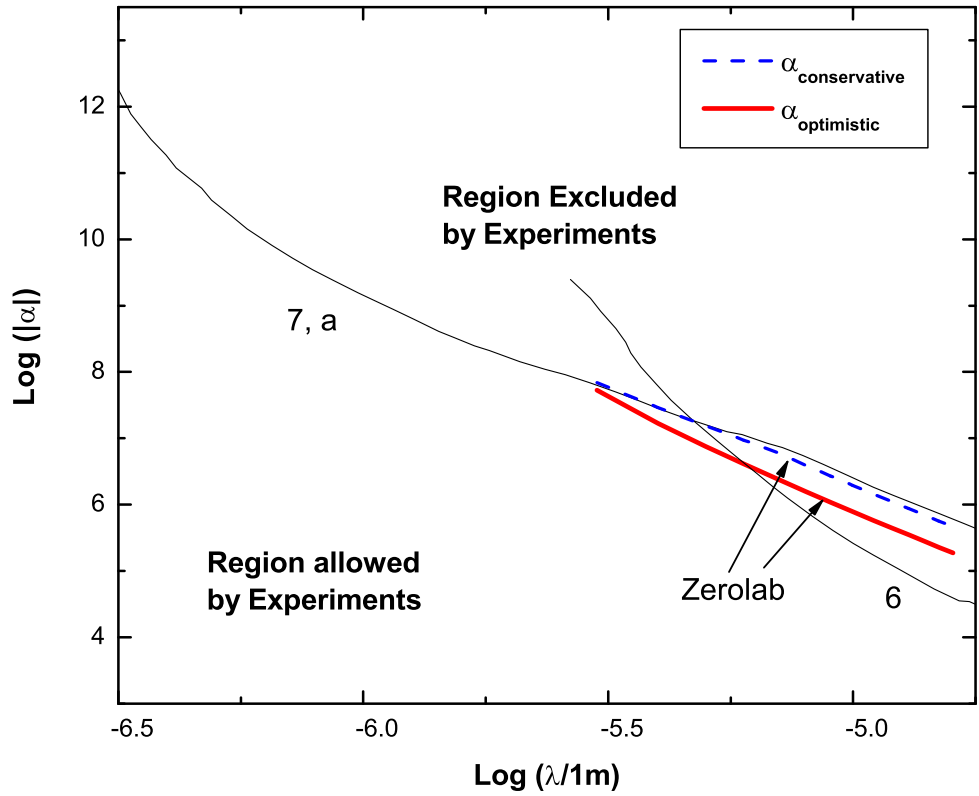}}
  \end{center}
  \vspace*{-1.3cm}
  \caption{Constraints on the inverse square law violating Yukawa
 interactions obtained from data presented in the thesis (`Zerolab')
 along with the best constraints in the same range of $\lambda$ from
 ~\cite{Most2003}}
\label{constraints-zerolab}
\end{figure}

In Fig.~\ref{constraints-zerolab}, the lowest values of
$\alpha_{conservative}(\lambda)$ and
$\alpha_{optimistic}(\lambda)$ estimated as described above are
plotted  along with those from other experiments as presented by
Mostepanenko in ~\cite{Most2003}.

The lines marked `Zerolab' are constraints as calculated from data
presented in this thesis. Curve $6$ is from the Canvendish-type
experiment \cite{Price2003}. Curve $7, a$ is from the torsion
balance experiment to measure Casimir force \cite{Lamor97}. This
is an upper limit of the constraint we can obtain from our
experiment. The constraints from the conservative estimate are
already at the level of those obtained from \cite{Lamor97}. The
optimistic estimates give the lowest values of $\alpha$ obtained
so far in the range $3~\mu$m $ \leq \lambda < 6~\mu$m and also the
best ever constraints from Casimir force measurements in the range
$3~\mu$m to $16~\mu$m. By calculating the theoretical forces
precisely, the constraints can be improved and also extended to
shorter distances. This also means that by choosing the materials
of the pendulum and attracting plate appropriately (for example,
thicker coatings, or higher density materials) we may be able to
improve the present constraints by more than an order of magnitude
in the distance range $3~\mu$m to $10~\mu$m.

\section{Future Directions}
We are in the process of performing a next generation experiment,
with the same general configuration but improved precision by
reducing the systematic and statistical noise. In the new scheme
we will perform a null experiment where the deflection of the
pendulum due to the torque from the lens is balanced by the
capacitive torques from a pair of metal plates located close to
the disc of the pendulum (see \S2.3). The voltages on capacitor
plates are controlled through a servo-loop and the servo-signal is
a direct measure of the torques acting on the balance. This mode
of experiment has several advantages. The gain in the feed-back
helps to reduce the noise in the system (see for example
~\cite{Rey2000-fluc}). The statistical accuracy can be
considerably improved by increasing the integration time. The
errors associated with the estimation of acceleration from the
limited time series data are avoided. The experiment can be
performed at closer separations as well. The torque due the lens
can be modulated by moving the lens back and forth, further
reducing the noise in the bandwidth of the observation. Efforts
are also on to characterize the background
electrostatic forces better and hence decrease systematic errors.\\

With the lens-plate combination, Casimir force at separations
greater than $30~\mu$m will be difficult to measure as
electrostatic and gravitational forces will start to dominate. A
similar experiment would be repeated with a flat plate instead of
a lens for separation up to about $100~\mu$m. The parallelism
between this plate and suspension disc is less critical at these
separations. The dependance of the Casimir force on separation
will be $1/d^{3}$ as opposed to the typical $1/d^2$ or $1/d$
dependance of the other background forces. Such an experiment will
be able to put stronger constraints on the new forces with
strength close to that of gravity in the sub-millimeter range as
the parallel plate configuration maximizes the sensitivity to
these forces as opposed to the plate-sphere experiments.\\

We expect a $5\%$ accurate measurement of the Casimir force in the
range $10~\mu$m to $100~\mu$m, and also improved constraints on
hypothetical Yukawa-type forces with range $10~\mu$m to
$100~\mu$m, from these future experiments.

\bibliographystyle{plain}
\bibliography{reference}

\clearpage
\appendix

\chapter{Casimir Force between infinite Parallel Plates} The
Casimir force is a manifestation of zero point energy of the
electromagnetic field. In the quantum mechanical description of
the electromagnetic field, the allowed energy levels of the
electromagnetic wave of angular frequency $\omega$ are given by
the Planck relation $E_n\ =\ (n+\frac{1}{2})\hbar \omega, n\ =\
0,1,2,3 \ldots$ . The integer $n$, for the electromagnetic field,
corresponds to the number of photons. The $n=0$ or the `vacuum'
state also has an energy of $1/2 \hbar \omega$ associated with it.
Thus, vacuum is not empty and contains fluctuations of the
electromagnetic field.
\begin{figure}
\begin{center}
\includegraphics{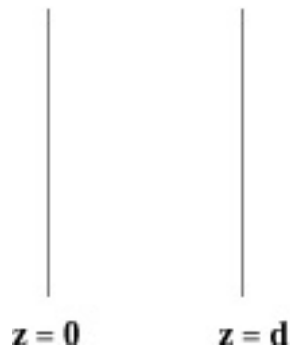}
\end{center}
\caption{Two perfectly conducting infinite plates, placed distance
d apart.} \label{plates}
\end{figure}
\noindent

Let us now calculate the effect of two parallel conducting plates
separated by a distance $d$, placed in such a field using the
`mode counting' technique~\cite{Miloni94}. The field modes within
this cavity (Fig.~\ref{plates}) are different from the free space
modes. In addition to satisfying the wave equation,
$\nabla^2\vec{A}\ =\
\frac{1}{c^2}\frac{\partial^2\vec{A}}{\partial t^2}$, together
with, $\nabla.\vec{A}\ =\ 0$, the field inside should have the
tangential component of electric field vanishing at the perfectly
conducting boundaries. The solution that satisfies these
conditions is given by,
\begin{eqnarray}
\vec{A}(\vec{r})& = & A_x(r) \hat{x} +  A_y(r) \hat{y} + A_z(r) \hat{z}; \\
where \quad
A_x(r) & = & \sqrt{\frac{8}{V}}\ a_x\ \cos{k_xx}\ \sin{k_yy}\ \sin{k_zz}, \\
A_y(r) & = & \sqrt{\frac{8}{V}}\ a_x\  \sin{k_xx}\ \cos{k_yy}\ \sin{k_zz}, \\
A_z(r) & = & \sqrt{\frac{8}{V}}\ a_x\  \sin{k_xx}\ \sin{k_yy}\
\cos{k_zz};
\end{eqnarray}
    with,
\begin{eqnarray}
a_x^2 + a_y^2 + a_z^2  =  1 ;   V = L^2L_z \ & \& &\\
k_x =\frac{\pi l}{L};k_y =\frac{\pi m}{L}; k_z =\frac{\pi n}{L}; & &
 \mathrm{where}\ l,m,n=0,1,2,3,\ldots
\end{eqnarray}

     $\nabla.\vec{A}\ =\ 0$ also requires that
\begin{eqnarray}
k_x A_x + k_y A_y + k_z A_z & = & 0 \ \ \mathrm{or}\\
\frac{\pi}{L}(l A_x + m A_x) + \frac{\pi}{L_z}(nA_z)& = &0.
\label{pol-cond}
\end{eqnarray}
There are two polarizations possible for each value of $l,m,n$
unless one of them is zero. In that case, only one polarization is
possible due to Eqn.~\ref{pol-cond}.   Thus, the allowed modes
within the cavity are,
\begin{equation}
\omega_{lmn} = k_{lmn}c = \pi c \left( \frac{l^2}{L^2} +
        \frac{m^2}{L^2} + \frac{n^2}{L^2}\right)^{\frac{1}{2}}.
\end{equation}
    The zero-point energy of the field inside is,
\begin{equation}
E(d)= \sum_{lmn}\hspace*{-1.0mm}'\ 2\frac{1}{2} \hbar
\omega_{lmn}= \sum_{lmn}\hspace*{-1.0mm}'\ \pi \hbar c %
  \left( \frac{l^2}{L^2} + \frac{m^2}{L^2} + %
          \frac{n^2}{L^2}\right)^{\frac{1}{2}}.
\end{equation}

    The factor 2 arises from the 2 independent polarizations for
$l,m,n \neq 0$ and the prime on $\sum$ indicates that when $l\
\mathrm{or}\ m\ \mathrm{or}\ n = 0$, a factor of half has to be
inserted.

    When $L >> L_z = d$,
\vspace*{-5.0mm}
\begin{eqnarray}
\nonumber
\sum_{lmn}\hspace*{-1.0mm}' & \longrightarrow &
\sum_n\hspace*{-1.0mm}' \left( \frac{L}{\pi} \right)^2 \int\int dk_x dk_y \ \
\mathrm{and} \\
E(d) & = & \sum_{lmn}\hspace*{-1.0mm}'\ 2\frac{1}{2} \hbar \omega_{lmn}\\ %
& = &  \left( \frac{L}{\pi} \right)^2 \hbar c
\sum_{n=0}^\infty \hspace*{-1.0mm}' \int_0^\infty \int_0^\infty
dk_x dk_y \left( k_x^2 + k_y^2 + \frac{n^2
\pi^2}{d^2}\right)^\frac{1}{2}
\end{eqnarray}
This is {\it infinite}. Outside the cavity, $\sum_n$ will also be replaced by
$ \frac{d}{\pi} \int dk_z$ and
\begin{equation}
E(\infty) = \left( \frac{L}{\pi} \right)^2 \hbar c \frac{d}{\pi}
\int_0^\infty \int_0^\infty \int_0^\infty  dk_x dk_y dk_z \left(
k_x^2 + k_y^2 + k_z^2\right)^\frac{1}{2}.
\end{equation}
This is also {\it infinite}. But the potential energy of the system when the plates
are separated by a distance $d$ is $ U(d) = E(d) - E(\infty)$.
\begin{eqnarray}
\nonumber
U(d) & = &\left( \frac{L}{\pi} \right)^2 \hbar c \left[
\sum_{n=0}^\infty \hspace*{-1.0mm}' \int_0^\infty \int_0^\infty  dk_x dk_y
\left( k_x^2 + k_y^2 + \frac{n^2 \pi^2}{d^2}\right)^\frac{1}{2}\right. \\
&&- \left. \frac{d}{\pi} \int_0^\infty \int_0^\infty \int_0^\infty
dk_x dk_y dk_z \left( k_x^2 + k_y^2 + k_z^2\right)^\frac{1}{2}
\right].
\end{eqnarray}
In polar co-ordinates, $u,\theta$ in $k_x,k_y$ plane,
$dk_xdk_y=udud\theta,\ \theta$ ranges from $0$ to $\pi/2$ for
positive $k_x,\ k_y$, and we have,
\begin{eqnarray}
\nonumber
U(d) & = &\left( \frac{L}{\pi} \right)^2 \hbar c \frac{\pi}{2}
\left[\sum_n\hspace*{-1.0mm}' \int_0^\infty  du\ u
\left( u^2 + \frac{n^2 \pi^2}{d^2}\right)^\frac{1}{2}\right. \\
&&- \left. \frac{d}{\pi}  \int_0^\infty dk_z  \int_0^\infty du\ u
\left( u^2 + k_z^2\right)^\frac{1}{2}  \right] \label{U_polar}
\end{eqnarray}
This is still infinite. We now introduce a cut-off function
$f(k)=f([u^2+k^2]^{1/2})$ such that $f(k)=1\ \mathrm{for}\ k<<k_m$
and $f(k)=0\ \mathrm{for}\ k>>k_m$. We could say that for
wavelengths comparable to atomic dimensions, the assumption of
perfect conductivity breaks down and hence a cut-off is necessary
($k_m \approx 1/a_0,\ a_0$ is Bohr radius).  Thus,
Eqn.~\ref{U_polar} becomes,
\begin{eqnarray}
\nonumber
U(d) & = &\left( \frac{L}{\pi} \right)^2 \hbar c \frac{\pi}{2}
\left[\sum_{(n=0)}^\infty \hspace*{-1.0mm}' \int_0^\infty  du\ u
\left( u^2 + \frac{n^2 \pi^2}{d^2}\right)^\frac{1}{2}
f([u^2+k^2]^\frac{1}{2})\right. \\
&&- \left. \frac{d}{\pi}  \int_0^\infty dk_z  \int_0^\infty du\ u
\left( u^2 + k_z^2\right)^\frac{1}{2}f([u^2+k^2]^\frac{1}{2})\right] \\
\mathrm{Defining}\ x\equiv u^2d^2/\pi^2 &\ \&\ & \kappa=k_zd/\pi,\\
& = & \frac{L^2 \hbar c}{4\pi} \frac{\pi^3}{d^3}
\left[\sum_n\hspace*{-1.5mm}' \int_0^\infty  dx
\left( x + n^2\right)^\frac{1}{2}
f(\frac{\pi}{d}[x+k^2]^\frac{1}{2})\right. \nonumber\\
&&- \left. \int_0^\infty d\kappa  \int_0^\infty dx \left( x +
\kappa^2\right)^\frac{1}{2}
f(\frac{\pi}{d}[x+\kappa^2]^\frac{1}{2} )\right].
\end{eqnarray}
Now,
\begin{eqnarray}
U(d)& = & \frac{\pi^2\hbar c}{4d^3}\left[\frac{1}{2}F(0) +
      \sum_{n=1}^\infty F(n) -  \int_0^\infty d\kappa F(\kappa) \right];\\
\mathrm{where}\ F(\kappa)& \equiv & \int_0^\infty dx
    \left( x + \kappa^2\right)^\frac{1}{2}
    f(\frac{\pi}{d}[x+\kappa^2]^\frac{1}{2}).
\end{eqnarray}

According to Euler-Maclarin summation formula \cite{Abromo71},
\begin{eqnarray}
\sum_{n=1}^\infty F(n) -  \int_0^\infty d\kappa F(\kappa) =
-\frac{1}{2}F(0) - \frac{1}{12}F'(0)+ \frac{1}{720}F''(0) +
\cdots,
\end{eqnarray}
for $F(\infty)\rightarrow 0$. To evaluate $F^n(0)$, we note that,
\begin{equation}
F(\kappa)=\int_{\kappa^2}^\infty du\ \sqrt{u}
f(\frac{\pi}{d}\sqrt{u}), \hspace*{5mm}
F'(\kappa)=-2\kappa^2f(\frac{\pi}{d}\kappa).
\end{equation}
    Then $F'(0)=0,F'''(0)=-4\ \mathrm{and}\ F^n(0)=0$ for $n>3$, if
the cut-off function vanishes at $\kappa=0$. Thus,

\begin{eqnarray}
\sum_{n=1}^\infty F(n) -  \int_0^\infty d\kappa F(\kappa)& = &
-\frac{1}{2}F(0)  -\frac{4}{720} \\
U(d) &=& \frac{\pi^2\hbar c}{4d^3} L^2 \left(-\frac{4}{720}\right)\\
&=&\frac{\pi^2\hbar c}{720d^3} L^2
\end{eqnarray}
    which is finite and independent of the cut-off function. The attractive
force per unit area,
\begin{equation}
F_c(d) = - \frac{\pi^2\hbar c}{240d^3},
\end{equation}
which is the Casimir force. Thus, Casimir showed that changes in zero-point energy
can be finite and observable.

\section{Force between Dielectrics}

    In the previous section, we went through Casimir's derivation of the
force between two perfectly conducting parallel plates. In
experimental situations, the simplified assumption of perfect
conductivity at all field frequencies is unrealistic. The
dielectric properties of the media should also be included.
    Liftshitz~\cite{Lift56}, developed the first macroscopic theory of forces
between dielectrics. His results reduce to the Casimir force derived above
for the case of perfect conductors. We will now derive Lifshitz's results in a
manner similar to Casimir's approach of mode counting~\cite{Miloni94}.

\begin{figure}
\begin{center}
\includegraphics{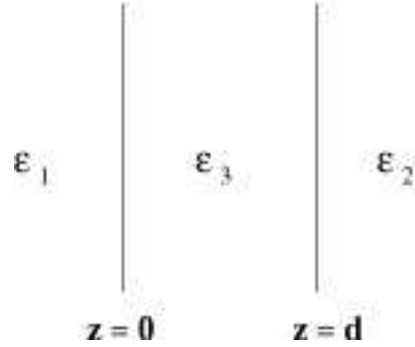}
\end{center}
\caption{Two semi-infinite dielectric slabs of dielectric
constants $\epsilon_1$ and $\epsilon_2$, placed distance d apart
in a medium of dielectric constants $\epsilon_3$.}
\label{dielectric}
\end{figure}
\noindent
    Consider the case of a medium with dielectric constant
$\epsilon_3(\omega)$ sandwiched between two semi-infinite media of
dielectric constants $\epsilon_1(\omega)$ and $\epsilon_2(\omega)$
[Fig.~\ref{dielectric}]. We will calculate the force per unit area
from the total zero-point energy of the modes $\omega_n$ that are
possible in this situation.
    The modes that are possible are the solutions of the Maxwell's equations,
\begin{eqnarray*}
\nabla.\vec{D} & = & 0, \\
\nabla \times \vec{E} & = & -\frac{1}{c}\frac{\partial \vec{B}}{\partial t}, \\
\nabla.\vec{B} & = & 0, \\
\nabla \times \vec{D} & = & \frac{1}{c}\frac{\partial \vec{D}}{\partial t},
\end{eqnarray*}
where, we have assumed the media to be isotropic, with magnetic
permeability $\mu =1$, and zero net charge density, so,
$\vec{D}(\vec{r},t) = \epsilon (\omega)\vec{E}(\vec{r},t)$. These
modes should also satisfy the appropriate boundary conditions.
    Consider solutions to the Maxwell's equations of the form
$\vec{E}(\vec{r},t)=\vec{E_0}(\vec{r},t)e^{-i\omega t}$,
$\vec{B}(\vec{r},t)=\vec{B_0}(\vec{r},t)e^{-i\omega t}$, such that in each
region, $\nabla.\vec{E_0},\nabla.\vec{B_0} = 0 $ and,
\begin{eqnarray}
\nabla^2 \vec{E_0} + \frac{\omega^2}{c^2}\epsilon (\omega )\vec{E_0}&=&0, \\
\nabla^2 \vec{B_0} + \frac{\omega^2}{c^2}\epsilon (\omega )\vec{B_0}&=&0.
\end{eqnarray}

    Across the boundaries,
\begin{description}
\item[(i)] normal component of $\vec{D}(=\epsilon\vec{E})$ should be continuous,
\item[(ii)] tangential component of $\vec{E}$ should be continuous,
\item[(iii)] normal component of $\vec{B}(=\vec{H})$ should be
continuous, and
\item[(iv)] tangential component of $\vec{B}$ should be continuous.
\end{description}
    Let us assume solutions of the form,
\begin{eqnarray}
\vec{E_0}(\vec{r})&=&\left[ e_x(z) \hat{x} + e_y(z) \hat{y} +e_z(z) \hat{z} \right]
            e^{i(k_x x+ k_y y)},\\
\vec{B_0}(\vec{r})&=&\left[ b_x(z) \hat{x} + b_y(z) \hat{y} +b_z(z) \hat{z} \right]
            e^{i(k_x x+ k_y y)}, \\
\mathrm{so\ that,}
\frac{d^2e_x}{dz^2} - K^2e_x = 0,&\hspace{5mm}& \frac{d^2b_x}{dz^2} - K^2b_x = 0,\\
\mathrm{where\ }K^2 &=& k_x^2 + k_y^2 - \frac{\omega^2}{c^2} \epsilon(\omega)
\label{d2e_x}
\end{eqnarray}
and likewise for $y$ and $z$ components.
    In free space, $\epsilon(\omega)=1$ everywhere for all $\omega$ and the
usual plane wave solutions are obtained for $K^2<0;\ k_x^2 +k_y^2
+k_z^2=\frac{\omega^2}{c^2},\ K^2=-k_z^2$.

    For the case under consideration, let us assume $\epsilon(\omega)$
to be real and $K^2>0$ in all three media. We can chose a
coordinate system in which $k_y=0$ and
$K^2=k^2-\epsilon(\omega)\frac{\omega^2}{c^2},\ k \equiv k_x$.
\vspace*{-2.0mm}
\begin{equation}
\nabla \vec{E_0}=0,\ \mathrm{implies\ that}\ ike_x +
\frac{de_z}{dz}=0. \label{e_x}
\end{equation}
\vspace*{-2.0mm}
$\nabla \times \vec{E_0}=i\frac{\omega}{c}\vec{B_0}$, gives
\begin{equation}
\vec{B_0}(\vec{r}) = \left[ i\frac{c}{\omega}\frac{de_y}{dz}\hat{x}
    - \frac{c}{\omega}\left(ke_z+i\frac{de_x}{dz}\right)\hat{y}
    + \frac{c}{\omega}ke_y \hat{z} \right] e^{ikx}. \label{B_0}
\end{equation}
    $\nabla.\vec{B_0}$ is also identically satisfied by Eqn.~\ref{B_0}.\\
    The boundary conditions mentioned above imply,
\begin{description}
\item[(i)]  $\epsilon(\omega)e_z(z)$ should be continuous,
\item[(ii)] $e_x$ should be continuous which is satisfied if
$\frac{de_x}{dz}$ is continuous (from Eqn.~\ref{e_x}), and
\item[(iii)]$e_y$ should be continuous, \item[(iv)]
$\frac{de_y}{dz}\ \mathrm{and}\ ke_z + i\frac{de_x}{dz}$ should be
continuous,
\end{description}
But, from Eqn.~ \ref{d2e_x} and ~\ref{e_x}, $ke_z +
i\frac{de_x}{dz} =
\frac{1}{k}\epsilon(\omega)\frac{\omega^2}{c^2}e_z$. Thus, all
conditions are satisfies if,
\begin{description}
\item[a.] $\epsilon(\omega)e_z(z)$ and $\frac{de_x}{dz}$ are
continuous, and
\item[b.] $e_y$ and $\frac{de_y}{dz}$ are continuous.
\end{description}
    Now, since $\frac{d^2e_z}{dz^2} - K^2e_z = 0$, in each region, ignoring
exponential growing solutions, we have,
\begin{eqnarray}
e_z(z) & =  & Ae^{k_1z}, \hspace*{23mm}\mathrm{for}\ z<0;\\
& = & Be^{k_3z}+Ce^{-k_3z,}  \hspace*{5.2mm}\mathrm{for}\ 0\leq z \leq 0; \\
& =  & De^{-k_2z,}  \hspace*{20mm}\mathrm{for}\ z>0;
\end{eqnarray}
where $K_j^2 \equiv k^2 - \epsilon_j(\omega)\frac{\omega^2}{c^2}$.
Applying the boundary conditions \textbf{`\texttt{a}'}, we have,
\begin{eqnarray}
-\epsilon_1 A + \epsilon_3 B + \epsilon_3 C & = & 0; \\
-K_1 A + K_3 B - K_3 C & = & 0; \\
\epsilon_3 e^{K_3 d} B + \epsilon_3 e^{-K_3 d} C - \epsilon_2 e^{K_2 d}D & = & 0;\\
K_3 e^{K_3 d} B - K_3 e^{-K_3 d} C + K_2 e^{K_2 d}D & = & 0.
\end{eqnarray}

For non-trivial solutions to exist,\\
\begin{center}
$\left|
\begin{array}{cccc}
-\epsilon_1 & \epsilon_3 & \epsilon_3 & 0\\
-K_1        &  K_3       & -K_3       & 0\\
0           &\epsilon_3 e^{K_3 d} & \epsilon_3 e^{-K_3 d}  & -\epsilon_2 e^{-K_2 d}\\
0           &K_3 e^{K_3 d} & -K_3 e^{-K_3 d}  & K_2 e^{-K_2 d}\\
\end{array}
\right| = 0 .$\\
\end{center}
This yields,
\begin{equation}
\frac{\epsilon_3K_1 + \epsilon_1 K_3}{\epsilon_3K_1 - \epsilon_1
K_3} . \frac{\epsilon_3K_2 + \epsilon_2 K_3}{\epsilon_3K_2 -
\epsilon_1 K_3}. e^{2K_3 d} - 1 = 0. \label{cond-a}
\end{equation}
Similarly, it can be shown that the boundary conditions
\textbf{`\texttt{b}'} are satisfied if,
\begin{equation}
\frac{K_1 + K_3}{K_1 - K_3} .\frac{K_2 + K_3}{K_2 - K_3}.e^{2K_3
d} - 1 = 0. \label{cond-b}
\end{equation}
Equations \ref{cond-a} and \ref{cond-b} are the conditions on the
allowed modes $\omega_n$. Both cannot in general be satisfied
simultaneously. For boundary conditions \textbf{\texttt{a}} and
\textbf{\texttt{b}} to be satisfied simultaneously; (a)
Eqn.~\ref{cond-a} should be satisfied with $e_y \equiv 0$ or (b)
Eqn.~\ref{cond-b} should be satisfied with $e_z \equiv 0$. Thus we
have two kinds of modes. Since these are exponentially decaying
functions of $z$ for $z<0$ and $z>d$, these are called  {\it
surface modes}~\cite{Barton79}.\\

Let us now calculate the zero-point energy of these modes. The
energy is given by,
\begin{equation}
E(d) = \sum_n \frac{1}{2}\hbar \omega_{na} + \sum_n \frac{1}{2}\hbar \omega_{nb},
\end{equation}
where $\omega_{na}$ are the modes of type (a) and $\omega_{nb}$ are the
modes of type (b). Since $k_x$ and $k_y$ are continuous, the summations over them
in $\omega_{n\alpha},\ \alpha=a,b$ can be replaced  with integrals.
\begin{equation}
\sum_x\ \rightarrow\ \left(\frac{L}{2\pi}\right)^2 \int dk_x \int dk_y \sum_N
= \left(\frac{L}{2\pi}\right)^2 \int 2\pi kdk \sum_N,
\end{equation}
 where $L$ is the length along $x$ and $y$ sides of the `quantization box' and $\sum_N$
denotes sum over the solutions of \ref{cond-a} and \ref{cond-b}. Thus,
\begin{equation}
E(d) = \frac{\hbar L^2}{4\pi} \int_0^\infty dk\ k
\left[ \sum_N \omega_{Na}(k) + \sum_N \omega_{Nb}(k)\right].
\end{equation}
In order to evaluate this summation, we employ the `argument
theorem' from the theory of functions of complex variables: for a
function $f(z)$ which is analytic except for poles on and inside a
simple closed circle $C$,
\begin{equation}
\frac{1}{2\pi i} \oint_C \frac{f'(z)}{f(z)} dz = N - P,
\end{equation}
where $N$ is the number of zeros and $P$ the number of poles $f(z)$ inside $C$
(see for example, \cite{Boas83}).
It can be shown from this that,
\begin{eqnarray}
\frac{1}{2\pi i} \oint_C z \frac{f'(z)}{f(z)} dz & =  &
\left[\sum_i z_i \right]_{f(z_i)=0} - \left[\sum_i z_i \right]_{f(z_i)=\infty};\\
& = & \mathrm{sum\ of\ zero's\ of}\ f(z_i)\ \mathrm{inside\ C}\\
&& -\ \mathrm{sum\ of\ poles\ of}\ f(z_i)\ \mathrm{inside\ C.}
\end{eqnarray}
Let $F_a(\omega)$ and $F_b(\omega)$ be the L.H.S of
Eqns.~\ref{cond-a} and \ref{cond-b} then,
\begin{equation}
\sum_N \omega_{N\alpha}(K) = \mathrm{sum\ of\ zero's\ of}\ F_{\alpha},\ \alpha= a,b.
\end{equation}
    The poles of $F_{\alpha}$, considered as function of complex variable
$\omega$, are independent of the boundaries and hence of $d$.
Therefore,
\begin{equation}
\frac{1}{2\pi i} \oint_C \omega \frac{F'_\alpha(\omega)}{F_\alpha(\omega)} d\omega
=  \sum_N \omega_{N\alpha}(K) -\ (\mathrm{Term\ independant\ of}\ d).
\end{equation}
The $\omega_{N\alpha}(K)$ of interest are those that are positive
and real. Hence, the closed curve $C$ is defined by the imaginary
axis of complex $\omega$ plane and a semicircle in the right half
of this plane, with radius extending to infinity. As the $d$
dependant term does not contribute to the force, for this
analysis, we can write,
\begin{equation}
E(d) = \frac{\hbar L^2}{4\pi} \frac{1}{2\pi i} \int_0^\infty dk\ k
\left[ \oint_C \omega \frac{F'_a(\omega)}{F_a(\omega)} d\omega +
 \oint_C \omega \frac{F'_b(\omega)}{F_b(\omega)} d\omega \right].
\end{equation}
Each of these contour integrals, can be written as a part along
the imaginary axis and a part along the infinite semicircle. The
later integral is $d$-independent and does not contribute to the
force. The former integral is,
\begin{eqnarray}
\int_\infty^{-\infty} i\xi \frac{1}{F_\alpha(i\xi)} \frac{\partial
F_\alpha(i\xi)}{\partial (i\xi)} i d\xi & = & -i
\int_\infty^{-\infty} d\xi\ \xi
\frac{G'_\alpha(\xi)}{G_\alpha(\xi)};\\
& = & -i \int_\infty^{-\infty} d\xi\ \xi \frac{d}{d\xi}\log{G_\alpha(\xi)};\\
& = & i \int_{-\infty}^\infty d\xi\ log{G_\alpha(\xi)};
\end{eqnarray}
where$\ G_\alpha(\xi) \equiv F_\alpha(i\xi),\ \alpha=a,b$. Writing explicitly,
\begin{eqnarray}
G_a(\xi) & = &
\frac{\epsilon_3K_1 + \epsilon_1 K_3}{\epsilon_3K_1 - \epsilon_1 K_3}.
\frac{\epsilon_3K_2 + \epsilon_2 K_3}{\epsilon_3K_2 - \epsilon_1 K_3}.
e^{2K_3 d} - 1 ;\label{G_a}\\
G_b(\xi) & = & \frac{K_1 + K_3}{K_1 - K_3} .\frac{K_2 + K_3}{K_2 -
K_3}.e^{2K_3 d} - 1; \label{G_b}
\end{eqnarray}
\vspace*{2mm} where, now $\epsilon_j = \epsilon_j(i\xi)$ and
$K_j^2 = k^2 + \epsilon_j(i\xi)\frac{\xi^2}{c^2}$. The zero-point
energy associated with the allowed modes is then,
\begin{equation}
E(d) = \frac{\hbar L^2}{8\pi^2}\int_0^\infty dk\ k
\left[ \int_{-\infty}^\infty d\xi\ log{G_a(\xi)} +
\int_{-\infty}^\infty d\xi\ log{G_a(\xi)} \right] +
\mathrm{d\ independant\ term}.
\end{equation}
The force $F(d) = - \frac{\partial}{\partial d}E(d)$,
\begin{equation}
F(d) = - \frac{\hbar L^2}{8\pi^2}\frac{\partial}{\partial d}
\left[ \int_0^\infty dk\ k \left\{ \int_{-\infty}^\infty d\xi\
log{G_a(\xi)} + \int_{-\infty}^\infty d\xi\ log{G_b(\xi)} \right\}
\right].
\end{equation}
But, from equation \ref{G_a} and \ref{G_b}, $\frac{\partial}{\partial d}
G_\alpha(\xi)= 2K_3\left( G_\alpha + 1 \right)$, and
\begin{eqnarray}
\frac{\partial}{\partial d}\int_{-\infty}^\infty d\xi\ log{G_\alpha(\xi)}
& = &  \int_{-\infty}^\infty d\xi\ \frac{\partial}{\partial d} log{G_\alpha(\xi)};\\
& = & \int_{-\infty}^\infty d\xi\ \frac{1}{G_\alpha(\xi)}
       \left[ 2K_3\left( G_\alpha + 1 \right) \right];\\
& = & 2 \int_{-\infty}^\infty d\xi\ K_3 +
      2 \int_{-\infty}^\infty d\xi\ \frac{K_3}{G_\alpha(\xi)}.
\end{eqnarray}
The first term is independent of the presence of the dielectric
slabs and hence is not related to the force between them.
Therefore, the force per unit area,
\begin{equation}
F(d) = -\frac{\hbar}{8\pi^2} \int_0^\infty dk\ k
\int_{-\infty}^\infty d\xi\ 2 K_3 \left[\frac{1}{G_a(\xi)} +
\frac{1}{G_b(\xi)}  \right].
\end{equation}
$K_j$ and $\epsilon_j$ are even functions of $\xi$, and so is
$G_\alpha$; Hence
\begin{equation}
F(d) = -\frac{\hbar}{2\pi^2} \int_0^\infty dk\ k \int_{0}^\infty
d\xi\ K_3 \left[\frac{1}{G_a(\xi)} + \frac{1}{G_b(\xi)}  \right].
\end{equation}
More explicitly,
\begin{eqnarray}
F(d)& = &-\frac{\hbar}{2\pi^2}
\int_0^\infty dk\ k \int_{0}^\infty d\xi\  K_3 \nonumber \\
&&\left[
\left\{ \frac{\epsilon_3K_1 + \epsilon_1 K_3}{\epsilon_3K_1 - \epsilon_1 K_3}.
        \frac{\epsilon_3K_2 + \epsilon_2 K_3}{\epsilon_3K_2 - \epsilon_1 K_3}.
        e^{2K_3 d} - 1 \right\}^{-1}  \right. \nonumber \\
&& + \left. \left\{ \frac{K_1 + K_3}{K_1 - K_3} .\frac{K_2 +
K_3}{K_2 - K_3}.e^{2K_3 d} - 1 \right\}^{-1} \right]
;\label{D-P(d)}
\end{eqnarray}
where, $\epsilon_j = \epsilon_j(i\xi)$ and $K_j^2 = k^2 +
\epsilon_j(i\xi)\frac{\xi^2}{c^2}$. \\
To compare with Lifshitz theory we introduce a variable $p$ such
that,
\begin{eqnarray}
k^2 & = &\epsilon_3 \frac{\xi^2}{c^2}\left(p^2 -1\right) .\\
\mathrm{Then,}\ K_3^2 = k^2 + \epsilon_3\frac{\xi^2}{c^2}
& = & \epsilon_3\frac{\xi^2}{c^2}p^2; \\
\mathrm{and}%
K_{1,2}^2 = k^2 + \epsilon_{1,2}\frac{\xi^2}{c^2} & =
& \epsilon_3\frac{\xi^2}{c^2}\left( p^2 - 1\right) +
    \epsilon_{1,2}\frac{\xi^2}{c^2} \nonumber ;\\
& = & \epsilon_3\frac{\xi^2}{c^2}
 \left( p^2 - 1 + \frac{\epsilon_{1,2}}{\epsilon_3} \right); \nonumber\\
& \equiv & \epsilon_3\frac{\xi^2}{c^2}s_{1,2}^2,\ \
\mathrm{where}\ s_{1,2}^2 = p^2 - 1 +
         \frac{\epsilon_{1,2}}{\epsilon_3}.
\end{eqnarray}
Changing to these set of variables, $dk\ k= \epsilon_3\frac{\xi^2}{c^2}dp\ p$ and,
\begin{eqnarray}
F(d)& = &-\frac{\hbar}{2\pi^2c^3}
\int_1^\infty dp\ p^2 \int_{0}^\infty d\xi\  \xi^3 \epsilon_3^{3/2} \nonumber \\
&&\left[
\left\{ \frac{\epsilon_3 s_1 + \epsilon_1 p}{\epsilon_3 s_1 - \epsilon_1 p}.
        \frac{\epsilon_3 s_2 + \epsilon_2 p}{\epsilon_3 s_2 - \epsilon_1 p}.
        e^{2\xi p\sqrt{\epsilon_3} d/c} - 1 \right\}^{-1}  \right. \nonumber \\
&& + \left.
\left\{ \frac{s_1 + p}{s_1 - p} .\frac{s_2 + p}{s_2 - p}.
     e^{2\xi p\sqrt{\epsilon_3} d/c} - 1 \right\}^{-1} \right]
\label{D1-P(d)}
\end{eqnarray}
    For the case when there is vacuum between the slabs, ($\epsilon_3 = 1$),
\begin{eqnarray}
F(d)& = &-\frac{\hbar}{2\pi^2c^3}
\int_1^\infty dp\ p^2 \int_{0}^\infty d\xi\  \xi^3  \nonumber \\
&&\left[
\left\{ \frac{ s_1 + \epsilon_1 p}{ s_1 - \epsilon_1 p}.
        \frac{ s_2 + \epsilon_2 p}{ s_2 - \epsilon_1 p}.
        e^{2\xi p d/c} - 1 \right\}^{-1}  \right. \nonumber \\
&& + \left.
\left\{ \frac{s_1 + p}{s_1 - p} .\frac{s_2 + p}{s_2 - p}.
     e^{2\xi p d/c} - 1 \right\}^{-1} \right].
\label{Lift-P(d)}
\end{eqnarray}
    which matches with Lifshitz's result \cite{Lift56}.

\section{Casimir force at finite temperature}
    When we derived the Casimir pressure between the dielectric slabs, we assumed
the vacuum state for the field and added only the zero-point energy of the allowed
modes to calculate the energy of the field. At any finite temperature,
\begin{equation}
E_n = \left( n(\omega) + \frac{1}{2}\right) \hbar \omega,\
\mathrm{where}\ n(\omega) =
\frac{1}{e^{\frac{\hbar\omega}{k_{\tiny B} T}}-1}.
\label{planck}
\end{equation}
Thus to calculate the energy of the allowed modes at a finite
temperature, we have to use Eqn.~\ref{planck}, rather than just
$\frac{1}{2}\hbar\omega$,
\begin{eqnarray}
E_{T>0}(d)& = &E(d).\frac{\left( n(\omega) + \frac{1}{2}\right)}{\frac{1}{2}};\\
& = & E(d). 2 .\left( \frac{1}{e^{\frac{\hbar\omega}{k_{\tiny B} T}}-1}
            + \frac{1}{2} \right);\\
& = & E(d)\coth{\frac{\hbar\omega}{2k_{\tiny B} T}}.
\end{eqnarray}
    The easiest way to now derive the force expression is to use $\omega = i \xi$
in Eqn.~\ref{Lift-P(d)} and multiply by the factor
$\coth{\frac{\hbar\omega} {2k_{\tiny B} T}}$. For simplicity, let
$\epsilon_3 =1, K_3 = -i\frac{\omega}{c}p$ and,
\begin{eqnarray}
F(d)& = &-\frac{\hbar}{2\pi^2c^3}
\int_1^\infty dp\ p^2 \int_{0}^\infty d\omega\  \omega^3  \nonumber \\
&&\left[
\left\{ \frac{ s_1 + \epsilon_1 p}{ s_1 - \epsilon_1 p}.
        \frac{ s_2 + \epsilon_2 p}{ s_2 - \epsilon_1 p}.
        e^{-2i\omega p d/c} - 1 \right\}^{-1}  \right. \nonumber \\
&& + \left.
\left\{ \frac{s_1 + p}{s_1 - p} .\frac{s_2 + p}{s_2 - p}.
     e^{-2i\omega p d/c} - 1 \right\}^{-1} \right]
\coth{\frac{\hbar\omega}{2k_{\tiny B} T}}. \label{Lift-PT(d)}
\end{eqnarray}
To evaluate this integral, we could use $\omega\rightarrow i\xi$
as before. But, $\coth{\frac{\hbar\omega}{2k_{\tiny B} T}}$ has
poles on the imaginary axis at $\omega_n=\frac{2\pi ink_BT}{\hbar}
\equiv i\xi_n$, for all integer $n$. So the closed curve $C$ will
now have semicircles around these poles in the path that runs
along imaginary axis. The integration along the semicircles about
$\omega_n$ contribute $i\pi$ times the residue of the integrand at
$\omega_n$ for $n>0$ and at $n=0$ it will contribute
$\frac{i\pi}{2}$ from the quarter circle.
\begin{eqnarray}
F(d)& = &-\frac{\hbar}{2\pi^2c^3}\frac{2\pi ik_BT}{\hbar} i^3
\sum_{n=0}^\infty \hspace*{-1.0mm}' \xi^3 \int_1^\infty dp\ p^2  \nonumber \\
&&\left[ \left\{
    \frac{\epsilon_{1n} p}{ s_{1n} - \epsilon_{1n} p}.
        \frac{ s_{2n} + \epsilon_{2n} p}{ s_{2n} - \epsilon_{1n} p}.
        e^{2\xi_n p d/c} - 1 \right\}^{-1}  \right. \nonumber \\
&& + \left.
\left\{ \frac{s_{1n} + p}{s_{1n} - p} .\frac{s_{2n} + p}{s_{2n} - p}.
     e^{2\xi_n p d/c} - 1 \right\}^{-1} \right];
\end{eqnarray}
where $s_{jn}=\sqrt{p^2-1+\epsilon_{jn}},\
\epsilon_{jn}=\epsilon_{j}(i\xi_n),\ j=1,2$ and the prime on the
summation sign indicates that a factor half must be included in
$n=0$ term. This result was first derived by Lifshitz
\cite{Lift56}. For the case when $\epsilon_{1,2}\rightarrow
\infty$,
\begin{eqnarray}
F(d)& = &-\frac{k_BT}{\pi c^3} \sum_{n=0}^\infty \hspace*{-1.0mm}'
\xi^3\int_1^\infty dp\ p^2
\left( \frac{2}{e^{2\xi_n p d/c} - 1}  \right); \nonumber \\
&=&-\frac{2k_BT}{\pi c^3}
\sum_{n=0}^\infty \hspace*{-1.0mm}' \xi^3 \int_1^\infty dp\
\frac{p^2}{e^{2\xi_n p d/c} - 1};\\
&=& -\frac{2k_BT}{\pi c^3}\frac{c^3}{8d^3} \sum_{n=0}^\infty
\hspace*{-1.0mm}' \int_{nx}^\infty dy\ \frac{y^2}{e^y - 1};
\end{eqnarray}
where $x\equiv \frac{4\pi k_B Td}{\hbar c}$.
When $x\gg 1$, the dominant contribution if from the $n=0$ term

\begin{eqnarray}
F(d) & = & -\frac{k_BT}{4\pi d^3}\frac{1}{2} \int_{0}^\infty dy\ \frac{y^2}{e^y - 1};\\
& = & -\frac{\zeta(3)k_BT}{4\pi d^3}.
\end{eqnarray}
    Note that the distance dependance changes from $1/d^4$ at `low temperature'
to $1/d^3$ at `high temperature'. Also, the high and low
temperature regime are defined by the dimensionless parameter
$x\equiv \frac{4\pi k_B Td}{\hbar c}$. At any given temperature
$T$, we can always choose a large enough $d$, such that $x \gg 1$.
Thus, low or high temperature is equivalent to smaller or larger
separations.

\bibliographystyle{plain}
\bibliography{reference}

\clearpage

\chapter{Casimir Force between spherical lens and flat plate}

The Casimir force per unit area between two parallel conducting
plates, at $T=0$ K is given by,
\begin{eqnarray}
f_{c0}(z)& = &-\frac{\pi^2 \hbar c}{240z^4} \label{fcas0}\\
      & = & -\frac{0.013}{z_\mu^4} \quad \mathrm{dyn.cm}^{-2}\ \
          \mathrm{where} \ z_\mu \equiv z\ \mathrm{in\ microns}.
\end{eqnarray}

For a finite temperature $T$ we approximate the distance
dependence as
\begin{eqnarray}
f_{cT}(z)  & = & f_{c0}(z), \hspace*{1.25cm}\mathrm{for}\ \ z < \lambda_T;\\
f_{cT}(z) & = & f_{c0}(z)\frac{z}{\lambda_T },
  \hspace*{0.7cm} \mathrm{for}\ \ z > \lambda_T. \label{fcasT}
\end{eqnarray}

\begin{figure}
\begin{center}
  \resizebox{10cm}{!}{\includegraphics[10mm,10mm][223mm,173mm]
  {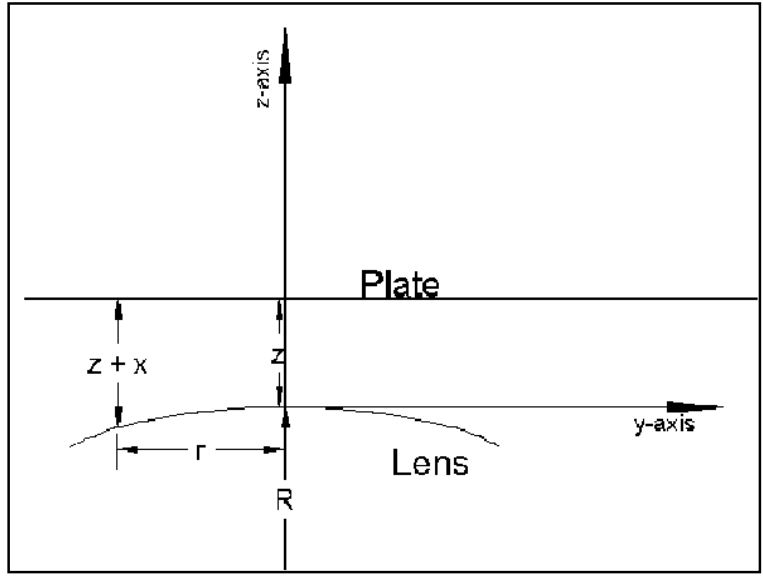}}\\
\end{center}\vspace*{-1cm}
  \caption{Reference axes for the system.}\label{axis}
\end{figure}

For convenience, we write Eqns.~\ref{fcas0} and \ref{fcasT} as,
\begin{eqnarray}
f_{c0}(z) & \equiv & -\frac{A_c}{z^4} \quad \mathrm{dyn.cm}^{-2},\\
f_{cT}(z) & \equiv & -\frac{A_c}{\lambda_T z^3}  \quad
\mathrm{dyn.cm}^{-2}, \\
&& \mathrm{where\ }\quad A_c = \frac{\pi^2 \hbar c}{240}.
\end{eqnarray}

The Casimir force between the lens and plate can now be estimated
using the ``Derjaguin (proximity) approximation", often called
``proximity theorem"~\cite{Proximity}. Here the total force is
calculated by integrating over concentric rings on the lens
surface that are equidistant from the plate. Thus, the  Casimir
force between the lens and plate is given by,
\begin{eqnarray}
f_{cl}(z)& = &\int_0^{r_{max}} f(z + x(r)). 2\pi r. dr, \\
\mathrm{with}\quad  (2R -x) x & = & r^2; \\
x & \approx & \frac{r^2}{2 R},
\end{eqnarray}
where $z$ is the separation between the lens and the plate, $R$ is
the radius of curvature of the lens and $r_{max}$ is the aperture
radius of the lens [Fig.~\ref{axis}]. The function, $f(z)$, takes
different forms depending on $z$ as compared with $\lambda_T$. For
$\mathbf{z < \lambda_T}$,
\begin{eqnarray}
f_{cl}(z)& = & \int_0^{\sqrt{2R\lambda_T}} f_{c0}(z + x(r)). 2\pi r. dr +%
            \int_{\sqrt{2R\lambda_T}}^{r_{max}} f_{cT}(z + x(r)). 2\pi r.dr\\%
            &=& \int_0^{\sqrt{2R\lambda_T}}
                    2 \pi A_c \frac{r dr}{\left( z + \frac{r^2}{2R}\right)^4} + %
                \int_{\sqrt{2R\lambda_T}}^{r_{max}}
                   \frac{2 \pi A_c}{\lambda_T}
                      \frac{r dr}{\left( z + \frac{r^2}{2R}\right)^3} \\
            & =& -\frac{2 \pi R A_c}{3} \left\{ \frac{1}{z^3} - \frac{1}{(z + \lambda_T )^3}%
                                          \right\} \\%
     & &    \quad -\frac{2 \pi R A_c}{2 \lambda_T} \left\{ \frac{1}{(z + \lambda_T )^2}-%
                            \frac{1}{\left (z + \frac{r_{max}^2}{2R} \right)^2}%
                                                   \right\}\\
f_{cl}(z)& = & -\frac{2 \pi R A_c}{3}
            \left [ \frac{1}{z^3}-\frac{1}{\left(z +\lambda_T\right)^3}%
         + \frac{3}{2 \lambda_T}\ \left \{ \frac{1}{(z + \lambda_T)^2}- %
          \frac{1}{(z + x_{max})^2} \right \} \right ];
\end{eqnarray}
where $x_{max} = \frac{r_{max}^2}{2R}$. For $\mathbf{z >
\lambda_T}$,
\begin{eqnarray}
f_{cl}(z)& = & \int_0^{r_{max}} f_{cT}(z + x(r)). 2\pi r.dr\\%
         & = & -\frac{\pi R A_c}{\lambda_T} \left[%
                   \frac{1}{z^2} - \frac{1}{(z + x_{max})^2}
                                             \right].
\end{eqnarray}

\bibliographystyle{plain}
\bibliography{reference}

\begin{thebibliography}{1}

\bibitem{Abromo71}
M.~{Abramowitz} and {Stegun}~I. A.
\newblock {\em {Handbook of Mathematical Functions}}.
\newblock Dover Books, Newyork, 1971.

\bibitem{Barton79}
G.~{Barton}.
\newblock {Some Surface Effects in the Hydrodynamic Model of Metals}.
\newblock {\em Rep. Prog. Phys.}, 42:963, 1979.

\bibitem{Boas83}
M.~L {Boas}.
\newblock {\em {Mathematical Methods in the Physical Sciences}}.
\newblock Wiley, Newyork, 1983.

\bibitem{Lift56}
E.~M. {Lifshitz}.
\newblock Theory of molecular attraction between solids.
\newblock {\em Sov. Phys. JETP}, 2:73, 1956.

\bibitem{Miloni94}
W.~P {Milonni}.
\newblock {\em {The Quantum Vacuum: An Introduction to Quantum
  Electrodynamics}}.
\newblock Academic Press, Newyork, 1994.

\end{thebibliography}


\begin{thebibliography}{1}

\bibitem{Proximity}
J.~{Blocki} et~al.
\newblock {}.
\newblock {\em Ann. Phys. (N.Y.)}, 105:427, 1977.

\end{thebibliography}


\begin{thebibliography}{10}

\bibitem{Adel2002}
E.~G. Adelberger.
\newblock {Sub-millimeter Tests of the Gravitational Inverse Square Law}.
\newblock {\em http://arXiv.org/abs}, hep-ex/0202008, 2002.

\bibitem{Adel2003}
E.~G. {Adelberger}, {Heckel}, and {Nelson} A.~E. B.~R.
\newblock {Tests of the Gravitational Inverse sqare law}.
\newblock {\em Ann. Rev. of Nuclear and Particle Science}, 53:77, December
  2003.

\bibitem{ADD2}
I.~{Antoniadis}, N.~{Arkani-Hamed}, S.~{Dimopoulos}, and G.~{Dvali}.
\newblock {New Dimension at a Millimeter to a Fermi and Superstrings at a TeV}.
\newblock {\em Phys. Lett. B}, 436:257, 1998.

\bibitem{Hamed02}
N.~{Arkani-Hamed}.
\newblock {Large Extra Dimensions: A new Arena for Particle Physics}.
\newblock {\em Physics Today}, page~35, Feb. 2002.

\bibitem{ADD1}
N.~{Arkani-Hamed}, S.~{Dimopoulos}, and G.~{Dvali}.
\newblock {The Hierarchy Problem and New Dimension at a Millimeter}.
\newblock {\em Phys. Lett. B}, 429:268, 1998.

\bibitem{ADD1999}
N.~{Arkani-Hamed}, S.~{Dimopoulos}, and G.~{Dvali}.
\newblock {Phenomenology, astrophysics, and cosmology of theories with
  submillimeter dimensions and TeV scale quantum gravity}.
\newblock {\em Phys. Rev. D}, 59:086004, April 1999.

\bibitem{Bordag2000}
M.~{Bordag}, B.~{Geyer}, G.~L. {Klimchitskaya}, and V.~M. {Mostepanenko}.
\newblock {Casimir Force at Both Nonzero Temperature and Finite Conductivity}.
\newblock {\em Physical Review Letters}, 85:503--506, July 2000.

\bibitem{Bordag01}
M.~{Bordag}, U.~{Mohideen}, and V.~M. {Mostepanenko}.
\newblock {New Developements in Casimir Force}.
\newblock {\em Phys. Rep.}, 353:1, 2001.

\bibitem{Ruoso02}
G.~{Bressi} et~al.
\newblock {Measurement of the Casimir Force between Parallel Metallic
  Surfaces}.
\newblock {\em Phys. Rev. Lett}, 88:041804--1, 2002.

\bibitem{Stein2000}
R.~R {Caldwell} and P~J {Steinhardt}.
\newblock Quintessence.
\newblock {\em Physics Worlds}, page~31, Nov 2000.

\bibitem{Cas48}
H.~B. {Casimir}.
\newblock {On the Attraction Between Two Perfectly Conducting Plates}.
\newblock {\em Proc. K. Ned. Akad. Wet.}, 51:793, 1948.

\bibitem{Chan01}
H.~B. {Chan} et~al.
\newblock {Quantum Mechanical Actuation of Micromechanical Systems by the
  Casimir Force}.
\newblock {\em Science}, 291:9, 2001.

\bibitem{RC1981}
R~{Cowsik}.
\newblock {A new torsion balance for studies in gravitation and cosmology.}
\newblock {\em Indian Journal of Physics}, 55B:487, 1981.

\bibitem{RC1982-1}
R~{Cowsik}.
\newblock {Challenges in experimental gravitation and cosmology.}
\newblock Technical report, Dept. of Science \& Technology, New Delhi, 1982.

\bibitem{RC1990}
R~{Cowsik}.
\newblock {Search for new forces in nature.}
\newblock In {\em From Mantle to Meteorites (Prof. D.Lal Festschrift)}, 1990.

\bibitem{RC1988-2}
R.~{Cowsik}, N.~{Krishnan}, P.~{Sarawat}, S.~N {Tandon}, and S.~{Unnikrishnan}.
\newblock {The Fifth Force Experiment at the TIFR}.
\newblock In {\em Gravitational Measurements, Fundamental Metrology and
  Constants}, 1988.

\bibitem{RC1989-3}
R.~{Cowsik}, N.~{Krishnan}, P.~{Sarawat}, S.~N {Tandon}, and S.~{Unnikrishnan}.
\newblock {Limits on the strength of the fifth-force.}
\newblock In {\em Advances in Space Research (Proc. XXI COSPAR, Espoo)}, 1989.

\bibitem{RC1988-PRL}
R.~{Cowsik}, N.~{Krishnan}, S.~N. {Tandon}, and C.~S. {Unnikrishnan}.
\newblock {Limit on the strength of intermediate-range forces coupling to
  isospin}.
\newblock {\em Physical Review Letters}, 61:2179--2181, November 1988.

\bibitem{RC1982-2}
R.~{Cowsik}, S.~N {Tandon}, and N.~{Krishnan}.
\newblock {Sensitive test of Equivalence Principle.}
\newblock Technical report, Dept. of Science \& Technology, New Delhi, 1982.

\bibitem{Cullen1999}
S.~{Cullen} and M.~{Perelstein}.
\newblock {SN 1987A Constraints on Large Compact Dimensions}.
\newblock {\em Physical Review Letters}, 83:268--271, July 1999.

\bibitem{Derj60}
B.~V {Derjaguin}.
\newblock {The Force Between Molecules}.
\newblock {\em Sci. Am.}, 203:47, 1960.

\bibitem{Derj54}
B.~V {Derjaguin} and J.~J. {Abrikossova}.
\newblock {\em Disc. Faraday Soc.}, 18:33, 1954.

\bibitem{Dimo96}
S.~{Dimopoulos} and G.~F. {Giudice}.
\newblock {Macroscopic Forces from Supersymmetry}.
\newblock {\em Phys. Lett. B}, 379:105, 1996.

\bibitem{Genet-T}
C.~{Genet}.
\newblock {\em La Force de Casimir Entre Deux Miroirs Metalliques \`{A}
  t\'{e}mperature Non Nulle}.
\newblock PhD thesis, Laboratoire Kastler Brossel, University of Paris, Paris,
  Italy, 1999.

\bibitem{Reynaud2000}
C.~{Genet}, A.~{Lambrecht}, and S.~{Reynaud}.
\newblock {Temperature dependence of the Casimir effect between metallic
  mirrors}.
\newblock {\em Phy.Rev. A}, 62:012110, July 2000.

\bibitem{Hanhart2001-1}
C.~{Hanhart}, D.~R. {Phillips}, S.~{Reddy}, and M.~{Savage}.
\newblock {Extra dimensions, SN1987a, and nucleon-nucleon scattering data}.
\newblock {\em Nuclear Physics B}, 595:335--359, February 2001.

\bibitem{Hanhart2001-2}
C.~{Hanhart}, J.~A. {Pons}, D.~R. {Phillips}, and S.~{Reddy}.
\newblock {The likelihood of GODs' existence: improving the SN 1987a constraint
  on the size of large compact dimensions}.
\newblock {\em Physics Letters B}, 509:1--2, June 2001.

\bibitem{Hoyle2001}
C.~D. {Hoyle}, U.~{Schmidt}, B.~R. {Heckel}, E.~G. {Adelberger}, J.~H.
  {Gundlach}, D.~J. {Kapner}, and H.~E. {Swanson}.
\newblock {Submillimeter Test of the Gravitational Inverse-Square Law: A Search
  for ``Large'' Extra Dimensions}.
\newblock {\em Physical Review Letters}, 86:1418--1421, February 2001.

\bibitem{Rey2002}
M.~{Jaekel}, A.~{Lambrecht}, and S.~{Reynaud}.
\newblock {Quantum vacuum, inertia and gravitation}.
\newblock {\em New Astronomy Review}, 46:727--739, November 2002.

\bibitem{Kap00}
D.~B. {Kaplan} and M.~B. {Wise}.
\newblock {Coupling of the light Dialoton and Violations of the Equivalance
  Principle}.
\newblock {\em J. High Energy Phys.}, 08:37, 2000.

\bibitem{Krishnan89-T}
N.~Krishnan.
\newblock {\em Search for Intermidiate Range forces Weaker than Gravity}.
\newblock PhD thesis, Tata Institute of Fundamental Research, Mumbai, India,
  1989.

\bibitem{Kulza21}
T.~{Kulza}.
\newblock {\em Preuss. Akad. Wiss.}, page 966, 1921.

\bibitem{Lamor97}
S.~K. {Lamoreaux}.
\newblock {Demonstration of the Casimir Force in the 0.6 to 6 $\mu$ Range}.
\newblock {\em Phys. Rev. Lett}, 78:5, 1997.

\bibitem{Lift56}
E.~M. {Lifshitz}.
\newblock Theory of molecular attraction between solids.
\newblock {\em Sov. Phys. JETP}, 2:73, 1956.

\bibitem{Price2003}
J.~C. {Long} et~al.
\newblock {Upper Limit on Submillimeter-range Forces from Extra Space-time
  dimensions}.
\newblock {\em Nature}, 421:924, 2003.

\bibitem{Miloni94}
W.~P {Milonni}.
\newblock {\em {The Quantum Vacuum: An Introduction to Quantum
  Electrodynamics}}.
\newblock Academic Press, Newyork, 1994.

\bibitem{Mohi98}
U.~{Mohideen} and A~{Roy}.
\newblock {Precision Measurement of the Casimir Force from 0.1 to 0.9 $\mu$m
  Range}.
\newblock {\em Phys. Rev. Lett}, 81:4549, 1998.

\bibitem{Most2003}
V.~M. {Mostepanenko}.
\newblock {Constraints on Non-Newtonian Gravity from Recent Casimir Force
  Measurements}.
\newblock {\em ArXiv General Relativity and Quantum Cosmology e-prints},
  November 2003.

\bibitem{Sparnaay52}
J.~T.~G. {Overbeek} and M.~J. {Sparnaay}.
\newblock {\em Proc. K. Ned. Akad. Wet.}, 54:387, 1952.

\bibitem{Paddy2003}
T.~{Padmanabhan}.
\newblock {Cosmological constant-the weight of the vacuum}.
\newblock {\em Phys. Rep.}, 380:235--320, July 2003.

\bibitem{Perlm1999}
S.~{Perlmutter} et~al.
\newblock {Measurements of Omega and Lambda from 42 High-Redshift Supernovae}.
\newblock {\em Astrophys. J}, 517:565--586, June 1999.

\bibitem{Randall02}
L.~{Randall}.
\newblock {Extra Dimensions and Warped Geometries}.
\newblock {\em Science}, 296:1422, 2002.

\bibitem{Riess1998}
A.~G. {Riess} et~al.
\newblock {Observational Evidence from Supernovae for an Accelerating Universe
  and a Cosmological Constant}.
\newblock {\em Astron. J}, 116:1009--1038, September 1998.

\bibitem{Sahni2002}
V.~{Sahni}.
\newblock {The cosmological constant problem and quintessence}.
\newblock {\em Classical and Quantum Gravity}, 19:3435--3448, July 2002.

\bibitem{Schwin78}
J.~{Schwinger}, L.~L. {DeRaad, Jr.}, and K.~A. {Milton}.
\newblock {Casimir Effect in Dielectrics}.
\newblock {\em Ann. Phys. NY}, 115:1, 1978.

\bibitem{Sparnaay57}
M.~J. {Sparnaay}.
\newblock {Attractive Forces Between Flat Plates}.
\newblock {\em Nature}, 180:334, 1957.

\bibitem{Sparnaay89}
M.~J. {Sparnaay}.
\newblock {Historical Background of Casimir force}.
\newblock In A.~{Sarlemijn} and M.~J. {Sparnaay}, editors, {\em {Physics in the
  Making}}, North-Holland, 1989. Elsevier Science Publishers B V.

\bibitem{Straumann2002}
N.~{Straumann}.
\newblock {The History of the Cosmological COnstant Problem}.
\newblock {\em ArXiv General Relativity and Quantum Cosmology e-prints}, August
  2008.

\bibitem{Unni92}
C.~S. Unnikrishnan.
\newblock {\em Torsion Balance Experiments to Search for New Composition
  Dependant Forces}.
\newblock PhD thesis, Tata Institute of Fundamental Research, Mumbai, India,
  1992.

\bibitem{Overbeek78}
P.~H.~G.~M {van Blokland} and J.~T.~G. {Overbeek}.
\newblock {van der Waals Forces Between Objects Covered with Chromium Layer}.
\newblock {\em J. Chem. Soc. Faraday Trans. 72}, 72:2637, 1978.

\bibitem{Wein1989}
S.~{Weinberg}.
\newblock {The cosmological constant problem}.
\newblock {\em Reviews of Modern Physics}, 61:1--23, January 1989.

\bibitem{Wein2001}
S.~{Weinberg}.
\newblock {The Cosmological Constant Problems}.
\newblock In {\em Sources and Detection of Dark Matter and Dark Energy in the
  Universe}, page~18, 2001.

\end{thebibliography}


\begin{thebibliography}{1}

\bibitem{RC1997}
R~{Cowsik}.
\newblock {Torsion balances and their application to the study of gravitation
  at Gauribidanur.}
\newblock In {\em New Challenges in Astrophysics. Special Volume of the IUCAA
  Dedication Seminar}, 1997.

\bibitem{MG8}
R.~{Cowsik}, B.~P. {Das}, N.~{Krishnan}, G.~{Rajalakshmi}, D.~{Suresh}, and
  C.~S. {Unnikrishnan}.
\newblock {High Sensitivity Measurement of Casimir Force and Observability of
  Finite Temperature Effects}.
\newblock In {\em {Proceedings of the Eighth Marcell-Grossmann Meeting on
  General Relativity}}, page 949. World Scientific, 1998.

\bibitem{Gillies93}
G.~T. {Gillies} and R.~C. {Ritter}.
\newblock {Torsion balances, torsion pendulums, and related devices}.
\newblock {\em Review of Scientific Instruments}, 64:283--309, 1993.

\bibitem{Unni95}
C.~S. {Unnikrishnan}.
\newblock {Observability of the Casimir force at macroscopic distances: A
  proposal}.
\newblock {\em Unpublished, TIFR Preprint}, G-EXP/95/11, 1995.

\end{thebibliography}


\begin{thebibliography}{1}

\bibitem{Krishnan89}
R.~{Cowsik} et~al.
\newblock {Torsion Balance Experiment for Measurement of Weak Forces in
  Nature}.
\newblock {\em {Indian Journal of Pure \& Applied Physics}}, 27:691--709, 1989.

\bibitem{Jones1951}
R.~V. {Jones}.
\newblock {Some Points in the Design of Optical Levers and Amplifiers}.
\newblock {\em Proceedings of the Physical Society B}, 64:469--482, June 1951.

\bibitem{Jones1959}
R.~V. {Jones} and J.~C.~S. {Richards}.
\newblock {CORRESPONDENCE: Recording optical lever}.
\newblock {\em Journal of Scientific Instruments}, 36:199, April 1959.

\end{thebibliography}


\begin{thebibliography}{10}

\bibitem{Adel2003}
E.~G. {Adelberger}, B.~R. {Heckel}, and A.~E. {Nelson}.
\newblock {Tests of the Gravitational Inverse-Square Law}.
\newblock {\em ArXiv High Energy Physics - Phenomenology e-prints}, July 2003.

\bibitem{Proximity}
J.~{Blocki} et~al.
\newblock {}.
\newblock {\em Ann. Phys. (N.Y.)}, 105:427, 1977.

\bibitem{Ruoso02}
G.~{Bressi} et~al.
\newblock {Measurement of the Casimir Force between Parallel Metallic
  Surfaces}.
\newblock {\em Phys. Rev. Lett}, 88:041804--1, 2002.

\bibitem{Chan01}
H.~B. {Chan} et~al.
\newblock {Quantum Mechanical Actuation of Micromechanical Systems by the
  Casimir Force}.
\newblock {\em Science}, 291:9, 2001.

\bibitem{Derj60}
B.~V {Derjaguin}.
\newblock {The Force Between Molecules}.
\newblock {\em Sci. Am.}, 203:47, 1960.

\bibitem{Derj54}
B.~V {Derjaguin} and J.~J. {Abrikossova}.
\newblock {\em Disc. Faraday Soc.}, 18:33, 1954.

\bibitem{Reynaud2000}
C.~{Genet}, A.~{Lambrecht}, and S.~{Reynaud}.
\newblock {Temperature dependence of the Casimir effect between metallic
  mirrors}.
\newblock {\em Phy.Rev. A}, 62:012110, July 2000.

\bibitem{Lamor97}
S.~K. {Lamoreaux}.
\newblock {Demonstration of the Casimir Force in the 0.6 to 6 $\mu$ Range}.
\newblock {\em Phys. Rev. Lett}, 78:5, 1997.

\bibitem{Mohi98}
U.~{Mohideen} and A~{Roy}.
\newblock {Precision Measurement of the Casimir Force from 0.1 to 0.9 $\mu$m
  Range}.
\newblock {\em Phys. Rev. Lett}, 81:4549, 1998.

\bibitem{Smythe39}
W.~R {Smythe}.
\newblock {\em {Static and Dynamic Electricity}}.
\newblock McGraw Hill, 1939.

\bibitem{Sparnaay57}
M.~J. {Sparnaay}.
\newblock {Attractive Forces Between Flat Plates}.
\newblock {\em Nature}, 180:334, 1957.

\bibitem{Speake1996}
C.~C. {Speake}.
\newblock {Forces and force gradients due to patch fields and contact-potential
  differences}.
\newblock {\em Classical and Quantum Gravity}, 13:291--, November 1996.

\bibitem{Speake2003}
C.~C. {Speake} and C.~{Trenkel}.
\newblock {Forces between Conducting Surfaces due to Spatial Variations of
  Surface Potential}.
\newblock {\em Physical Review Letters}, 90:160403, 2003.

\bibitem{Overbeek78}
P.~H.~G.~M {van Blokland} and J.~T.~G. {Overbeek}.
\newblock {van der Waals Forces Between Objects Covered with Chromium Layer}.
\newblock {\em J. Chem. Soc. Faraday Trans. 72}, 72:2637, 1978.

\end{thebibliography}


\begin{thebibliography}{10}

\bibitem{Bordag01}
M.~{Bordag}, U.~{Mohideen}, and V.~M. {Mostepanenko}.
\newblock {New Developements in Casimir Force}.
\newblock {\em Phys. Rep.}, 353:1, 2001.

\bibitem{Ruoso02}
G.~{Bressi} et~al.
\newblock {Measurement of the Casimir Force between Parallel Metallic
  Surfaces}.
\newblock {\em Phys. Rev. Lett}, 88:041804--1, 2002.

\bibitem{Cas48}
H.~B. {Casimir}.
\newblock {On the Attraction Between Two Perfectly Conducting Plates}.
\newblock {\em Proc. K. Ned. Akad. Wet.}, 51:793, 1948.

\bibitem{Chan01}
H.~B. {Chan} et~al.
\newblock {Quantum Mechanical Actuation of Micromechanical Systems by the
  Casimir Force}.
\newblock {\em Science}, 291:9, 2001.

\bibitem{Mg8}
R.~{Cowsik}, B.~P. {Das}, N.~{Krishnan}, G.~{Rajalakshmi}, D.~{Suresh}, and
  C.~S. {Unnikrishnan}.
\newblock {High Sensitivity Measurement of Casimir Force and Observability of
  Finite Temperature Effects}.
\newblock In {\em {Proceedings of the Eighth Marcell-Grossmann Meeting on
  General Relativity}}, page 949. World Scientific, 1998.

\bibitem{Decca03E}
R.~{Decca}, D.~{L{\' o}pez}, E.~{Fischbach}, and D.~{Krause}.
\newblock {Measurement of the Casimir Force between Dissimilar Metals}.
\newblock {\em Phys. Rev. Lett.}, 91:050402, July 2003.

\bibitem{Derj60}
B.~V {Derjaguin}.
\newblock {The Force Between Molecules}.
\newblock {\em Sci. Am.}, 203:47, 1960.

\bibitem{Derj54}
B.~V {Derjaguin} and J.~J. {Abrikossova}.
\newblock {\em Disc. Faraday Soc.}, 18:33, 1954.

\bibitem{Tabor1972b}
{Israelachvili} and {Tabor}.
\newblock {The Measurement of van der Waals dispersion forces in the range 1.5
  to 130nm}.
\newblock {\em Proc. R. Soc. Lond. A}, 331:39--55, 1972.

\bibitem{Tabor1972a}
{Israelachvili} and {Tabor}.
\newblock {Van der Waals dispersion forces-Measurements for curved mica
  surfaces separated by between 1.4 and 70 nm}.
\newblock {\em Nature Physical Science}, 236:106, April 1972.

\bibitem{Lamor97}
S.~K. {Lamoreaux}.
\newblock {Demonstration of the Casimir Force in the 0.6 to 6 $\mu$ Range}.
\newblock {\em Phys. Rev. Lett}, 78:5, 1997.

\bibitem{Lift56}
E.~M. {Lifshitz}.
\newblock Theory of molecular attraction between solids.
\newblock {\em Sov. Phys. JETP}, 2:73, 1956.

\bibitem{Miloni94}
W.~P {Milonni}.
\newblock {\em {The Quantum Vacuum: An Introduction to Quantum
  Electrodynamics}}.
\newblock Academic Press, Newyork, 1994.

\bibitem{Mohi98}
U.~{Mohideen} and A~{Roy}.
\newblock {Precision Measurement of the Casimir Force from 0.1 t0 0.9 $\mu$m
  Range}.
\newblock {\em Phys. Rev. Lett}, 81:4549, 1998.

\bibitem{Sparnaay52}
J.~T.~G. {Overbeek} and M.~J. {Sparnaay}.
\newblock {\em Proc. K. Ned. Akad. Wet.}, 54:387, 1952.

\bibitem{Sparnaay57}
M.~J. {Sparnaay}.
\newblock {Attractive Forces Between Flat Plates}.
\newblock {\em Nature}, 180:334, 1957.

\bibitem{Sparnaay89}
M.~J. {Sparnaay}.
\newblock {Historical Background of Casimir force}.
\newblock In A.~{Sarlemijn} and M.~J. {Sparnaay}, editors, {\em {Physics in the
  Making}}, North-Holland, 1989. Elsevier Science Publishers B V.

\bibitem{Overbeek78}
P.~H.~G.~M {van Blokland} and J.~T.~G. {Overbeek}.
\newblock {van der Waals Forces Between Objects Covered with Chromium Layer}.
\newblock {\em J. Chem. Soc. Faraday Trans. 72}, 72:2637, 1978.

\end{thebibliography}


\begin{thebibliography}{10}

\bibitem{Adel2003}
E.~G. {Adelberger}, B.~R. {Heckel}, and A.~E. {Nelson}.
\newblock {Tests of the Gravitational Inverse-Square Law}.
\newblock {\em Ann. Rev. of Nuclear and Particle Science}, 53:77-121 2003.

\bibitem{ADD1999}
N.~{Arkani-Hamed}, S.~{Dimopoulos}, and G.~{Dvali}.
\newblock {Phenomenology, astrophysics, and cosmology of theories with
  submillimeter dimensions and TeV scale quantum gravity}.
\newblock {\em Phys. Rev. D}, 59:086004, April 1999.

\bibitem{Bordag1998}
M.~{Bordag}, B.~{Geyer}, G.~L. {Klimchitskaya}, and V.~M. {Mostepanenko}.
\newblock {Constraints for hypothetical interactions from a recent
  demonstration of the Casimir force and some possible improvements}.
\newblock {\em Phys. Rev. D}, 58:075003, October 1998.

\bibitem{Bordag01}
M.~{Bordag}, U.~{Mohideen}, and V.~M. {Mostepanenko}.
\newblock {New Developements in Casimir Force}.
\newblock {\em Phys. Rep.}, 353:1, 2001.

\bibitem{RC1981}
R~{Cowsik}.
\newblock {A new torsion balance for studies in gravitation and cosmology.}
\newblock {\em Indian Journal of Physics}, 55B:487, 1981.

\bibitem{RC1982-1}
R~{Cowsik}.
\newblock {Challenges in experimental gravitation and cosmology.}
\newblock Technical report, Dept. of Science \& Technology, New Delhi, 1982.

\bibitem{RC1990}
R~{Cowsik}.
\newblock {Search for new forces in nature.}
\newblock In {\em From Mantle to Meteorites (Prof. D.Lal Festschrift)}, 1990.

\bibitem{RC1988-2}
R.~{Cowsik}, N.~{Krishnan}, P.~{Sarawat}, S.~N {Tandon}, and S.~{Unnikrishnan}.
\newblock {The Fifth Force Experiment at the TIFR}.
\newblock In {\em Gravitational Measurements, Fundamental Metrology and
  Constants}, 1988.

\bibitem{RC1989-3}
R.~{Cowsik}, N.~{Krishnan}, P.~{Sarawat}, S.~N {Tandon}, and S.~{Unnikrishnan}.
\newblock {Limits on the strength of the fifth-force.}
\newblock In {\em Advances in Space Research (Proc. XXI COSPAR, Espoo)}, 1989.

\bibitem{RC1988-PRL}
R.~{Cowsik}, N.~{Krishnan}, S.~N. {Tandon}, and C.~S. {Unnikrishnan}.
\newblock {Limit on the strength of intermediate-range forces coupling to
  isospin}.
\newblock {\em Physical Review Letters}, 61:2179--2181, November 1988.

\bibitem{RC1982-2}
R.~{Cowsik}, S.~N {Tandon}, and N.~{Krishnan}.
\newblock {Sensitive test of Equivalence Principle.}
\newblock Technical report, Dept. of Science \& Technology, New Delhi, 1982.

\bibitem{Cullen1999}
S.~{Cullen} and M.~{Perelstein}.
\newblock {SN 1987A Constraints on Large Compact Dimensions}.
\newblock {\em Physical Review Letters}, 83:268--271, July 1999.

\bibitem{Decca03C}
R.~S. {Decca}, E.~{Fischbach}, G.~L. {Klimchitskaya}, D.~E. {Krause}, D.~{L{\'
  o}pez}, and V.~M. {Mostepanenko}.
\newblock {Improved tests of extra-dimensional physics and thermal quantum
  field theory from new Casimir force measurements}.
\newblock {\em Phys. Rev. D}, 68:116003, December 2003.

\bibitem{Fischbach01}
E.~{Fischbach}, D.~E. {Krause}, V.~M. {Mostepanenko}, and M.~{Novello}.
\newblock {New constraints on ultrashort-ranged Yukawa interactions from atomic
  force microscopy}.
\newblock {\em Phys. Rev. D}, 64:075010, October 2001.

\bibitem{Fischbach1999}
E.~{Fischbach} and C.~L. {Talmadge}.
\newblock {\em {The Search for Non-Newtonian Gravity}}.
\newblock The Search for Non-Newtonian Gravity, XVII, 305 pp.~58 figs..~
  Springer-Verlag New York, 1999.

\bibitem{Genet-T}
C.~{Genet}.
\newblock {\em La Force de Casimir Entre Deux Miroirs Metalliques \`{A}
  t\'{e}mperature Non Nulle}.
\newblock PhD thesis, Laboratoire Kastler Brossel, University of Paris, Paris,
  Italy, 1999.

\bibitem{Reynaud2000}
C.~{Genet}, A.~{Lambrecht}, and S.~{Reynaud}.
\newblock {Temperature dependence of the Casimir effect between metallic
  mirrors}.
\newblock {\em Phy.Rev. A}, 62:012110, July 2000.

\bibitem{Rey2000-fluc}
F.~{Grassia}, J.-M. {Courty}, S.~{Reynaud}, and P.~{Touboul}.
\newblock {Quantum theory of fluctuations in a cold damped accelerometer}.
\newblock {\em European Physical Journal D}, 8:101--110, 2000.

\bibitem{Adel1997}
J.~H. {Gundlach}, G.~L. {Smith}, E.~G. {Adelberger}, B.~R. {Heckel}, and H.~E.
  {Swanson}.
\newblock {Short-Range Test of the Equivalence Principle}.
\newblock {\em Physical Review Letters}, 78:2523--2526, March 1997.

\bibitem{Hanhart2001-1}
C.~{Hanhart}, D.~R. {Phillips}, S.~{Reddy}, and M.~{Savage}.
\newblock {Extra dimensions, SN1987a, and nucleon-nucleon scattering data}.
\newblock {\em Nuclear Physics B}, 595:335--359, February 2001.

\bibitem{Hanhart2001-2}
C.~{Hanhart}, J.~A. {Pons}, D.~R. {Phillips}, and S.~{Reddy}.
\newblock {The likelihood of GODs' existence: improving the SN 1987a constraint
  on the size of large compact dimensions}.
\newblock {\em Physics Letters B}, 509:1--2, June 2001.

\bibitem{Hoyle2001}
C.~D. {Hoyle}, U.~{Schmidt}, B.~R. {Heckel}, E.~G. {Adelberger}, J.~H.
  {Gundlach}, D.~J. {Kapner}, and H.~E. {Swanson}.
\newblock {Submillimeter Test of the Gravitational Inverse-Square Law: A Search
  for ``Large'' Extra Dimensions}.
\newblock {\em Physical Review Letters}, 86:1418--1421, February 2001.

\bibitem{Krishnan89-T}
N.~Krishnan.
\newblock {\em Search for Intermidiate Range forces Weaker than Gravity}.
\newblock PhD thesis, Tata Institute of Fundamental Research, Mumbai, India,
  1989.

\bibitem{Lamor97}
S.~K. {Lamoreaux}.
\newblock {Demonstration of the Casimir Force in the 0.6 to 6 $\mu$ Range}.
\newblock {\em Phys. Rev. Lett}, 78:5, 1997.

\bibitem{Price2003}
J.~C. {Long} et~al.
\newblock {Upper Limit on Submillimeter-range Forces from Extra Space-time
  dimensions}.
\newblock {\em Nature}, 421:924, 2003.

\bibitem{Most2003}
V.~M. {Mostepanenko}.
\newblock {Constraints on Non-Newtonian Gravity from Recent Casimir Force
  Measurements}.
\newblock {\em ArXiv General Relativity and Quantum Cosmology e-prints},
  November 2003.

\bibitem{Unni92}
C.~S. Unnikrishnan.
\newblock {\em Torsion Balance Experiments to Search for New Composition
  Dependant Forces}.
\newblock PhD thesis, Tata Institute of Fundamental Research, Mumbai, India,
  1992.

\end{thebibliography}


\begin{thebibliography}{10}

\bibitem{Abromo71}
M.~{Abramowitz} and {Stegun}~I. A.
\newblock {\em {Handbook of Mathematical Functions}}.
\newblock Dover Books, Newyork, 1971.

\bibitem{Adel2002}
E.~G. Adelberger.
\newblock {Sub-millimeter Tests of the Gravitational Inverse Square Law}.
\newblock {\em http://arXiv.org/abs}, hep-ex/0202008, 2002.

\bibitem{Adel2003}
E.~G. {Adelberger}, {Heckel}, and {Nelson} A.~E. B.~R.
\newblock {Tests of the Gravitational Inverse sqare law}.
\newblock {\em Ann. Rev. of Nuclear and Particle Science}, 53:77, December
  2003.

\bibitem{ADD2}
I.~{Antoniadis}, N.~{Arkani-Hamed}, S.~{Dimopoulos}, and G.~{Dvali}.
\newblock {New Dimension at a Millimeter to a Fermi and Superstrings at a TeV}.
\newblock {\em Phys. Lett. B}, 436:257, 1998.

\bibitem{Hamed02}
N.~{Arkani-Hamed}.
\newblock {Large Extra Dimensions: A new Arena for Particle Physics}.
\newblock {\em Physics Today}, page~35, Feb. 2002.

\bibitem{ADD1}
N.~{Arkani-Hamed}, S.~{Dimopoulos}, and G.~{Dvali}.
\newblock {The Hierarchy Problem and New Dimension at a Millimeter}.
\newblock {\em Phys. Lett. B}, 429:268, 1998.

\bibitem{ADD1999}
N.~{Arkani-Hamed}, S.~{Dimopoulos}, and G.~{Dvali}.
\newblock {Phenomenology, astrophysics, and cosmology of theories with
  submillimeter dimensions and TeV scale quantum gravity}.
\newblock {\em Phys. Rev. D}, 59:086004, April 1999.

\bibitem{Barton79}
G.~{Barton}.
\newblock {Some Surface Effects in the Hydrodynamic Model of Metals}.
\newblock {\em Rep. Prog. Phys.}, 42:963, 1979.

\bibitem{Proximity}
J.~{Blocki} et~al.
\newblock {}.
\newblock {\em Ann. Phys. (N.Y.)}, 105:427, 1977.

\bibitem{Boas83}
M.~L {Boas}.
\newblock {\em {Mathematical Methods in the Physical Sciences}}.
\newblock Wiley, Newyork, 1983.

\bibitem{Bordag1998}
M.~{Bordag}, B.~{Geyer}, G.~L. {Klimchitskaya}, and V.~M. {Mostepanenko}.
\newblock {Constraints for hypothetical interactions from a recent
  demonstration of the Casimir force and some possible improvements}.
\newblock {\em Phys. Rev. D}, 58:075003, October 1998.

\bibitem{Bordag2000}
M.~{Bordag}, B.~{Geyer}, G.~L. {Klimchitskaya}, and V.~M. {Mostepanenko}.
\newblock {Casimir Force at Both Nonzero Temperature and Finite Conductivity}.
\newblock {\em Physical Review Letters}, 85:503--506, July 2000.

\bibitem{Bordag01}
M.~{Bordag}, U.~{Mohideen}, and V.~M. {Mostepanenko}.
\newblock {New Developements in Casimir Force}.
\newblock {\em Phys. Rep.}, 353:1, 2001.

\bibitem{Ruoso02}
G.~{Bressi} et~al.
\newblock {Measurement of the Casimir Force between Parallel Metallic
  Surfaces}.
\newblock {\em Phys. Rev. Lett}, 88:041804--1, 2002.

\bibitem{Stein2000}
R.~R {Caldwell} and P~J {Steinhardt}.
\newblock Quintessence.
\newblock {\em Physics Worlds}, page~31, Nov 2000.

\bibitem{Cas48}
H.~B. {Casimir}.
\newblock {On the Attraction Between Two Perfectly Conducting Plates}.
\newblock {\em Proc. K. Ned. Akad. Wet.}, 51:793, 1948.

\bibitem{Chan01}
H.~B. {Chan} et~al.
\newblock {Quantum Mechanical Actuation of Micromechanical Systems by the
  Casimir Force}.
\newblock {\em Science}, 291:9, 2001.

\bibitem{RC1981}
R~{Cowsik}.
\newblock {A new torsion balance for studies in gravitation and cosmology.}
\newblock {\em Indian Journal of Physics}, 55B:487, 1981.

\bibitem{RC1982-1}
R~{Cowsik}.
\newblock {Challenges in experimental gravitation and cosmology.}
\newblock Technical report, Dept. of Science \& Technology, New Delhi, 1982.

\bibitem{RC1990}
R~{Cowsik}.
\newblock {Search for new forces in nature.}
\newblock In {\em From Mantle to Meteorites (Prof. D.Lal Festschrift)}, 1990.

\bibitem{RC1997}
R~{Cowsik}.
\newblock {Torsion balances and their application to the study of gravitation
  at Gauribidanur.}
\newblock In {\em New Challenges in Astrophysics. Special Volume of the IUCAA
  Dedication Seminar}, 1997.

\bibitem{MG8}
R.~{Cowsik}, B.~P. {Das}, N.~{Krishnan}, G.~{Rajalakshmi}, D.~{Suresh}, and
  C.~S. {Unnikrishnan}.
\newblock {High Sensitivity Measurement of Casimir Force and Observability of
  Finite Temperature Effects}.
\newblock In {\em {Proceedings of the Eighth Marcell-Grossmann Meeting on
  General Relativity}}, page 949. World Scientific, 1998.

\bibitem{Krishnan89}
R.~{Cowsik} et~al.
\newblock {Torsion Balance Experiment for Measurement of Weak Forces in
  Nature}.
\newblock {\em {Indian Journal of Pure \& Applied Physics}}, 27:691--709, 1989.

\bibitem{RC1988-2}
R.~{Cowsik}, N.~{Krishnan}, P.~{Sarawat}, S.~N {Tandon}, and S.~{Unnikrishnan}.
\newblock {The Fifth Force Experiment at the TIFR}.
\newblock In {\em Gravitational Measurements, Fundamental Metrology and
  Constants}, 1988.

\bibitem{RC1989-3}
R.~{Cowsik}, N.~{Krishnan}, P.~{Sarawat}, S.~N {Tandon}, and S.~{Unnikrishnan}.
\newblock {Limits on the strength of the fifth-force.}
\newblock In {\em Advances in Space Research (Proc. XXI COSPAR, Espoo)}, 1989.

\bibitem{RC1988-PRL}
R.~{Cowsik}, N.~{Krishnan}, S.~N. {Tandon}, and C.~S. {Unnikrishnan}.
\newblock {Limit on the strength of intermediate-range forces coupling to
  isospin}.
\newblock {\em Physical Review Letters}, 61:2179--2181, November 1988.

\bibitem{RC1982-2}
R.~{Cowsik}, S.~N {Tandon}, and N.~{Krishnan}.
\newblock {Sensitive test of Equivalence Principle.}
\newblock Technical report, Dept. of Science \& Technology, New Delhi, 1982.

\bibitem{Cullen1999}
S.~{Cullen} and M.~{Perelstein}.
\newblock {SN 1987A Constraints on Large Compact Dimensions}.
\newblock {\em Physical Review Letters}, 83:268--271, July 1999.

\bibitem{Decca03E}
R.~{Decca}, D.~{L{\' o}pez}, E.~{Fischbach}, and D.~{Krause}.
\newblock {Measurement of the Casimir Force between Dissimilar Metals}.
\newblock {\em Phys. Rev. Lett.}, 91:050402, July 2003.

\bibitem{Decca03C}
R.~S. {Decca}, E.~{Fischbach}, G.~L. {Klimchitskaya}, D.~E. {Krause}, D.~{L{\'
  o}pez}, and V.~M. {Mostepanenko}.
\newblock {Improved tests of extra-dimensional physics and thermal quantum
  field theory from new Casimir force measurements}.
\newblock {\em Phys. Rev. D}, 68:116003, December 2003.

\bibitem{Derj60}
B.~V {Derjaguin}.
\newblock {The Force Between Molecules}.
\newblock {\em Sci. Am.}, 203:47, 1960.

\bibitem{Derj54}
B.~V {Derjaguin} and J.~J. {Abrikossova}.
\newblock {\em Disc. Faraday Soc.}, 18:33, 1954.

\bibitem{Dimo96}
S.~{Dimopoulos} and G.~F. {Giudice}.
\newblock {Macroscopic Forces from Supersymmetry}.
\newblock {\em Phys. Lett. B}, 379:105, 1996.

\bibitem{Fischbach01}
E.~{Fischbach}, D.~E. {Krause}, V.~M. {Mostepanenko}, and M.~{Novello}.
\newblock {New constraints on ultrashort-ranged Yukawa interactions from atomic
  force microscopy}.
\newblock {\em Phys. Rev. D}, 64:075010, October 2001.

\bibitem{Fischbach1999}
E.~{Fischbach} and C.~L. {Talmadge}.
\newblock {\em {The Search for Non-Newtonian Gravity}}.
\newblock The Search for Non-Newtonian Gravity, XVII, 305 pp.~58 figs..~
  Springer-Verlag New York, 1999.

\bibitem{Genet-T}
C.~{Genet}.
\newblock {\em La Force de Casimir Entre Deux Miroirs Metalliques \`{A}
  t\'{e}mperature Non Nulle}.
\newblock PhD thesis, Laboratoire Kastler Brossel, University of Paris, Paris,
  Italy, 1999.

\bibitem{Reynaud2000}
C.~{Genet}, A.~{Lambrecht}, and S.~{Reynaud}.
\newblock {Temperature dependence of the Casimir effect between metallic
  mirrors}.
\newblock {\em Phy.Rev. A}, 62:012110, July 2000.

\bibitem{Gillies93}
G.~T. {Gillies} and R.~C. {Ritter}.
\newblock {Torsion balances, torsion pendulums, and related devices}.
\newblock {\em Review of Scientific Instruments}, 64:283--309, 1993.

\bibitem{Rey2000-fluc}
F.~{Grassia}, J.-M. {Courty}, S.~{Reynaud}, and P.~{Touboul}.
\newblock {Quantum theory of fluctuations in a cold damped accelerometer}.
\newblock {\em European Physical Journal D}, 8:101--110, 2000.

\bibitem{Adel1997}
J.~H. {Gundlach}, G.~L. {Smith}, E.~G. {Adelberger}, B.~R. {Heckel}, and H.~E.
  {Swanson}.
\newblock {Short-Range Test of the Equivalence Principle}.
\newblock {\em Physical Review Letters}, 78:2523--2526, March 1997.

\bibitem{Hanhart2001-1}
C.~{Hanhart}, D.~R. {Phillips}, S.~{Reddy}, and M.~{Savage}.
\newblock {Extra dimensions, SN1987a, and nucleon-nucleon scattering data}.
\newblock {\em Nuclear Physics B}, 595:335--359, February 2001.

\bibitem{Hanhart2001-2}
C.~{Hanhart}, J.~A. {Pons}, D.~R. {Phillips}, and S.~{Reddy}.
\newblock {The likelihood of GODs' existence: improving the SN 1987a constraint
  on the size of large compact dimensions}.
\newblock {\em Physics Letters B}, 509:1--2, June 2001.

\bibitem{Hoyle2001}
C.~D. {Hoyle}, U.~{Schmidt}, B.~R. {Heckel}, E.~G. {Adelberger}, J.~H.
  {Gundlach}, D.~J. {Kapner}, and H.~E. {Swanson}.
\newblock {Submillimeter Test of the Gravitational Inverse-Square Law: A Search
  for ``Large'' Extra Dimensions}.
\newblock {\em Physical Review Letters}, 86:1418--1421, February 2001.

\bibitem{Tabor1972b}
{Israelachvili} and {Tabor}.
\newblock {The Measurement of van der Waals dispersion forces in the range 1.5
  to 130nm}.
\newblock {\em Proc. R. Soc. Lond. A}, 331:39--55, 1972.

\bibitem{Tabor1972a}
{Israelachvili} and {Tabor}.
\newblock {Van der Waals dispersion forces-Measurements for curved mica
  surfaces separated by between 1.4 and 70 nm}.
\newblock {\em Nature Physical Science}, 236:106, April 1972.

\bibitem{Rey2002}
M.~{Jaekel}, A.~{Lambrecht}, and S.~{Reynaud}.
\newblock {Quantum vacuum, inertia and gravitation}.
\newblock {\em New Astronomy Review}, 46:727--739, November 2002.

\bibitem{Jones1951}
R.~V. {Jones}.
\newblock {Some Points in the Design of Optical Levers and Amplifiers}.
\newblock {\em Proceedings of the Physical Society B}, 64:469--482, June 1951.

\bibitem{Jones1959}
R.~V. {Jones} and J.~C.~S. {Richards}.
\newblock {CORRESPONDENCE: Recording optical lever}.
\newblock {\em Journal of Scientific Instruments}, 36:199, April 1959.

\bibitem{Kap00}
D.~B. {Kaplan} and M.~B. {Wise}.
\newblock {Coupling of the light Dialoton and Violations of the Equivalance
  Principle}.
\newblock {\em J. High Energy Phys.}, 08:37, 2000.

\bibitem{Krishnan89-T}
N.~Krishnan.
\newblock {\em Search for Intermidiate Range forces Weaker than Gravity}.
\newblock PhD thesis, Tata Institute of Fundamental Research, Mumbai, India,
  1989.

\bibitem{Kulza21}
T.~{Kulza}.
\newblock {\em Preuss. Akad. Wiss.}, page 966, 1921.

\bibitem{Lamor97}
S.~K. {Lamoreaux}.
\newblock {Demonstration of the Casimir Force in the 0.6 to 6 $\mu$ Range}.
\newblock {\em Phys. Rev. Lett}, 78:5, 1997.

\bibitem{Lift56}
E.~M. {Lifshitz}.
\newblock Theory of molecular attraction between solids.
\newblock {\em Sov. Phys. JETP}, 2:73, 1956.

\bibitem{Price2003}
J.~C. {Long} et~al.
\newblock {Upper Limit on Submillimeter-range Forces from Extra Space-time
  dimensions}.
\newblock {\em Nature}, 421:924, 2003.

\bibitem{Miloni94}
W.~P {Milonni}.
\newblock {\em {The Quantum Vacuum: An Introduction to Quantum
  Electrodynamics}}.
\newblock Academic Press, Newyork, 1994.

\bibitem{Mohi98}
U.~{Mohideen} and A~{Roy}.
\newblock {Precision Measurement of the Casimir Force from 0.1 to 0.9 $\mu$m
  Range}.
\newblock {\em Phys. Rev. Lett}, 81:4549, 1998.

\bibitem{Most2003}
V.~M. {Mostepanenko}.
\newblock {Constraints on Non-Newtonian Gravity from Recent Casimir Force
  Measurements}.
\newblock {\em ArXiv General Relativity and Quantum Cosmology e-prints},
  November 2003.

\bibitem{Sparnaay52}
J.~T.~G. {Overbeek} and M.~J. {Sparnaay}.
\newblock {\em Proc. K. Ned. Akad. Wet.}, 54:387, 1952.

\bibitem{Paddy2003}
T.~{Padmanabhan}.
\newblock {Cosmological constant-the weight of the vacuum}.
\newblock {\em Phys. Rep.}, 380:235--320, July 2003.

\bibitem{Perlm1999}
S.~{Perlmutter} et~al.
\newblock {Measurements of Omega and Lambda from 42 High-Redshift Supernovae}.
\newblock {\em Astrophys. J}, 517:565--586, June 1999.

\bibitem{Randall02}
L.~{Randall}.
\newblock {Extra Dimensions and Warped Geometries}.
\newblock {\em Science}, 296:1422, 2002.

\bibitem{Riess1998}
A.~G. {Riess} et~al.
\newblock {Observational Evidence from Supernovae for an Accelerating Universe
  and a Cosmological Constant}.
\newblock {\em Astron. J}, 116:1009--1038, September 1998.

\bibitem{Sahni2002}
V.~{Sahni}.
\newblock {The cosmological constant problem and quintessence}.
\newblock {\em Classical and Quantum Gravity}, 19:3435--3448, July 2002.

\bibitem{Schwin78}
J.~{Schwinger}, L.~L. {DeRaad, Jr.}, and K.~A. {Milton}.
\newblock {Casimir Effect in Dielectrics}.
\newblock {\em Ann. Phys. NY}, 115:1, 1978.

\bibitem{Smythe39}
W.~R {Smythe}.
\newblock {\em {Static and Dynamic Electricity}}.
\newblock McGraw Hill, 1939.

\bibitem{Sparnaay57}
M.~J. {Sparnaay}.
\newblock {Attractive Forces Between Flat Plates}.
\newblock {\em Nature}, 180:334, 1957.

\bibitem{Sparnaay89}
M.~J. {Sparnaay}.
\newblock {Historical Background of Casimir force}.
\newblock In A.~{Sarlemijn} and M.~J. {Sparnaay}, editors, {\em {Physics in the
  Making}}, North-Holland, 1989. Elsevier Science Publishers B V.

\bibitem{Speake1996}
C.~C. {Speake}.
\newblock {Forces and force gradients due to patch fields and contact-potential
  differences}.
\newblock {\em Classical and Quantum Gravity}, 13:291--, November 1996.

\bibitem{Speake2003}
C.~C. {Speake} and C.~{Trenkel}.
\newblock {Forces between Conducting Surfaces due to Spatial Variations of
  Surface Potential}.
\newblock {\em Physical Review Letters}, 90:160403, 2003.

\bibitem{Straumann2002}
N.~{Straumann}.
\newblock {The History of the Cosmological COnstant Problem}.
\newblock {\em ArXiv General Relativity and Quantum Cosmology e-prints}, August
  2008.

\bibitem{Unni92}
C.~S. Unnikrishnan.
\newblock {\em Torsion Balance Experiments to Search for New Composition
  Dependant Forces}.
\newblock PhD thesis, Tata Institute of Fundamental Research, Mumbai, India,
  1992.

\bibitem{Unni95}
C.~S. {Unnikrishnan}.
\newblock {Observability of the Casimir force at macroscopic distances: A
  proposal}.
\newblock {\em Unpublished, TIFR Preprint}, G-EXP/95/11, 1995.

\bibitem{Overbeek78}
P.~H.~G.~M {van Blokland} and J.~T.~G. {Overbeek}.
\newblock {van der Waals Forces Between Objects Covered with Chromium Layer}.
\newblock {\em J. Chem. Soc. Faraday Trans. 72}, 72:2637, 1978.

\bibitem{Wein1989}
S.~{Weinberg}.
\newblock {The cosmological constant problem}.
\newblock {\em Reviews of Modern Physics}, 61:1--23, January 1989.

\bibitem{Wein2001}
S.~{Weinberg}.
\newblock {The Cosmological Constant Problems}.
\newblock In {\em Sources and Detection of Dark Matter and Dark Energy in the
  Universe}, page~18, 2001.

\end{thebibliography}

\clearpage

\chapter{Yukawa Force on a Gold Coated Lens due to Gold Coated Plate}
Let the Yukawa potential due to a point mass $m$ at a distance $r$
from the mass be given by,
\begin{eqnarray}
Y(r)  & = &- m G \alpha \rho \frac{e^{-r/\lambda}}{r}.
\end{eqnarray}

\begin{figure}
\begin{center}
  \resizebox{10cm}{!}{\includegraphics
  {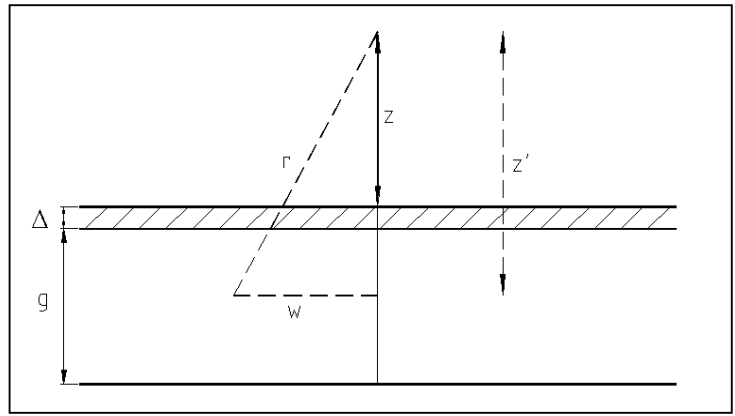}}
\end{center}\vspace*{-1cm}
  \caption{Schematic of the plate with coating.}\label{plate-Yu}
\end{figure}

Consider a glass plate of thickness, $g$ and density, $\rho_v$,
coated with a thin layer of gold of thickness, $\Delta$ and
density, $\rho_s$. The Yukawa potential due to the plate at a
height, $z$ [Fig.~\ref{plate-Yu}] from the surface of the plate is
obtained by integrating the contribution due to various mass
elements on the plate,
\begin{eqnarray}
  U(z)  &=& - G \alpha \int \rho(r)\ \frac{e^{-r/\lambda}}{r} dV;\\
  \mathrm{where} \quad r^2 &=& w^2 + z'^2. \\
  U(z) &=& - G \alpha
             \int \rho(z')\ \frac{e^{-\sqrt{(w^2 + z'^2)}/\lambda}}{\sqrt{(w^2 +
             z'^2)}} 2 \pi w\ dw\ dz' \\
     & = & - G \alpha 2 \pi  \int_{z+g+\Delta}^{z} dz'\ \rho(z')
                \int_{z'}^{\infty} dr\ e^{-r/\lambda}\\
     & = & - 2 \pi G \alpha \lambda \int_{z+g+\Delta}^{z} dz'
     \rho(z') e^{-z'/\lambda}\\
     & = & - 2 \pi G \alpha \lambda
        \left(  \rho_v\ \int_{z+g+\Delta}^{z+\Delta} dz'\ e^{-z'/\lambda} %
         + \rho_s\ \int_{z+\Delta}^{z} dz'\ e^{-z'/\lambda} \right) \\
      & = &  2 \pi G \alpha \lambda^2%
           \left[ \rho_v \left\{ e^{-(z+\Delta)/\lambda} -%
                            e^{-(z+g+\Delta)/\lambda} \right\} \right. + \nonumber\\%
      & & \hspace*{2cm}   \left.    \rho_s \left\{ e^{-z/\lambda} -%
                            e^{-(z+\Delta)/\lambda} \right\}%
           \right].\\
      \mathrm{For}\ \lambda  << g , \\
      U(z) &\approx&  2 \pi G \alpha \lambda^2%
                     \left[ \rho_v e^{-(z+\Delta)/\lambda} %
                          + \rho_s \left\{ e^{-z/\lambda} - e^{-(z+\Delta)/\lambda} \right\}%
                     \right]. \label{Uz}
\end{eqnarray}

Keeping in mind that our interest is in the region $\lambda  << g$
we use Eqn.~\ref{Uz} in further analysis. The force due to this
potential is,
\begin{eqnarray}
f(z) & = & - \frac{\partial U}{\partial z} \\
& = & 2 \pi G \alpha \lambda \left[ \rho_v e^{-(z+\Delta)/\lambda} %
                    + \rho_s \left\{ e^{-z/\lambda} - e^{-(z+\Delta)/\lambda} \right\}%
                     \right]\\
& = & 2 \pi G \alpha \lambda \left[ \rho_s -
                                   (\rho_s - \rho_v)e^{-\Delta/\lambda}\right]
                                   e^{-z/\lambda}\\
& = &  f_c e^{-z/\lambda}; \\
\mathrm{where}\ f_c & = & 2 \pi G \alpha \lambda \left[ \rho_s -
                                   (\rho_s -
                                   \rho_v)e^{-\Delta/\lambda}\right].
\end{eqnarray}
\begin{figure}
\begin{center}
  \resizebox{10cm}{!}{\includegraphics
  {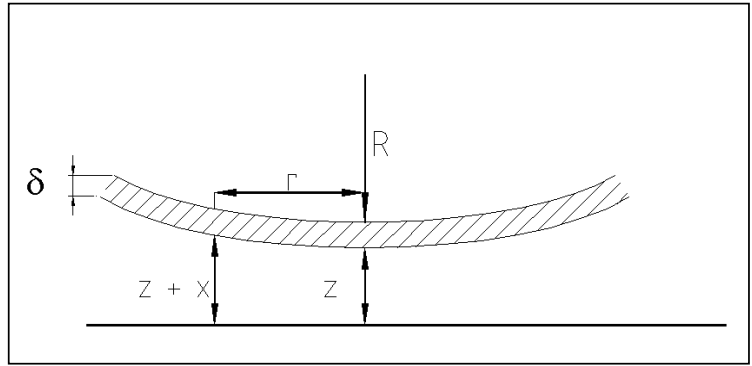}}
\end{center}\vspace*{-0.5cm}
  \caption{Schematic of the lens with coating.}\label{plate-lens-Yu}
\end{figure}
The lens has a radius of curvature $R$, a volume density $\rho_l$
and a surface coating of uniform thickness $\delta$ with density
$\rho_c$. We now proceed to derive the force on the lens in two
limiting cases:
(a) $\lambda << \delta $ and (b)  $\lambda >> \delta $. \\

\underline{Case (a): $\lambda << \delta $} Here the lens may be
assumed to be completely made of the coating material alone.\\
\begin{eqnarray}
\mathcal{F}_a(z)& = & \int f(z')\ \rho_c\ dV_l; \\
& = & \int f_c e^{-(z+x)/\lambda} \rho_c\ \delta\ dA(x);\\
&& \mathrm{where}\ r^2 = 2 R x.\\
\mathcal{F}_a(z)& = & f_c\ \rho_c\ \delta\ e^{-z/\lambda}%
                \int_0^{r_{max}} e^{-r^2/2R\lambda} 2 \pi r dr\\
& = & f_c\ \rho_c\ \delta\ e^{-z/\lambda}\ 2 \pi R \lambda %
         \left ( 1 - e^{-r_{max}^2/2R\lambda} \right )\\
\mathrm{For}\quad r_{max}^2/2R >> \lambda, && \\
\mathcal{F}_a(z)& \approx & f_c\ 2 \pi R \lambda \rho_c\
                             \delta\ e^{-z/\lambda}. \label{f-coating}
\end{eqnarray}

\underline{Case (b): $\lambda >> \delta $} This case contains two
terms one for the contribution from the coating and the other for
the body of the lens. \\
\textit{b1. Contribution from the coating:} This is same as in
case (a) and is given by Eqn.\ref{f-coating}.
\begin{eqnarray}
\mathcal{F}_{b1}(z)& = & f_c\ 2 \pi R \lambda \rho_c\
                             \delta\ e^{-z/\lambda}.
\end{eqnarray}

\textit{b2. Contribution from the lens:}
\begin{eqnarray}
\mathcal{F}_{b2}(z)& = & \int f(z')\ \rho_l\ dV_l; \\
& = & \int f_c\ e^{-(z+\delta+x)/\lambda}\ \rho_l\ \pi r^2 dx,
\end{eqnarray}
\begin{eqnarray}%
& = & f_c\ e^{-(z+\delta)/\lambda}\ \rho_l %
            \int_0^{r_{max}^2/2R} e^{-x/\lambda}\ 2 \pi R x dx\\
& = & f_c\ e^{-(z+\delta)/\lambda}\ \rho_l\ 2 \pi R \lambda
\nonumber\\ && \hspace*{0.75cm}
    \left [ - \frac{r_{max}^2}{2R}\ e^{r_{max}^2/2R\lambda}
            +  \lambda \left ( 1 - e^{r_{max}^2/2R\lambda} \right )%
    \right ].\\
\mathrm{For}\quad r_{max}^2/2R >> \lambda ,&& \\
\mathcal{F}_{b2}(z)& \approx & f_c\ e^{-(z+\delta)/\lambda}\
                         \rho_l\ 2 \pi R \lambda.\label{f-lens}
\end{eqnarray}
The net force due to Yukawa potential between the lens and the
plate for $\lambda >> \delta $ is given by,
\begin{eqnarray}
\mathcal{F}_b & = & \mathcal{F}_{b1} + \mathcal{F}_{b2} \\
& = & f_c\ 2 \pi R \lambda%
         \left(\ \rho_c\ \delta + \rho_l\ \lambda\ e^{-\delta/\lambda}%
       \  \right)\ e^{-z/\lambda}. \label{f-total}
\end{eqnarray}

In the case our experimental set up, the lens and the plate are
both made of the same material (BK7 glass) and coated with the
gold of equal thickness ($1~\mu$m). Hence,
\begin{equation}
\rho_l = \rho_v \ ; \ \rho_c = \rho_s\  \mathrm{and} \ \Delta =
\delta
\end{equation}
Thus, the force in the two cases is given by,
\begin{eqnarray}
\mathcal{F}_a(z) & = &  4 \pi^2 G \alpha R \lambda^2%
                     \left[ \rho_s - (\rho_s - \rho_v)%
                       e^{-\delta/\lambda}\right]\ %
                       \rho_s\ \delta\ e^{-z/\lambda},\quad %
                       \mathrm{for}\ \lambda  << \delta
                       \label{f-a}.\\
\mathcal{F}_{b}(z)& = & 4 \pi^2 G \alpha R \lambda^2%
                     \left[ \rho_s - (\rho_s - \rho_v)%
                       e^{-\delta/\lambda}\right]%
       \left(\ \rho_s\ \delta + \rho_v\ \lambda\ e^{-\delta/\lambda}%
       \  \right)\ e^{-z/\lambda}, \label{f-b}\\
&& \hspace*{8.5cm} \mathrm{for}\ \lambda >> \delta. \nonumber
\end{eqnarray}

\end{document}